\PassOptionsToPackage{table}{xcolor}
\documentclass[twocolumn]{aastex63}

\newcommand\aastex{AAS\TeX}

\usepackage{multirow,amsfonts,natbib,color,gensymb,longtable}

\usepackage{savesym}
\savesymbol{tablenum}
\usepackage{siunitx}
\restoresymbol{SIX}{tablenum}

\accepted{November 10, 2021}

\usepackage{graphicx,epsf}
\usepackage{subfigure}
\usepackage{graphicx}	
\usepackage{amsmath}	
\usepackage{amssymb}	
\usepackage{units}
\usepackage{newtxtext,newtxmath}
\usepackage{siunitx}
\usepackage{footnote}

\def \sn{SN\,2020pni}

\def \lam {$\lambda$}

\newcommand{\Ha}{H$\alpha$}
\newcommand{\Hb}{H$\beta$}

\newcommand{\Msun}{$M_\odot$}

\newcommand{\kms}{\textrm{km}\,\textrm{s}^{-1}}

\newcommand{\msunyr}{\,$M_{\sun}$\,yr$^{-1}$}

\def\arcmin{\hbox{$^\prime$}}
\def\arcsec{\hbox{$^{\prime\prime}$}}

\DeclareRobustCommand{\ion}[2]{\relax\ifmmode\ifx\testbx\f@series{\mathbf{#1\,\mathsc{#2}}}\else{\mathrm{#1\,\mathsc{#2}}}\fi\else\textup{#1\,{\mdseries\textsc{#2}}}\fi}

\graphicspath{{./}{figures/}}

 \shorttitle{Early Phases of SN 2020pni}
\shortauthors{Terreran et al.}

\begin{document}

\title{Template \aastex Article with Examples: 
v6.3\footnote{Released on June, 10th, 2019}}

\title{The Early Phases of Supernova 2020pni: Shock Ionization of the Nitrogen-enriched Circumstellar Material}

\correspondingauthor{Giacomo Terreran}
\email{giacomo.terreran@northwestern.edu}

\author[0000-0003-0794-5982]{G. Terreran}
\affiliation{Center for Interdisciplinary Exploration and Research in Astrophysics (CIERA), and Department of Physics and Astronomy, Northwestern University, Evanston, IL 60208, USA}

\author[0000-0003-1103-3409]{W.~V. Jacobson-Gal\'{a}n}
\affiliation{Center for Interdisciplinary Exploration and Research in Astrophysics (CIERA), and Department of Physics and Astronomy, Northwestern University, Evanston, IL 60208, USA}

\author[0000-0001-7675-3381]{J.~H. Groh}
\affil{Trinity College Dublin, The University of Dublin, Dublin, Ireland}

\author[0000-0003-4768-7586]{R. Margutti}
\affiliation{Center for Interdisciplinary Exploration and Research in Astrophysics (CIERA), and Department of Physics and Astronomy, Northwestern University, Evanston, IL 60208, USA}
\affiliation{Department of Astronomy, University of California, Berkeley, CA 94720-3411, USA}

\author[0000-0001-5126-6237]{D.~L. Coppejans}
\affiliation{Center for Interdisciplinary Exploration and Research in Astrophysics (CIERA), and Department of Physics and Astronomy, Northwestern University, Evanston, IL 60208, USA}

\author[0000-0001-9494-179X]{G. Dimitriadis}
\affiliation{Department of Astronomy and Astrophysics, University of California, Santa Cruz, CA 95064, USA}

\author[0000-0002-5740-7747]{C.~D. Kilpatrick}
\affiliation{Center for Interdisciplinary Exploration and Research in Astrophysics (CIERA), and Department of Physics and Astronomy, Northwestern University, Evanston, IL 60208, USA}

\author[0000-0002-4513-3849]{D.~J. Matthews}
\affiliation{Center for Interdisciplinary Exploration and Research in Astrophysics (CIERA), and Department of Physics and Astronomy, Northwestern University, Evanston, IL 60208, USA}

\author[0000-0003-2445-3891]{M.~R. Siebert}
\affiliation{Department of Astronomy and Astrophysics, University of California, Santa Cruz, CA 95064, USA}

\author[0000-0002-4269-7999]{C.~R. Angus}
\affiliation{DARK, Niels Bohr Institute, University of Copenhagen, Jagtvej 128, DK-2200 Copenhagen, Denmark}

\author[0000-0001-5955-2502]{T.~G. Brink}
\affiliation{Department of Astronomy, University of California, Berkeley, CA 94720-3411, USA}

\author[0000-0003-3460-0103]{A.~V. Filippenko}
\affiliation{Department of Astronomy, University of California, Berkeley, CA 94720-3411, USA}
\affiliation{Miller Institute for Basic Research in Science, University of California, Berkeley, CA 94720, USA.}

\author[0000-0002-2445-5275]{R.~J. Foley}
\affiliation{Department of Astronomy and Astrophysics, University of California, Santa Cruz, CA 95064, USA}

\author[0000-0002-6230-0151]{D.~O. Jones}
\affiliation{Department of Astronomy and Astrophysics, University of California, Santa Cruz, CA 95064, USA}

\author[0000-0002-1481-4676]{S. Tinyanont}
\affiliation{Department of Astronomy and Astrophysics, University of California, Santa Cruz, CA 95064, USA}

\author[0000-0002-8526-3963]{C.~Gall}
\affiliation{DARK, Niels Bohr Institute, University of Copenhagen, Jagtvej 128, DK-2200 Copenhagen, Denmark}

\author[0000-0003-0841-5182]{H. Pfister}
\affiliation{DARK, Niels Bohr Institute, University of Copenhagen, Jagtvej 128, DK-2200 Copenhagen, Denmark}
\affiliation{Department of Physics, The University of Hong Kong, Pokfulam Road, Hong Kong, China}

\author[0000-0002-0632-8897]{Y. Zenati}
\affil{Department of Physics and Astronomy, Johns Hopkins University, Baltimore, MD 21218, USA}

\author[0000-0002-4775-9685]{Z.~Ansari}
\affiliation{DARK, Niels Bohr Institute, University of Copenhagen, Jagtvej 128, DK-2200 Copenhagen, Denmark}

\author[0000-0002-4449-9152]{K.~Auchettl}
\affiliation{School of Physics, The University of Melbourne, VIC 3010, Australia}
\affiliation{Department of Astronomy and Astrophysics, University of California, Santa Cruz, CA 95064, USA}
\affiliation{DARK, Niels Bohr Institute, University of Copenhagen, Jagtvej 128, DK-2200 Copenhagen, Denmark}

\author[0000-0002-6871-1752]{K. El-Badry}
\affiliation{Department of Astronomy, and Theoretical Astrophysics Center, University of California, Berkeley, CA 94720-3411, USA}

\author[0000-0002-7965-2815]{E.~A.~Magnier}
\affiliation{Institute for Astronomy, University of Hawaii, 2680 Woodlawn Drive, Honolulu, HI 96822, USA}

\author[0000-0002-2636-6508]{W. Zheng}
\affiliation{Department of Astronomy, University of California, Berkeley, CA 94720-3411, USA}
\begin{abstract}

We present multiwavelength observations of the Type II SN\,2020pni. Classified at $\sim 1.3$\,days after explosion, the object showed narrow (FWHM intensity $<250\,\kms$) recombination lines of ionized helium, nitrogen, and carbon, as typically seen in flash-spectroscopy events. Using the non-LTE radiative transfer code CMFGEN to model our first high-resolution spectrum, we infer a progenitor mass-loss rate of $\dot{M}=(3.5-5.3)\times10^{-3}$\,\msunyr (assuming a wind velocity of $v_w=200\,\kms$), estimated at a radius of $R_{\rm in}=2.5\times10^{14}\,\rm{cm}$. In addition, we find that the progenitor of SN\,2020pni was enriched in helium and nitrogen (relative abundances in mass fractions of 0.30--0.40 and $8.2\times10^{-3}$, respectively). Radio upper limits are also consistent with dense circumstellar material (CSM) and a mass-loss rate of $\dot M>5 \times 10^{-4}$\,\msunyr. During the initial 4 days after first light, we also observe an increase in velocity of the hydrogen lines (from $\sim 250$ to $\sim 1000\,\kms$), suggesting complex CSM. The presence of dense and confined CSM, as well as its inhomogeneous structure, indicates a phase of enhanced mass loss of the progenitor of SN\,2020pni during the last year before explosion. Finally, we compare SN\,2020pni to a sample of other shock-photoionization events. We find no evidence of correlations among the physical parameters of the explosions and the characteristics of the CSM surrounding the progenitors of these events. This favors the idea that the mass loss experienced by massive stars during their final years could be governed by stochastic phenomena and that, at the same time, the physical mechanisms responsible for this mass loss must be common to a variety of different progenitors.

\end{abstract}


\section{Introduction} \label{sec:intro}
Strong winds or eruptive events are typical phenomena that lead massive stars to lose large amounts of material during the ark of their lives \citep[e.g.,][]{deJager1988,Vink2001,Mauron2011,Smith2014}. This phenomenon can lead to regions of high-density circumstellar material (CSM) in the immediate surroundings of the star. When the star then explodes as a core-collapse supernova (SN), the ejecta ram through this material and a double-shock structure is formed. Energetic photons are thus produced, which ionize the unshocked CSM in front of the ejecta \citep{Chevalier1994}. This material then recombines, emitting narrow lines, reflecting the low velocities of the CSM --- FWHM intensity on the order of a few hundred kilometers per second. These types of SNe are usually called SNe~IIn \citep{Schlegel1990,filippenko97}. 

Recent observations have brought to light a growing number of peculiar transitional objects that bridge the gap between SNe~IIn and more normal core-collapse explosions \citep[i.e., H-rich Type II and H-poor Type Ibc; e.g.,][]{Foley2007,Roming2012,Ofek2013,Gal-Yam2014,Margutti1409ip}. Some objects discovered very soon after explosion (hours to days) exhibit recombination signatures for only a few days, followed by a transition to normal SNe~II (e.g., \citealt{leonard2000,terreran16,Yaron17}). These events are sometimes dubbed \textit{flash-spectroscopy} SNe or \textit{shock-photoionization} SNe \citep{Gal-Yam2014,Khazov2016}, as the spectra typically show recombination lines of highly-ionized helium, carbon, nitrogen, and oxygen. In addition, the velocities measured from these narrow features have been found to be considerably larger than those typically associated with red supergiant (RSG) winds \citep[e.g.,][]{Groh14,Yaron17}, which are usually on the order of a few tens of kilometers per second \cite[e.g.,][]{Mauron2011}. Further modeling of these lines shows that mass-loss rates on the order of $10^{-4}$--$10^{-2}$\,\msunyr are necessary to reproduce the shock-ionization features \cite[e.g.,][]{Boian20}, in contrast to typical mass-loss rates of RSGs, $\sim 10^{-5}$\,\msunyr \citep[e.g.,][]{Smith2014}. The presence of dense CSM surrounding RSGs is also suggested by hydrodynamical models of hydrogen-rich SN light curves \citep{Morozova2018}. The interaction of the SN ejecta with this CSM is sometimes inferred by boxy and flat-topped \Ha{} and \Hb{} profiles that start to appear a few months after explosion \citep[e.g.,][]{Inserra2011,terreran16,Jerkstrand2017}. In addition, this material is found to be confined within a radius of $\sim$800--3000\,R$_\odot$ \citep[e.g.,][]{Morozova2018}. All of these facts provide strong evidence for a period of enhanced mass loss in RSG progenitors approaching their demise. 

The traditionally accepted scenario of stellar evolution envisions a single massive star (like an RSG) evolving unperturbed during its final $\sim 1000$\,yr prior to explosion \citep[e.g.,][]{Woosley2002}. The neutrino-cooled core keeps violently burning, while the outer envelope stays unperturbed. However, the shock-ionization features exhibited by some SNe, produced by the interaction of the ejecta with confined and nearby CSM, suggests otherwise, hinting toward a period of enhanced mass loss preceding the explosion.

What physical mechanism is responsible for this phenomenon is not clear, and several scenarios have been proposed. Evolved RSGs could eject part of the loosely bound envelope through nuclear flashes \citep{Weaver1979,Dessart2010,Woosley2015}. These are expected to be caused by dynamical burning triggered by oxygen, neon, or silicon igniting off-center. Such late-stage burning instabilities could easily explain the ejection of material in the months and years preceding an SN explosion. However, this phenomenon can occur only for stars of 8--12\,\Msun\ \citep{Weaver1979}. Different studies have shown that some of the objects presenting shock-ionization features have progenitors with zero-age main-sequence (ZAMS) masses above 12\,\Msun{} \citep[e.g.,][]{terreran16,Morozova2017,Morozova2018,Tartaglia21} so nuclear flashes cannot be invoked to explain all of the objects showing flash-spectroscopy features. Eruptive mass loss during the late evolution of RSGs can also be caused by large-amplitude pulsations, induced by partial ionization of hydrogen in the envelope \citep[e.g.,][]{Li1994,Heger1997,Yoon2010}. Alternatively, gravity waves could be a viable mechanism to unbind up to a few \Msun\ of material \citep{Quataert2012,Shiode2014,Fuller2017,Linial2021,Wu2021}. These waves are supposed to be created by the vigorous convection during carbon fusion and beyond. Consequently, the timing of this mechanism makes it a very good candidate to explain the enhanced mass loss in evolved massive stars. An alternative explanation is that the material that is interacting with the SN ejecta is not the result from a super-wind phase of the progenitor star; rather, it is rather a ``cocoon'' of stagnating material, composed actually by gas that does not reach the escape velocity \citep[e.g.,][]{Dessart2017,Soker2021}. This scenario removes the requirement for the fine-tuned stellar activity in the years immediately preceding the SN explosion, although this cannot explain the presence of dense material at larger distances, like the one showed by some of the shock-ionization objects \citep[e.g., SN~1998S;][]{Mauerhan2012}.

\begin{figure}[t]
\centering
\includegraphics[width=\columnwidth]{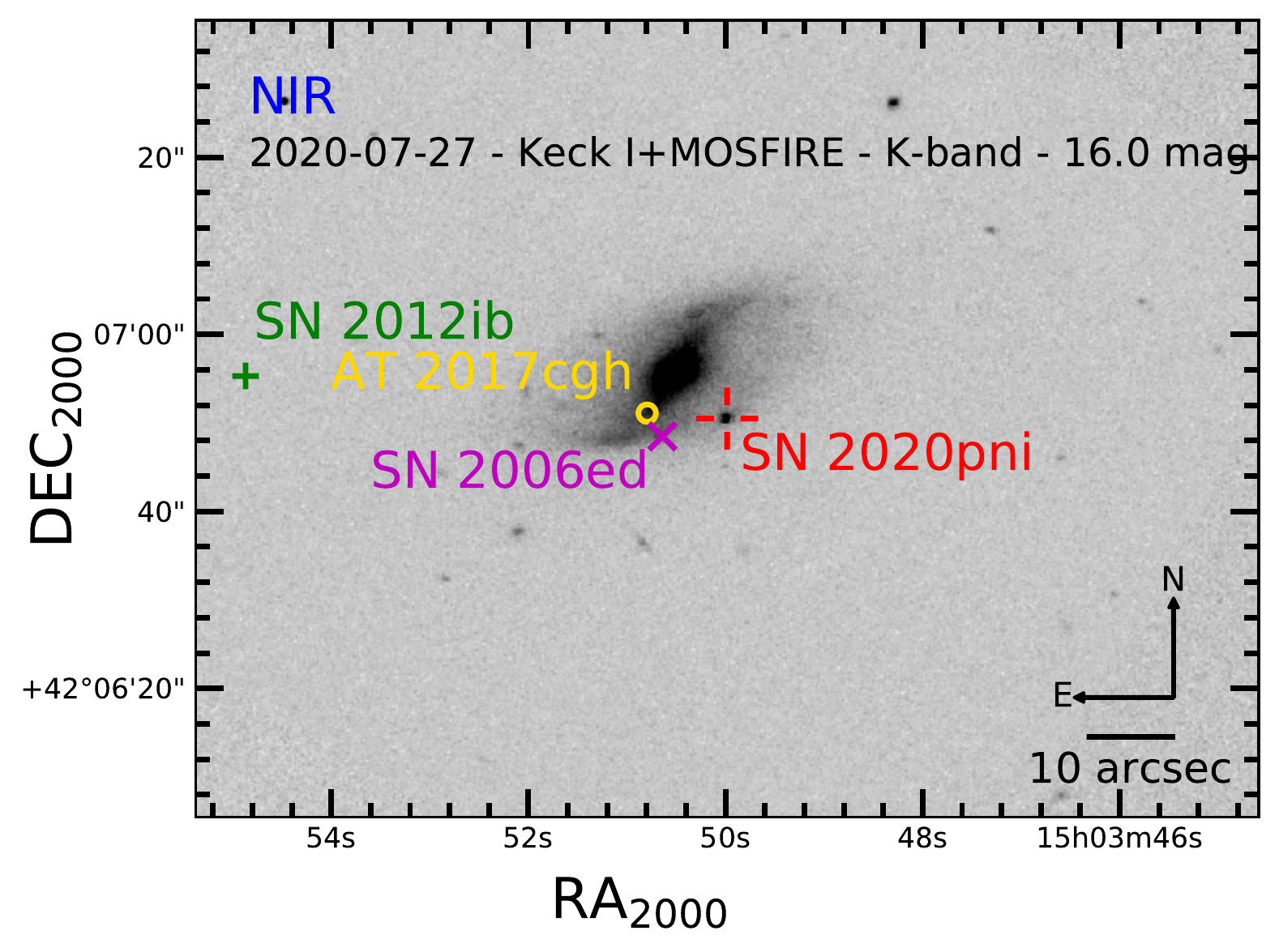} 
\caption{$K_s$-band image of \sn{} (marked with the red cross) and its host galaxy UGC 09684. We also report the positions of the other three transients discovered in UGC 09684: SN 2006ed (magenta), SN 2012ib (green cross), and AT 2017cgh (yellow circle). See the main text for references.}
\label{fig:host}
\end{figure}

The number of SNe showing narrow lines within the first week after explosion has been growing rapidly in recent times \cite[e.g.,][]{Boian20,Bruch21,Gangopadhyay20,Zhang20,Tartaglia21}. According to \cite{Khazov2016}, $\sim 20$\% of core-collapse SNe discovered within 5 days from explosion show flash-ionization features, while \cite{Bruch21} find that the fraction is 30\% for SNe observed within 2 days of explosion. In this context, we present a new addition to the class. \sn{} (also known as ATLAS20sxl, Gaia20dus, PS20fyg, ZTF20ablygyy) was discovered on 2020 July 16.19 (UT dates are used throughout this paper; MJD 59,046.19) by ALeRCE \citep{Forster2020}, using the Zwicky Transient Facility \citep[ZTF;][]{Kulkarni2018} data stream. The transient is located at coordinates \mbox{$\alpha_{\rm{J}2000}\,=\,15^{\rm{h}}03^{\rm{m}}49^{\rm{s}}.964$}, \mbox{$\delta_{\rm{J}2000}\,=\,+42\degr06\arcmin50\farcs52$}, sitting in the outskirts of the spiral galaxy UGC 09684 (see Figure~\ref{fig:host}). The last nondetection (by ZTF) was $< 24$\,hr earlier (MJD 59,045.25), placing a strong constraint on the time of first light (see Sec.~\ref{phot_analysis}). A spectroscopic classification was obtained by the ZTF collaboration on 2020 July 17.27 (MJD 59,047.27), $\sim 1$ day after discovery, using the Spectral Energy Distribution Machine \citep[SEDM;][]{Blagorodnova2018,Rigault2019} on the Palomar 60-inch (P60) telescope. Unresolved lines of hydrogen and ionized helium were detected \citep{Bruch2020b}, which are indicative of a young SN~II with flash-spectroscopy features. We started observing \sn{} on 2020 July 17.32 (MJD 59,047.32), $\sim 1.5$\,days since discovery, confirming the classification. In this paper we present the results from our optical observing campaign in the first $\sim 60$\,days since explosion, as well as our radio follow-up observations beyond 300\,days after first light.

In Section \ref{sec:host} we present the properties of the host galaxy UGC 09684. We describe our dataset and the data-reduction techniques in Section \ref{sec:obs}. The multiwavelength evolution of \sn{} is described in Section \ref{sec:analysis}, focusing in particular on the early spectroscopic evolution and modeling. Finally, we discuss the results in Section \ref{sec:discussion}, where we compare \sn{} to the population of flash-ionization events with the goal of constraining the progenitor properties.

\begin{table}
\caption{Parameters for \sn{} and Host Galaxy \mbox{UGC 09684}.}
\begin{tabular}{ll}
\hline
Discovery & MJD 59,046.19\\
Time of first light & MJD 59,045.80\\
Redshift $z$ & 0.01665\\
Distance (modulus $\mu$)&73.7\,Mpc (34.34\,mag)\\
$E(B-V)_{\rm{MW}}$&0.017\,mag\\
$E(B-V)_{\rm{host}}$&0.063\,mag\\
Host metallicity & $12 + \rm{log(O/H)}$ = 9.0\\
SFR & 0.25--0.39\,\Msun\,yr$^{-1}$\\
Stellar mass & (2.0--3.5) $\times 10^{10}$\,\Msun\\
\hline
\end{tabular}
\label{tab:params}
\end{table}

\newpage

\section{UGC 09684}
\label{sec:host}
The host of \sn{}, known as \mbox{UGC 09684}, is an SBac star-forming galaxy. It hosted at least three additional transients --- the Type II SN~2006ed \citep{Foley2006,Joubert2006}, the stripped-envelope SN~2012ib \citep{Lipunov2012,Tomasella2012}, and the unclassified transient AT~2017cgh \citep{Chambers2017}. \sn{} is the fourth confirmed SN-like event in \mbox{UGC 09684} during the past 15\,yr, marking a rate of SN production comparable to the most active ``SN factory'' galaxies.

From our highest-resolution spectrum of \sn{}, acquired on 2020 July 22.26 ($\sim 6.5$ days after first light) with \mbox{Keck~II}+ESI \citep{Sheinis2002}, we measure a redshift $z = 0.01665\pm0.00030$; corrected for Virgo infall, this corresponds to a distance of $73.7\pm1.3$~Mpc ($H_{0} = 73$\,km\,s$^{-1}$\,Mpc$^{-1}$, $\Omega_{\rm M} = 0.27$, $\Omega_{\rm \Lambda} = 0.73$), equivalent to a distance modulus $\mu =34.34\pm 0.15$\,mag (the errors are propagated from the redshift uncertainties). We can estimate the local reddening by measuring the equivalent width (EW) of the \ion{Na}{i} \lam\lam5890, 5896 doublet absorption at the redshift of the host galaxy \citep{Turatto2003b,Poznanski12}. From the \mbox{Keck~II}+ESI spectrum, we also infer an EW of $0.55\pm0.01$\,\AA\ for \ion{Na}{i}~D, which corresponds to a host-galaxy extinction of $E(B-V)_{\rm{host}}=0.063\pm0.010$\,mag. These values are consistent with measurements performed on spectra at later phases when the ejecta are no longer interacting with the nearby CSM (see Sec.~\ref{sec: line_ev}). We can thus confidently associate the \ion{Na}{i} absorption with an interstellar origin. The Milky Way color excess in the direction of \sn{} is $E(B-V)_{\rm{MW}}=0.017$\,mag \citep{schlafly11}.

We estimate the star-formation rate (SFR) of \mbox{UGC 09684} with the Fitting and Assessment of Synthetic Templates code \citep[FAST;][]{Kriek2009}. We used ultraviolet (UV, far-UV, near-UV), optical ($ugriz$), and near-infrared (NIR; $JHK_s$) luminosity measurements from the GALEX All-Sky Survey Source Catalog \citep[GASC;][]{Seibert2012}, the Sloan Digital Sky Survey (SDSS) Data Release 6,\footnote{\url{http://www.sdss.org/dr6/products/catalogs/index.html}} and the final release of the Two Micron All Sky Survey (2MASS) Extended Source Catalog \citep{Jarrett2000}. All of the data were retrieved from the NASA/IPAC Extragalactic Database (NED)\footnote{\url{http://ned.ipac.caltech.edu}}. In our initial grid of models, we considered both a \cite{Salpeter1955} and a \cite{Chabrier2003} stellar initial mass function (IMF). For the star-formation history (SFH), we employed an exponentially decreasing function (${\rm SFR} \propto {e}^{-t}$) and a delayed function as well (${\rm SFR} \propto t \times {e}^{-t}$). We also assumed a \cite{Calzetti2000} reddening law. Finally, we used the stellar population libraries of \cite{Bruzual2003} and \cite{Conroy2009}\footnote{Note that only the \cite{Chabrier2003} IMF was available for the latter library.}. Several metallicity estimates for \mbox{UGC 09684} have been published \citep[e.g.,][]{Prieto2008,Kelly2012}, the majority of them agreeing on a metallicity slightly above solar (oxygen abundance $12 + \rm{log(O/H)} \approx 9.0$, corresponding to $\sim${}$2\,\rm{Z}_\odot$). Therefore, we limited our search to stellar population libraries with above-solar metallicities ($Z>0.019$). The range of best-fitting SFRs for \mbox{UGC 09684} is 0.25--0.39\,\Msun\,yr$^{-1}$. We also infer a total stellar mass of $M_\star = (2.0$--3.5) $\times 10^{10}$\,\Msun, and hence a current specific SFR ${\rm sSFR} \approx 0.01$\,Gyr$^{-1}$. This is higher than what is found in the literature \citep[e.g.,][]{Kelly2012} but agrees with the relatively large number of recent events reported in this host galaxy.
We summarize all of the inferred and adopted parameters for \sn{} and \mbox{UGC 09684} in Table \ref{tab:params}.

\section{Observations of SN 2020pni} \label{sec:obs}

\subsection{UV/Optical/NIR Photometry}\label{photometry}

\begin{figure*}[t]
\centering
\includegraphics[width=\textwidth]{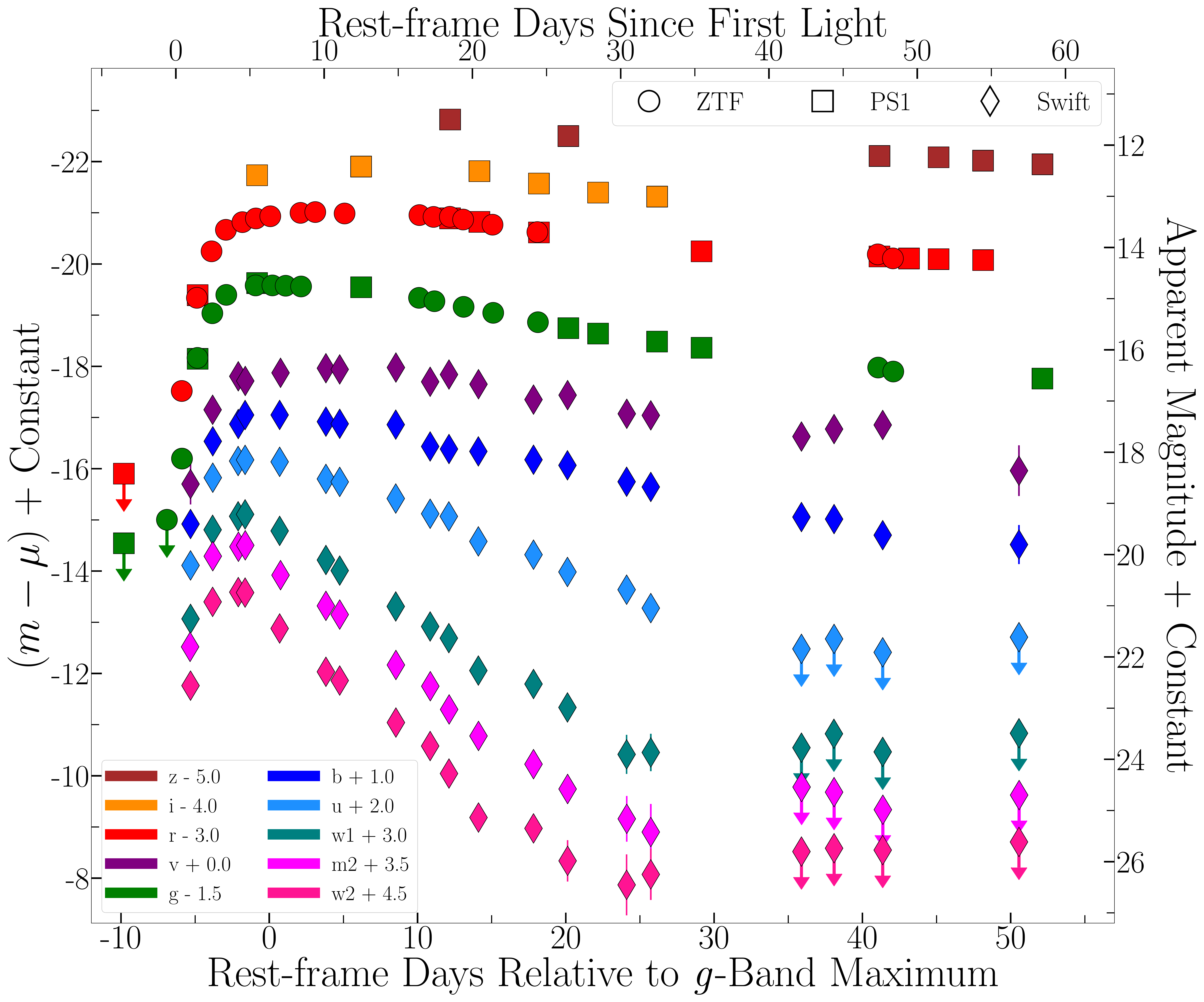} 
\caption{UV/optical/NIR light curve of \sn{} with respect to $g$-band maximum brightness. Observed photometry is presented in the AB magnitude system \citep{Oke83}. PS1 data and 3$\sigma$ upper limits are presented as squares, ZTF as circles, and Swift as diamonds.}
\label{fig:LC}
\end{figure*}

\begin{figure*}[t!]
\centering
\includegraphics[width=0.48\textwidth]{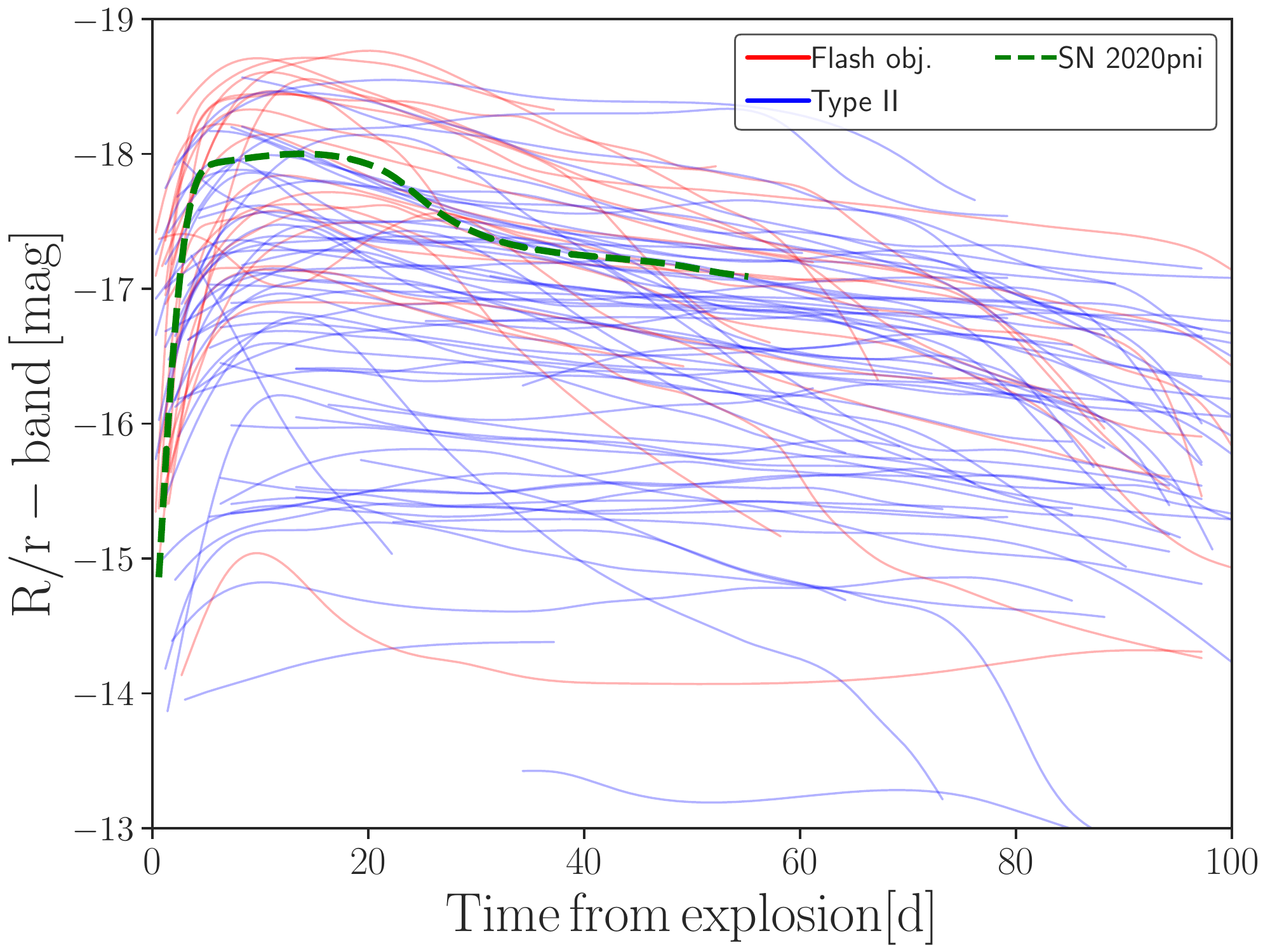}
\includegraphics[width=0.48\textwidth]{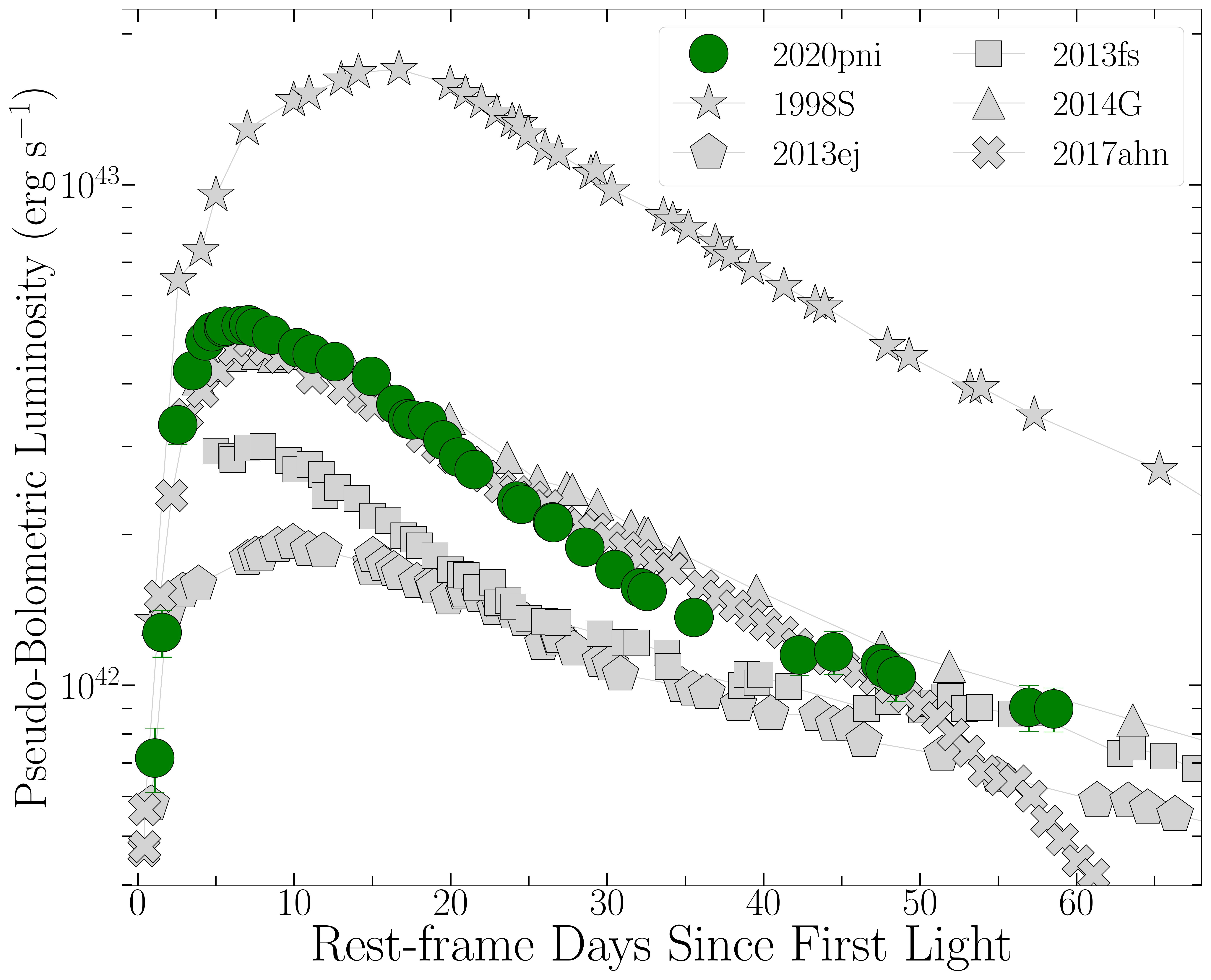}
\caption{
Left panel: comparison of the $r$-band light curve of \sn{} with those of other shock-ionization events (in red) and other ``normal'' SNe~II. The objects were taken from \cite{Khazov2016}, \cite{Bruch21}, and \cite{deJaeger2019}, with the addition of SN 1998S, SN 2016bkv, and SN 2017ahn \citep{leonard2000,Hosseinzadeh2018,Tartaglia21}. Right panel: pseudobolometric (3000--10,000\,\AA) light-curve comparison of \sn{} and other flash-spectroscopy SNe. \label{fig:bol_LCs}} 
\end{figure*}

We observed SN\,2020pni with the Ultraviolet Optical Telescope (UVOT; \citealt{Roming05}) on board the Neil Gehrels Swift Observatory \citep{Gehrels04} from 2020 July 16.8 until 2020 September 10.8 ($\delta t= 1.0$--56.9\,days since first light). We performed aperture photometry using a 3$\arcsec$-radius circular region with \texttt{uvotsource} within HEAsoft v6.26,\footnote{We used the calibration database (CALDB) version 20201008.} following the standard guidelines from \cite{Brown09}. In order to remove contamination from the host galaxy, we employed images acquired at $t\approx105$\,days after first light, assuming that the SN contribution is negligible. This is supported by visual inspection, in which we found no flux associated with \sn{}, although we cannot exclude some residuals given the relatively early phase of the SN, especially in the \textit{B} and \textit{V} bands. We subtracted the measured count rate at the location of the SN from the count rates in the SN images following the prescriptions of \cite{Brown14}. We detect UV emission from the earliest Swift epoch ($\delta t = 1.0$\,days; Figure~\ref{fig:LC}) until $\delta t\approx32$ days after first light. Subsequent nondetections in the $u$, $w1$, $m2$, and $w2$ bands indicate significant cooling of the photosphere.

Additional $griz$-band imaging of \sn{} was obtained through the Young Supernova Experiment (YSE) sky survey \citep{Jones2021} with the Pan-STARRS telescope \citep[PS1;][]{Kaiser2002} between 2020 July 17.3 and 2020 September 12.2 ($\delta t= 1.5$--58.4\,days since first light). The YSE photometric pipeline is based on {\tt photpipe} \citep{Rest05}. Each image template was taken from stacked PS1 exposures, with most of the input data from the PS1 3$\pi$ survey. All images and templates are resampled and astrometrically aligned to match a skycell in the PS1 sky tessellation. An image zero-point is determined by comparing point-spread function (PSF) photometry of the stars to updated stellar catalogs of PS1 observations \citep{Chambers2017}. The PS1 templates are convolved to match the nightly images, and the convolved templates are subtracted from the nightly images with {\tt HOTPANTS} \citep{becker15}. Finally, a flux-weighted centroid is found for each SN position, and PSF photometry is performed using ``forced photometry'': the centroid of the PSF is forced to be at the SN position. The nightly zero-point is applied to the photometry to determine the brightness of the SN for that epoch.

We obtained multiband NIR data for \sn{} on 2020 July 28 using the Multi-Object Spectrometer For Infra-Red Exploration (MOSFIRE; \citealt{mclean2012}) at Keck Observatory. We imaged the object using $JHK_s$ filters. Standard flat-fielding has been applied, and the instrumental magnitudes were extracted through PSF photometry. We used the 2MASS catalog\footnote{\url{http://www.ipac.caltech.edu/2mass/}} \citep{Skrutskie2006} for the flux calibration.

In addition to our own observations, we include $g$-band and $r$-band photometry from the ZTF public data stream \citep{bellm19, graham19}, which span from 2020 July 16.2 to 2020 September 2.2 ($\delta t= 0.4$--48.4\,days since first light). All photometric observations are listed in Tables \ref{tab: griz}--\ref{tab: NIR}.


\subsection{Optical/NIR Spectroscopy} \label{SubSec:SpecData}
The spectroscopic campaign of \sn{} started $\sim 1$\,day after discovery. Here we present the first 64\,days of evolution. All of the spectra were reduced using standard techniques, which included correction for bias, overscan, and flat field. Spectra of comparison lamps and standard stars acquired during the same night and with the same instrumental setting have been used for the wavelength and flux calibrations, respectively. When possible, we further removed the telluric bands using standard stars. Given the various instruments employed, the data-reduction steps described above have been applied using several instrument-specific routines. Data from Keck using the LRIS, DEIMOS, and MOSFIRE instruments were processed with the \textsc{PypeIt} software package \citep{Prochaska2020}. We used standard \textsc{IRAF}\footnote{IRAF is distributed by NOAO, which is operated by AURA, Inc., under cooperative agreement with the National Science Foundation (NSF).} commands to extract the spectra from GMOS, Binospec, and ESI data. The SEDM spectrum was downloaded directly from the Transient Name Server\footnote{\url{https://wis-tns.weizmann.ac.il/object/2020pni}} (TNS).

Spectra of \sn{} were obtained with the Kast spectrograph \citep{KAST} on the Shane 3\,m telescope at Lick Observatory on 2020 July 18, 19, and 28, August 9 and 11, and September 7 (programs 2020A-S008 and 2020B-S001, PI Filippenko; program 2020A-S011, PI Foley). Observations were made with the 452/3306 and the 600/4310 grisms on the blue arm and the 300/7500 grating on the red arm, using the 2.0\arcsec\ slit aligned along the parallactic angle to minimize the effects of atmospheric dispersion \citep{filippenko82}. Calibration observations (arc lamps, dome flats, and spectrophotometric standards) were performed on the same night. The data were reduced with standard IRAF/pyRAF and Python routines, including flat-fielding, determining the wavelength solution and small-scale wavelength corrections from night-sky lines, with flux calibration and telluric removal accomplished through the use of spectrophotometric standard stars. An additional spectrum was obtained on July 30 as part of the Lick Supernova Program (ToO) 2020A-S012 (PI Foley), using a different setup compared to the classical program (830/3460 grism, blue arm; 1200/5000 grating, red arm; 2.0\arcsec\ slit). The data were reduced in a similar manner.

We obtained spectra of \sn{} with the FLOYDS spectrograph on the Faulkes-N telescope at Haleakal\={a}, Hawaii, as part of the Las Cumbres Observatory \citep[LCO;][]{Brown13} network. The spectra were acquired on 2020 August 1, 4, 10, and 22 and on September 7, as shown in Table~\ref{tab: instr2}. All spectra were obtained with the 1.6\arcsec\ slit under nearly photometric conditions, and the slit was aligned along the parallactic angle. Comparison-lamp and dome-flat exposures were obtained immediately before and after each observation. Following standard procedures in {\tt pyraf}, we reduced all of these spectra using the FLOYDS pipeline \citep{Valenti2014}\footnote{\url{https://github.com/svalenti/FLOYDS_pipeline}}. This included standard image reductions, aperture extraction, flux calibration, wavelength calibration, telluric removal, and order combination.

A spectrum was also taken with the Alhambra Faint Object Spectrograph and Camera (ALFOSC) on the Nordic Optical Telescope (NOT) at La Palma on 2020 July 25, using grism 4 and a 1.0\arcsec\ slit, aligned along the parallactic angle, and under clear observing conditions and good seeing. The spectrum was reduced with a custom pipeline running standard {\tt pyraf} procedures. 

We obtained an NIR spectrum of SN\,2020pni in the $YJHK$ bands on 2020 July 28 using MOSFIRE at Keck Observatory.
The data were reduced using the MOSFIRE data-reduction pipeline,\footnote{\url{https://github.com/Keck-DataReductionPipelines/MosfireDRP}} which performed flat-field correction, wavelength calibration using night-sky lines and arc-lamp observations, and spectral extraction. 
We then used \texttt{xtellcor} \citep{vacca03} to perform flux calibration and telluric correction with observations of an A0V star HIP 71172.

A summary of all the telescopes, instruments, and configurations used for the spectroscopic observations of \sn{} is presented in Table \ref{tab: instr2}. All of the spectra shown will be available at the Weizmann Interactive Supernova data
REPository \citep[WISeREP;\footnote{\url{https://wiserep.weizmann.ac.il/}}][]{Yaron12}.

\subsection{X-rays}\label{SubSec:XrayData}
We started observing \sn{} with Swift-XRT \citep{burrows05} on 2020 July 16 until September 10 ($\delta t\approx1.3$--56 days since time of first light), for a total exposure time of 31.5\,ks. We reduced the data following standard practice using HEASOFT v6.28 and the latest Swift calibration files. We find no evidence for X-ray emission at the location of the SN in the individual exposures and in the merged event file, for which we infer a $3\sigma$ count-rate upper limit of $8.0\times 10^{-4}\,\rm{counts\,s^{-1}}$ (0.3--10\,keV). The neutral hydrogen column density in the direction of the SN is $1.40\times 10^{20}\,\rm{cm^{-2}}$ \citep{Kalberla05}. For a power-law spectrum $F_{\nu}\propto \nu^{-1}$, the corresponding unabsorbed flux limit is $F_x<2.9\times 10^{-14}\,\rm{erg\,s^{-1}cm^{-2}}$, which is $L_x<1.9\times 10^{40}\,\rm{erg\,s^{-1}}$ (0.3--10\,keV). Individual segments of observations have a typical exposure time of $\sim 1.6$\,ks, which leads to flux limits $F_x<2.0\times 10^{-13}\,\rm{erg\,s^{-1}\,cm^{-2}}$ ($L_x<1.2\times 10^{41}\,\rm{erg\,s^{-1}}$).

We emphasize that these limits are corrected only for the absorption component that originates in the Galaxy. However, the modeling of the optical spectra and early-time light curve (\S\ref{SubSec:environmentmodels}) indicates high densities in the local SN environment at distances $<10^{15}\,\rm{cm}$, from which we estimate large intrinsic absorption corresponding to $N_{\rm H-int} \gtrsim 10^{25}\,\rm{cm^{-2}}$ at the time of radiation breakout (assuming that a large fraction of the material is neutral). The lack of detected X-rays in \sn{} is thus most likely a consequence of the very large local absorption by the extended layer of CSM from which the H lines originate at early times. A later-time Swift-XRT observation was acquired on 2020 October 28 ($\delta t\approx 105$ days since time of first light, exposure time 3.5\,ks), from which we derive $L_x<9\times 10^{40}\,\rm{erg\,s^{-1}}$. The lack of detectable X-ray emission is consistent with the low-density, larger-scale environment inferred from the radio observations (\S\ref{SubSec:RadioData} and \S\ref{SubSubSec:RadioModeling}). 

\subsection{Radio}\label{SubSec:RadioData}
We observed \sn{} with the NSF's Karl G. Jansky Very Large Array (VLA) through our joint Fermi/VLA program SD1096/131096 (PI Margutti) on 2020 August 21.8 ($\delta t=37.0$ days after time of first light), 2020 November 21.5 ($\delta t=128.7$ days), and 2021 May 18.1 ($\delta t=306.3$ days). We carried out observations at a mean frequency of 10.0\,GHz (X band) with a bandwidth of 4.096\,GHz. The data were taken in standard phase-referencing mode, with 3C~286 as the bandpass and flux-density calibrator and 9C~J1506+4239 and B3~1456+375 as the complex-gain calibrators. We calibrated the data using the VLA pipeline in the Common Astronomy Software Applications package \citep[CASA;][]{McMullin07} v5.6.2, with additional flagging. For imaging, we used Briggs weighting with a robust parameter of 1. No self-calibration was performed. The details of these observations are given in Table~\ref{tbl:radio}. 

We find no evidence for radio emission at the SN location and infer a flux-density limit of $F_{\nu}<19\,\mu\rm{Jy}$ and $F_{\nu}<12\,\mu\rm{Jy}$ ($3\times$ image rms) for the first and second epochs, respectively, corresponding to $L_{\nu}<1.2\times 10^{26}\,\rm{erg\,s^{-1}\,Hz^{-1}}$ and $L_{\nu}<0.8\times 10^{26}\,\rm{erg\,s^{-1}\,Hz^{-1}}$ at the distance of \sn{}. In the third epoch the VLA was in its D configuration and there was a significant contribution from the host galaxy at the location of \sn{}, but we found no evidence of a point source at the SN location. After fitting and subtracting the host emission in the image plane using PyBDSM (Python Blob Detection and Source Measurement; \citealt{Mohan15}), we find a 3$\sigma$ upper limit of $F_{\nu}<30\,\mu\rm{Jy}$ ($L_{\nu}<2\times 10^{26}\,\rm{erg\,s^{-1}\,Hz^{-1}}$).

\section{Analysis} \label{sec:analysis}

\begin{figure*}[t]
\centering
\includegraphics[height=\dimexpr \textheight - 5\baselineskip\relax]{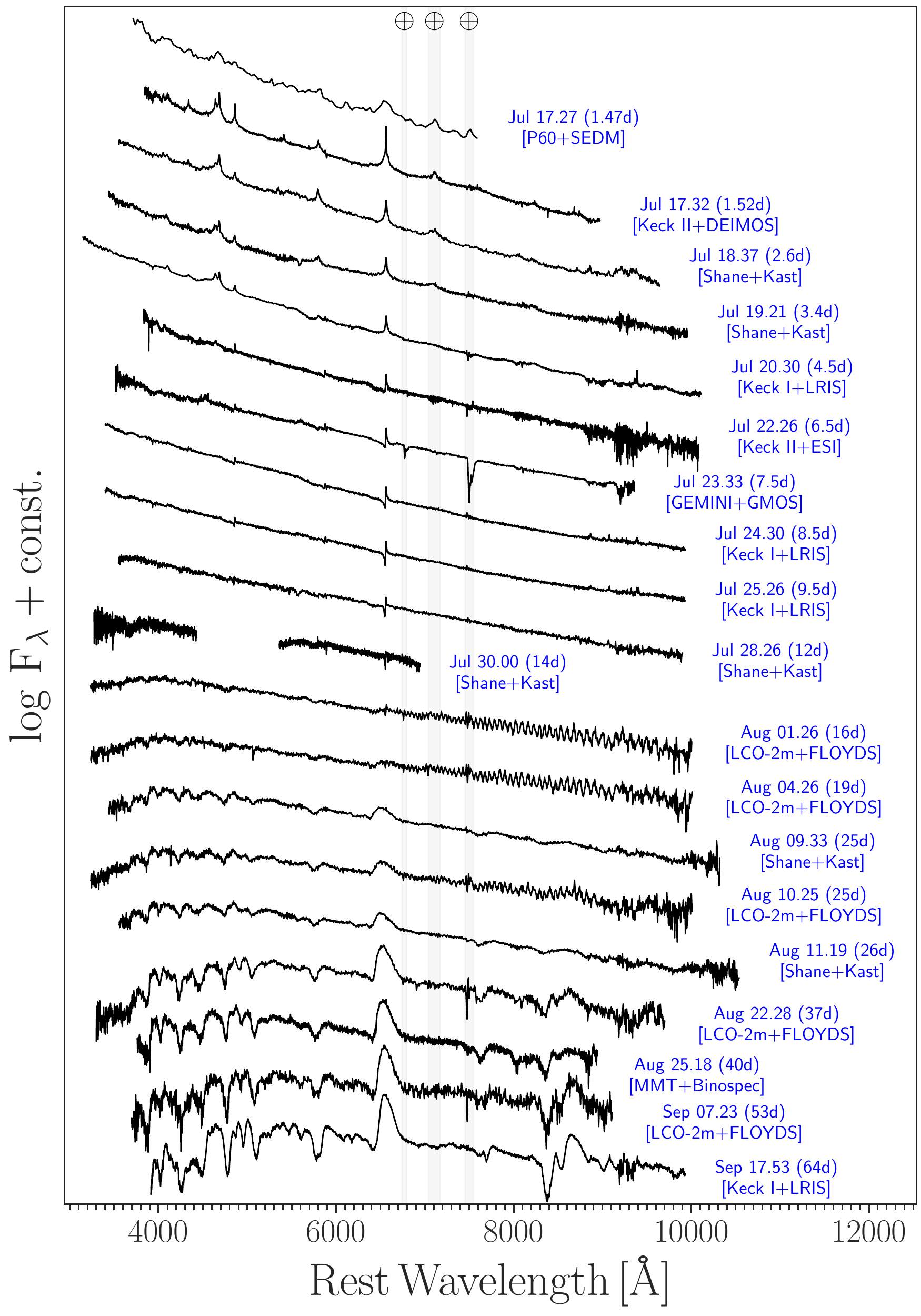} 
\caption{Optical spectral evolution of \sn{}. The spectra are presented in the rest frame ($z=0.01665$) and have been corrected for Galactic and host-galaxy extinction along the line of sight, and they are shifted vertically for display purposes. Spectra are labeled based on the epoch of their acquisition (from the time of first light at the top to the latest ones at the bottom) and by the telescope and instrument used. The FLOYDS spectra suffer from CCD interference fringes in the NIR region. The gray vertical bands mark the positions of the telluric O$_2$ (A and B) absorption bands.}
\label{fig:spec_ev}
\end{figure*}

\begin{figure*}[t]
\centering
\includegraphics[width=\textwidth]{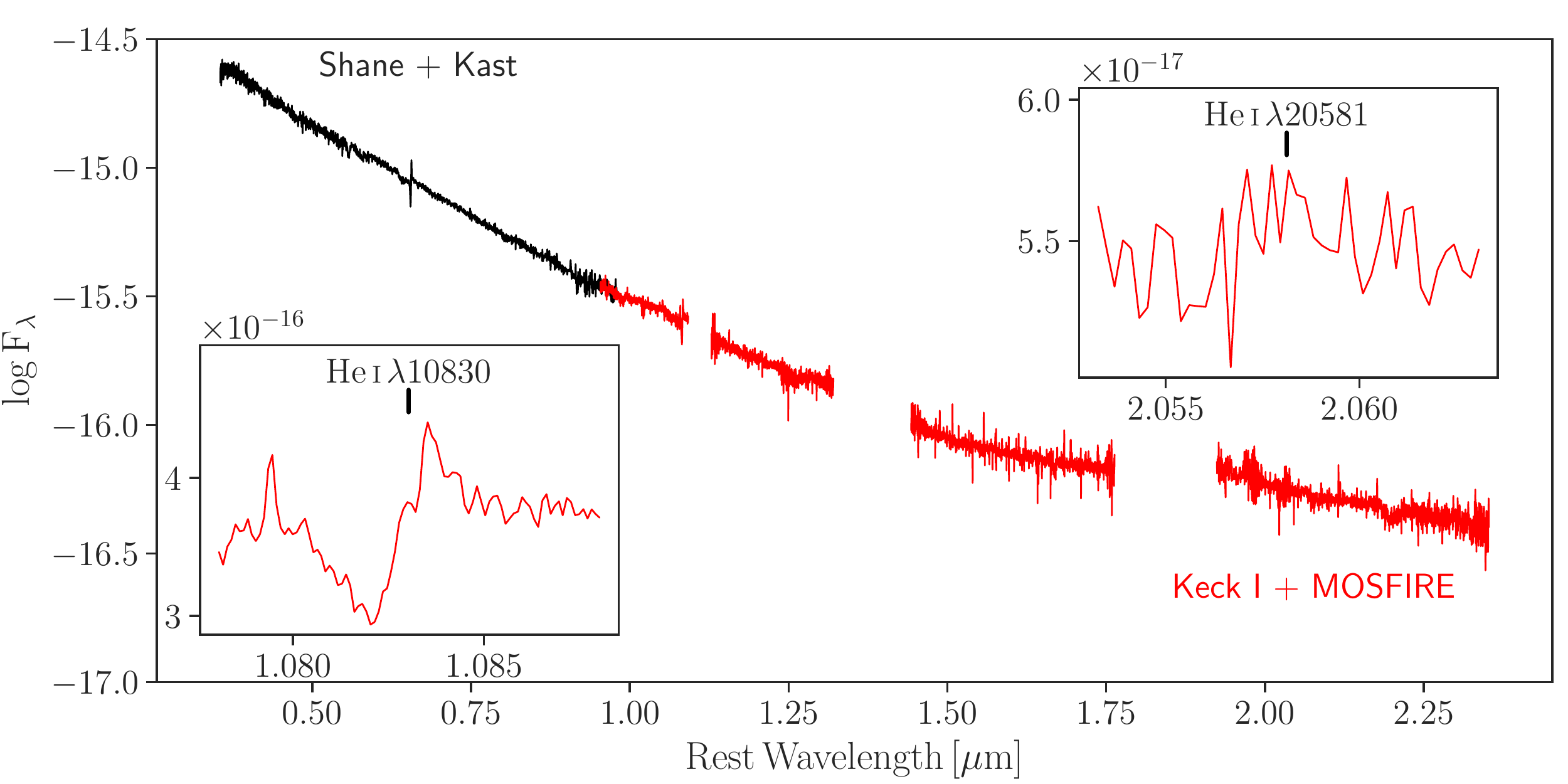} 
\caption{Shane+Kast optical spectrum (in black) combined with the Keck~I+MOSFIRE NIR spectrum of \sn{}. Both spectra were taken on MJD 59,058, corresponding to 12\,days after first light. In the two insets, we highlighted the regions of \ion{He}{I} \lam10830 and \ion{He}{I} \lam20581. The minimum of the 10830\,\AA\ helium lines corresponds to a velocity of $\sim320\,\kms$.}
\label{fig:NIR_spec}
\end{figure*}

\subsection{Photometric Properties and Pseudobolometric Light Curve} \label{phot_analysis}

The complete UV and optical light curves of \sn{} are presented in Figure \ref{fig:LC}. To estimate the time of escape of the first photons, we fit a three-parameter power-law function (e.g., $m = a + b t^c$) to the early-time \textit{g} and \textit{r} data. From this, we infer a time of first light of MJD $59,045.8 \pm 0.1$. The error on the time is estimated based on the nonlinear least-squares fitting routine, and we point out that it likely underestimates the true (systematic) uncertainty of the measurement. We use this value throughout the paper. In order to estimate the peak absolute magnitudes, we fit a low-order polynomial to the \sn{} light curve. We obtain $M_B = -18.28 \pm 0.10$\,mag at MJD $59,052.1\pm0.2$ and $M_r = -18.02 \pm 0.20$\,mag at MJD $59,055.6\pm0.2$. Using the adopted time of first light, the rise time of \sn{} is $t_r = 9.8 \pm 0.3$\,days with respect to $r$-band maximum and $t_r = 6.3 \pm 0.2$\,days with respect to $B$-band maximum. 

We then compare the $r$-band light curve of \sn{} to a sample of SNe~II with and without shock-ionization features at early times. We include the sample of flash-spectroscopy objects from \cite{Khazov2016} and \cite{Bruch21}, in addition to the sample of ``normal'' SNe~II from \cite{deJaeger2019}. We note that although SN 2013ej and SN 2014G were included in the latter collection, these objects showed flash-spectroscopy features in their early-time spectra \citep{Valenti2014,terreran16}, and they are therefore treated as such. Furthermore, we expand the shock-ionization sample with the addition of SN 1998S, SN 2016bkv, and SN 2017ahn \citep{leonard2000,Hosseinzadeh2018,Nakaoka2018,Tartaglia21}. The light curves of all the objects presented were retrieved from the OSC \citep[\url{https://sne.space};][]{Guillochon2017}, apart from those of \cite{Bruch21}, which were taken directly from their paper. For some objects the $r$ band was not available, and we used the $R$ band instead, without loss of generality.

We present the full comparison in the left panel of Figure \ref{fig:bol_LCs}. The great majority of flash-ionization objects populate the upper-end part of the plot, with \sn{} sitting right in the middle of the sample. Indeed, \cite{Khazov2016} claimed that the objects showing shock-ionization lines tended to be on average brighter than those that did not exhibit such features. On the other hand, \cite{Bruch21} did not find the same trend from the analysis of their sample. The addition of the sample of ``normal'' SNe~II from \cite{deJaeger2019} seems to support the former study, although the presence of one extreme outlier complicates the scenario. In fact, SN~2016bkv (the lowest red line on the plot) is the most striking example that objects presenting shock-ionization features can indeed have luminosities well below the average of SNe~II. There is some disagreement on the nature of this event, with claims for it to be an electron-capture SN rather than a core-collapse SN \citep{Hosseinzadeh2018}, but the presence of material around the progenitor star at the time of explosion should transcend the explosion mechanism responsible for the stellar demise.

The apparently higher luminosity shown on average by flash objects could naturally be explained by the extra energy injection from the early interaction responsible for the shock ionization. However, selection criteria could also play a factor in the luminosity distribution. Brighter objects are typically of more interest to the community, and therefore a classification spectrum could be sought with more urgency. More low-luminosity SNe~II could indeed exhibit flash-ionization features if observed sufficiently early. The size of the flash-spectroscopy sample is still quite small, preventing us from reaching a definitive conclusion on the matter.

We construct a pseudobolometric light curve of \sn{} using a combination of multicolor photometry from ZTF, PS1, and Swift observations.\footnote{The extremely blue UV colors and early-time ($t<5$\,days) color evolution of SN~2020pni impose nonnegligible deviations from the standard
Swift-UVOT count-to-flux conversion factors. We account for this effect following
the prescriptions by \cite{brown10}.} For each epoch, luminosities are calculated through trapezoidal integration of SN flux in the $ubvgri$ bands (3000--10000\,\AA). Uncertainties are estimated through a Monte Carlo simulation that includes 1000 realizations of the data. In time intervals without complete color information, we interpolate between light-curve data points using a low-order polynomial spline. The complete pseudobolometric light curve of \sn{} is presented in the right panel of Figure \ref{fig:bol_LCs} for phases $t<60$\,days after explosion. We choose to estimate luminosities for \sn{} using a trapezoidal integration method rather than fitting a blackbody model so as to better compare \sn{} to other SNe~II lacking UV or NIR photometric coverage. 

We also construct a complete bolometric light curve of \sn{} by fitting the broadband photometry with a blackbody model that is dependent on radius and temperature. Each spectral energy distribution (SED) was generated from the combination of multicolor UV/optical photometry in the $w2$, $m2$, $w1$, $u$, $b$, $v$, $g$, $r$, and $i$ bands (1500--9000\,\AA). In regions without complete color information, we extrapolate between light-curve data points using a low-order polynomial spline. This yields an initial radius of $R_{\rm bo} = (2.4 \pm 0.14) \times 10^{14}$\,cm, as well as a peak temperature and luminosity of $T = (2.5 \pm 0.16) \times 10^4$\,K and $L_{\rm bol} = (2.7 \pm 0.39) \times 10^{43}$\,erg \,s$^{-1}$. 

As shown in the left panel of Figure \ref{fig:bol_LCs}, we compare the $r$-band light-curve evolution of \sn{} to popular SNe~II discovered within a few days of explosion, many of which have flash-ionized spectral features detected in their early-time spectra, such as SN 1998S \citep{leonard2000,fassia01,shivvers15}, SN 2013fs \citep{Yaron17}, SN 2014G \citep{terreran16}, and SN 2017ahn \citep{Tartaglia21}. These objects were selected for reference to SN~2020pni because of their high-cadence optical sampling, complete pseudobolometric light curves, and distinct shock-ionization features in the early-time spectra that then disappeared at later phases. With respect to this sample, the peak $r$-band absolute magnitude of \sn{} is more luminous than that of SN 2013ej, SN 2013fs, and SN 2017ahn, but less luminous than that of SN 1998S and SN 2014G. Its $r$-band rise time and post-maximum decline rate are most similar to those of SN~2013fs. The photometric evolution of \sn{} is comparable to that of SN~2017ahn out to $t<30$\,days post-peak, until the plateau in the light curve of SN~2017ahn falls off at $\sim 40$\,days. 

In the right panel of Figure \ref{fig:bol_LCs}, we compare the pseudobolometric light-curve evolution of SNe~II discovered within a few days of explosion to that of \sn{}. We find that \sn{} has a lower overall luminosity than SN~1998S and a pseudobolometric luminosity comparable to or higher then the other presented SNe. Like the $r$-band light curve, the overall pseudobolometric evolution of \sn{} is most comparable to that of SN~2013fs at phases $t < 60$\,days after explosion. Out to phases $t<40$\,days post-explosion, its pseudobolometric rise time, light-curve decline rate, and peak luminosity are nearly identical to those of SN 2014G and SN 2017ahn. 

\subsection{Optical and NIR Spectroscopic Evolution}

\begin{figure*}[t]
\centering
\includegraphics[width=\textwidth]{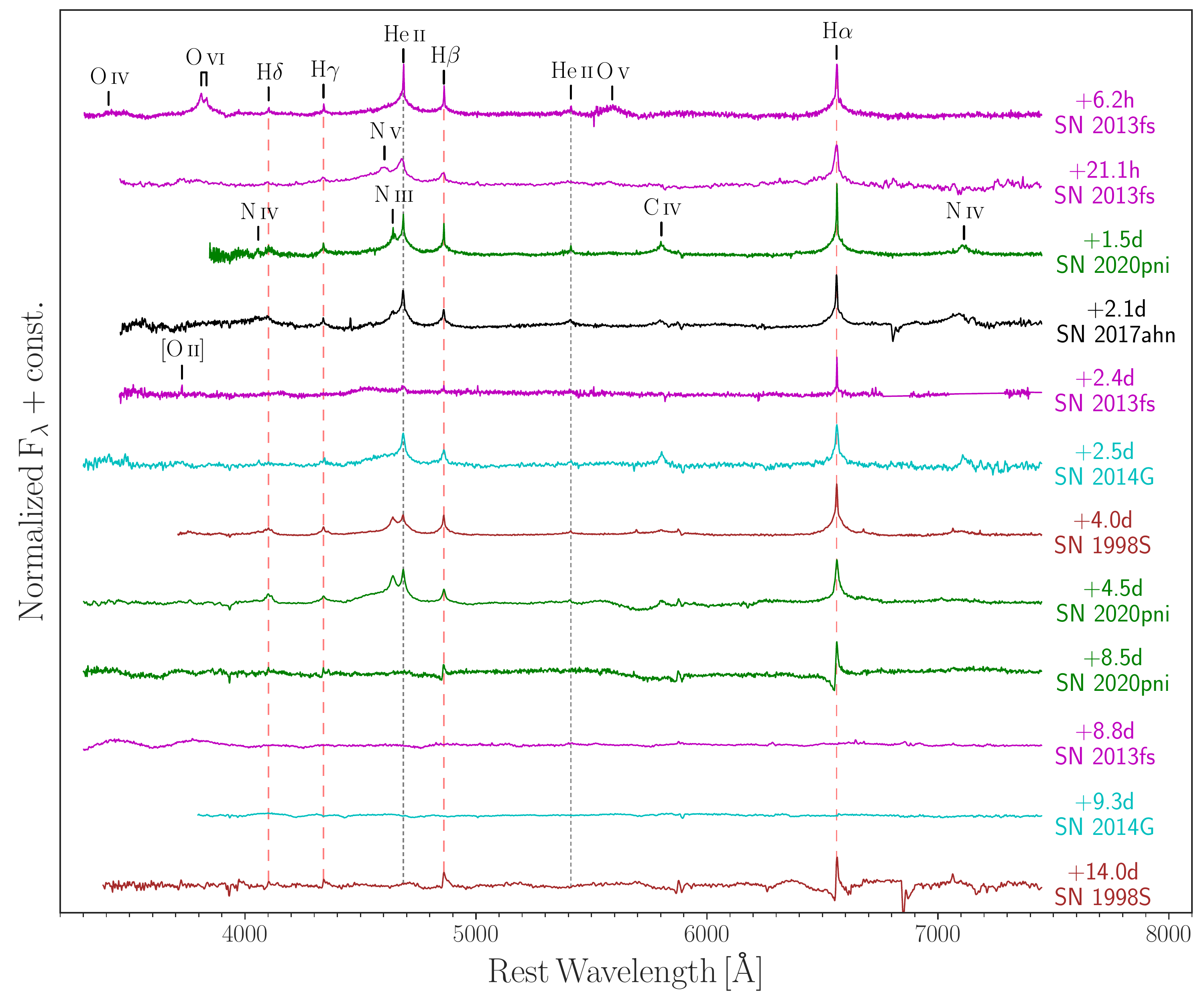} 
\caption{Comparison of early-phase spectra of \sn{} with those of SN 1998S \citep{fassia01}, SN 2013fs \citep{Yaron17}, SN 2014G \citep{terreran16}, and SN 2017ahn \citep{Tartaglia21}. The spectra have been continuum-subtracted and arbitrarily normalized for better display. While CSM-ionization features appear in all these objects, some of them already present a featureless continuum a week after first light. (Without loss of generality, we assume that the reported time of explosion for the targets found in the literature is actually the time of first light.) All of the spectra are in the rest frame and have been corrected for Galactic and host-galaxy reddening.}
\label{fig:early_spec}
\end{figure*}

\begin{figure*}[t]
\centering
\includegraphics[width=\textwidth]{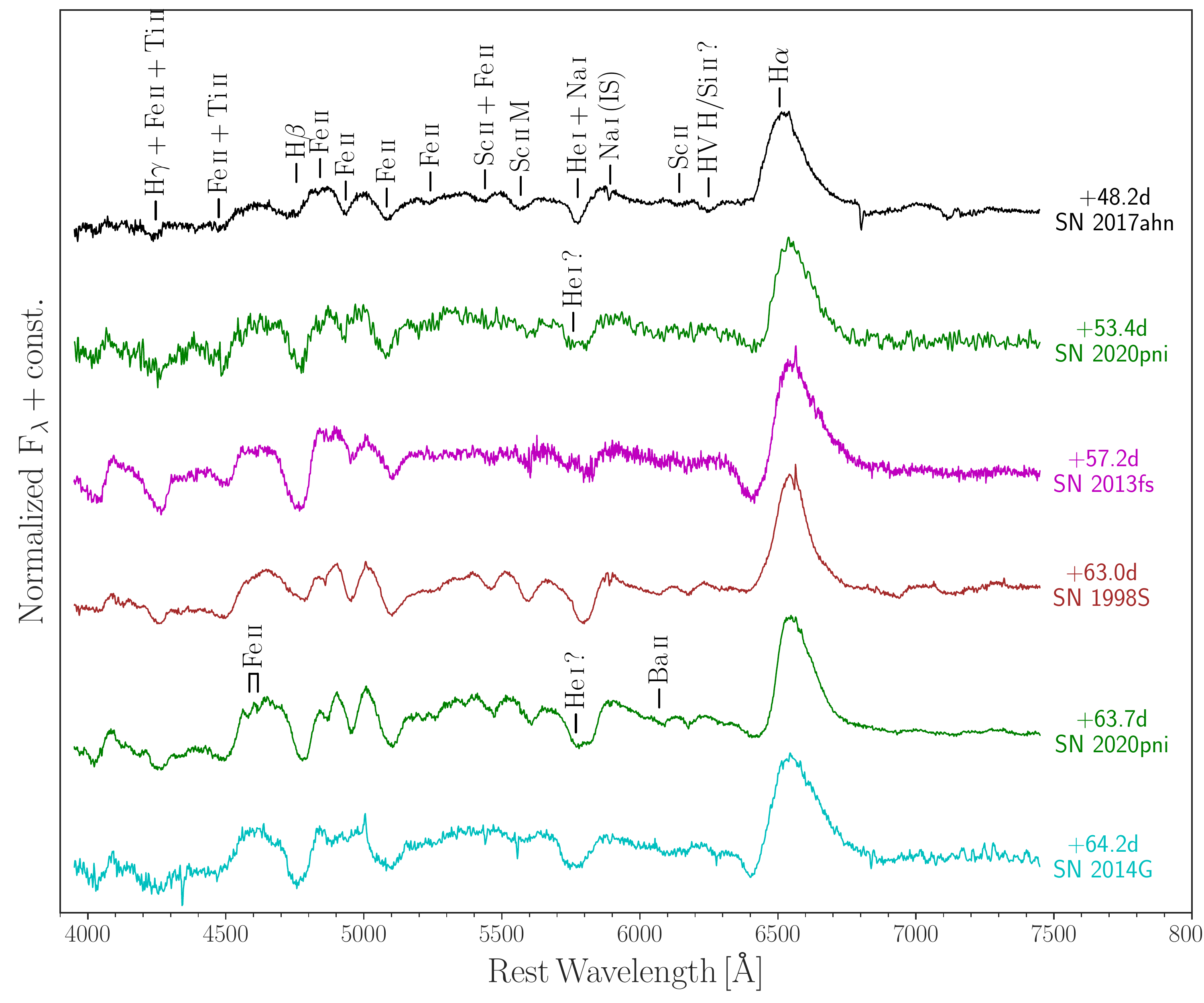} 
\caption{Spectral comparison of \sn{} with SN 1998S \citep{fassia01}, SN 2013fs \citep{Yaron17}, SN 2014G \citep{terreran16}, and SN 2017ahn \citep{Tartaglia21} at phase 48--64 days. The spectra have been continuum-subtracted and arbitrarily normalized for better display. All of the spectra are in the rest frame and corrected for Galactic and host-galaxy reddening.}
\label{fig:mid_spec}
\end{figure*}

The complete optical spectral evolution of \sn{} is presented in Figure \ref{fig:spec_ev}, and a log of the presented spectra is reported in Table \ref{fig: spec_log}. We also include the ZTF classification spectrum that was taken 1\,hr prior to our first spectrum \citep{Bruch2020b}. We started our spectroscopic campaign less than 2\,days after our estimated time of first light, and we kept monitoring \sn{} for $\sim 60$\,days, until constrained by the Sun. The early-time spectra show a blue continuum, with prominent narrow lines of H, \ion{He}{ii}, \ion{C}{iv}, \ion{N}{iii}, and \ion{N}{iv}. After 1 week, most of these features disappear, and only the H lines remain, which at this epoch are still narrow but with a P~Cygni profile. At a phase of 17\,days from first light, the spectrum seems completely featureless, while broad lines ($v\approx10,000\,\kms$) begin to appear a few days later. More typical features of SNe~II start to shape the spectrum \citep[see, e.g.,][]{Gutierrez2017}, including H, \ion{Fe}{ii}, \ion{Na}{i}, \ion{Sc}{ii}, \ion{Ba}{ii}, \ion{Ti}{ii}, and \ion{Ca}{ii}. By the time \sn{} became Sun constrained, it was still in the photospheric phase, and no nebular lines were visible yet.

At $\sim$12\,days after first light, we acquired a Keck~I+MOSFIRE NIR spectrum, in coordination with an optical spectrum with Shane+Kast (see Figure~\ref{fig:NIR_spec}). Although the H Balmer series is clearly visible at this phase in the optical spectrum, we do not detect any Paschen or Brackett lines in the NIR. However, we identify a P~Cygni profile of \ion{He}{I} \lam10830 and possibly \ion{He}{I} \lam20581.

In Figure \ref{fig:early_spec} we compare the early phases of \sn{} with those of other well-observed flash-spectroscopy events --- SN 1998S \citep{leonard2000,fassia01,shivvers15}, SN 2013fs \citep{Yaron2017}, SN 2014G \citep{terreran16}, and SN 2017ahn \citep{Tartaglia21}. The spectra were first normalized by the underlying continuum\footnote{The continuum is estimated using a spline fit to the region free of emission lines.}, and then scaled by an arbitrary factor for better display. A diversity in line intensities and timescales is clearly evident, and the presence of different elements with different ionization levels could be linked to variations in CSM composition as well. Prominent, highly ionized oxygen lines are present in the first spectrum of SN 2013fs, although no oxygen is detected in the first spectrum of \sn{}. These lines disappear very rapidly in SN 2013fs, so we cannot exclude that these lines were present in \sn{} at an earlier phase. 

The presence of a relatively prominent \ion{C}{iv} \lam5803 line in \sn{}, which is missing in the spectra of SN 2013fs, hints at a carbon-rich environment for \sn{} (see Figure~\ref{fig:early_spec}). This carbon line seems to also be quite prominent in SN 2014G, and possibly detected in SN 1998S and SN 2017ahn. The spectroscopic similarities between SN 1998S and \sn{} are remarkable (see spectra at $\sim 4$\,days). Indeed, among the presented sample, SN 1998S is the only object other than \sn{} that still shows narrow lines beyond a week after explosion. SN 2013fs and SN 2014G present featureless spectra at $\sim 9$\,days after explosion, while SN 2020pni and SN 1998S show narrow H lines with P~Cygni profiles, both with a low-velocity component in absorption (see also Sec.~\ref{sec: line_ev}). This indicates that at this phase the photosphere is still in the slow-moving CSM, suggesting a more radially extended, high-density environment compared to that of SN 2013fs and SN 2014G. 

We compare the later spectroscopic evolution of \sn{} with the same sample of objects. In Figure \ref{fig:mid_spec}, we display spectra at $\sim 2$ months after explosion. All of the objects present a very similar pattern, with the spectrum being dominated by P~Cygni profiles of hydrogen, sodium, iron, and other metal lines like scandium and barium. Focusing on the P~Cygni profile of \Ha, SN 2017ahn and SN 1998S exhibit very shallow to no absorption, SN 2013fs shows a relatively deep trough, and SN 2020pni and SN 2014G are intermediate. We find a certain range of velocities of ejected material in the sample. The minimum of the \Ha\ absorption feature of \sn{} sits at $\sim 6500\,\kms$, while that of SN 2014G is at $\sim 7600\,\kms$. Indeed, focusing on the position of the minimum of the absorption of the features in common, we notice that \sn{} has the slowest material. From the minimum of the \ion{Fe}{i} \lam5169 feature, we measure a velocity of $\sim 3700\,\kms$, which we can use as a proxy for the photospheric velocity \citep{Hamuy2001}. The slower material displayed by \sn{} manifests as a higher number of discernible features in the spectra of \sn{}, as a consequence of less severe line blending. This can be appreciated especially from the spectrum $\sim 64$\,days after explosion, where several \ion{Fe}{ii}, \ion{Sc}{ii}, and \ion{Ba}{ii} multiplets can be identified (see Figure~\ref{fig:mid_spec}). 

Focusing on the \ion{He}{i} and \ion{Na}{i} absorption blend around 5800\,\AA, it is clear from Figure \ref{fig:mid_spec} how this feature looks broad, and possibly with multiple components, in SN 2020pni, SN 2013fs, and SN 2014G. The same feature appears narrower in SN 1998S and SN 2013fs. It is tempting to attribute its reddest component to a resolved \ion{He}{i} line. In \sn{} this identification would correspond to a velocity of the helium material of $\sim 5000\,\kms$, which is considerably slower than what is shown by the \Ha{} minimum. \ion{Ba}{ii} has a multiplet at \lam\lam5854, 6142, and 6497. Assuming then that the reddest absorption in the feature at 5800\,\AA\ is barium, we obtain a velocity of $\sim 4000\,\kms$, while from the feature at 6100\,\AA\ we measure $\sim 3000\,\kms$. Therefore, the \ion{Ba}{ii} association also seems in conflict with other identifications of the same ion. It is likely that a combination of helium and barium is responsible for the broader feature at 5800\,\AA\ observed in some SNe~II, such as SN 2020pni and SN 2013fs.

\subsubsection{Line Evolution}\label{sec: line_ev}
\begin{figure}[t]
\centering
\includegraphics[width=\columnwidth]{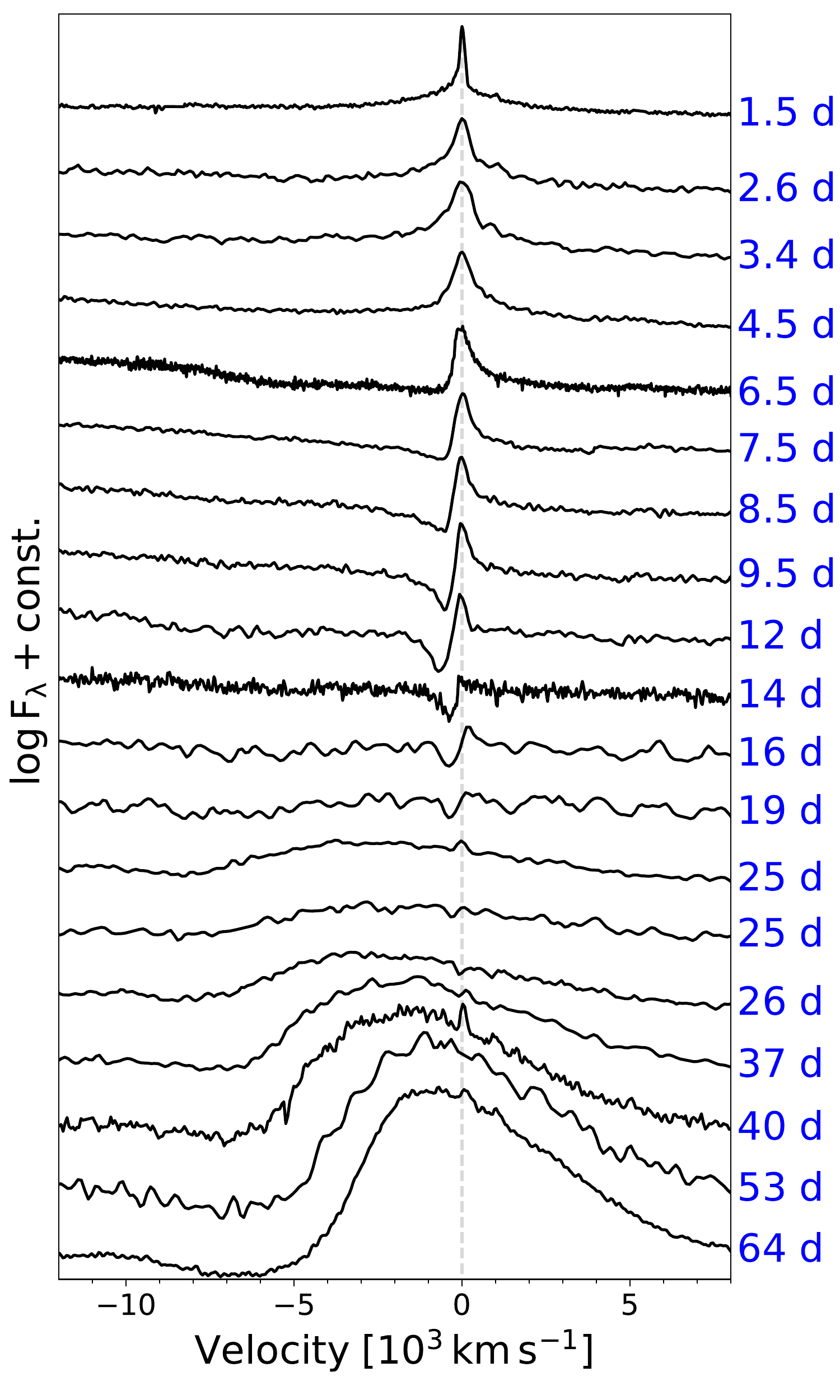} 
\caption{\Ha{} line evolution in \sn{}. The spectra are shifted vertically for display purposes. The phases are reported from the time of first light. Narrow lines persist until $\sim 15$\,days, while the spectra become ejecta dominated starting at $\sim 20$\,days.}
\label{fig:Ha_ef}
\end{figure}

\begin{figure}[t]
    \includegraphics[width=\columnwidth]{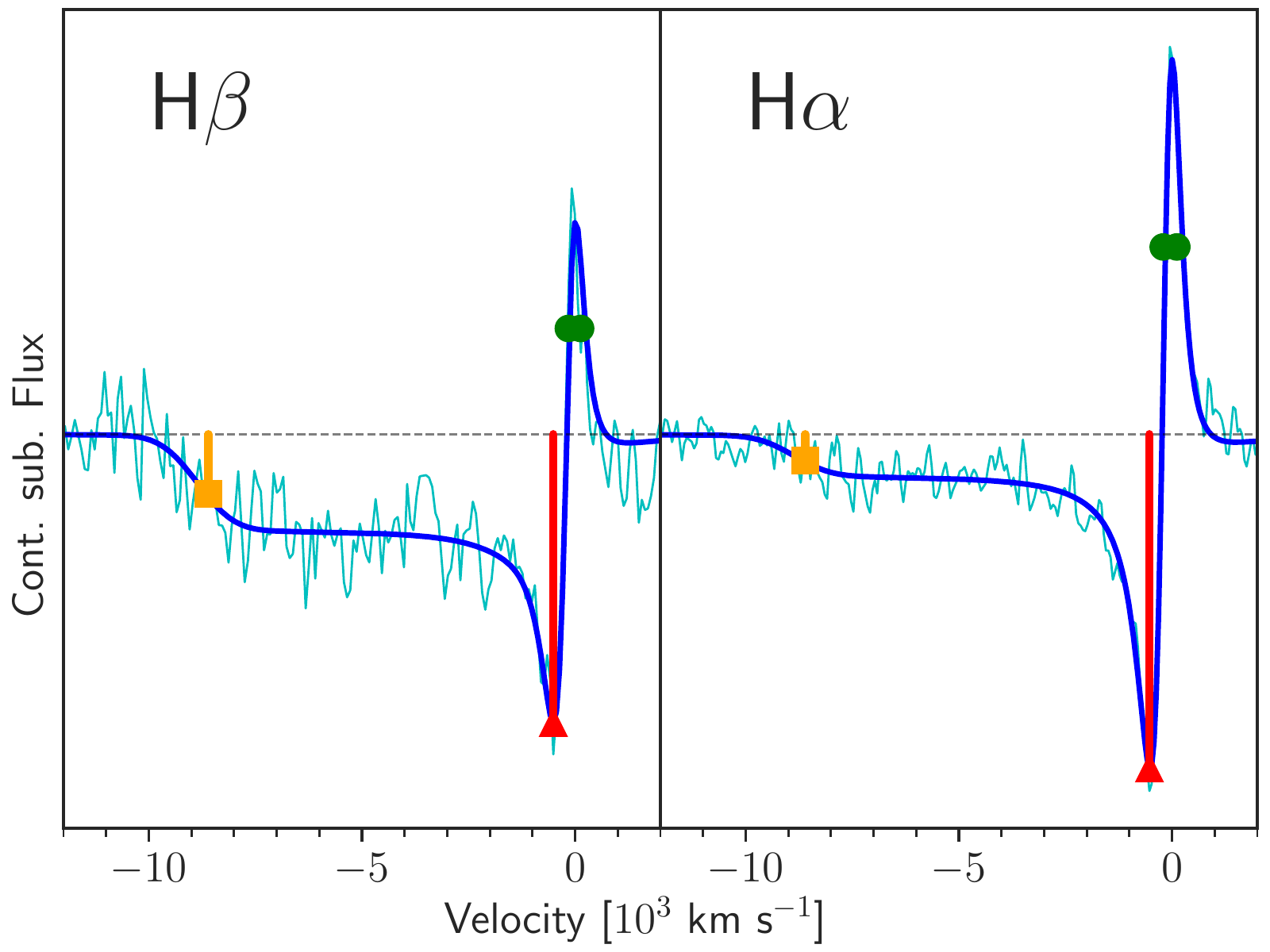} \\
    \includegraphics[width=\columnwidth]{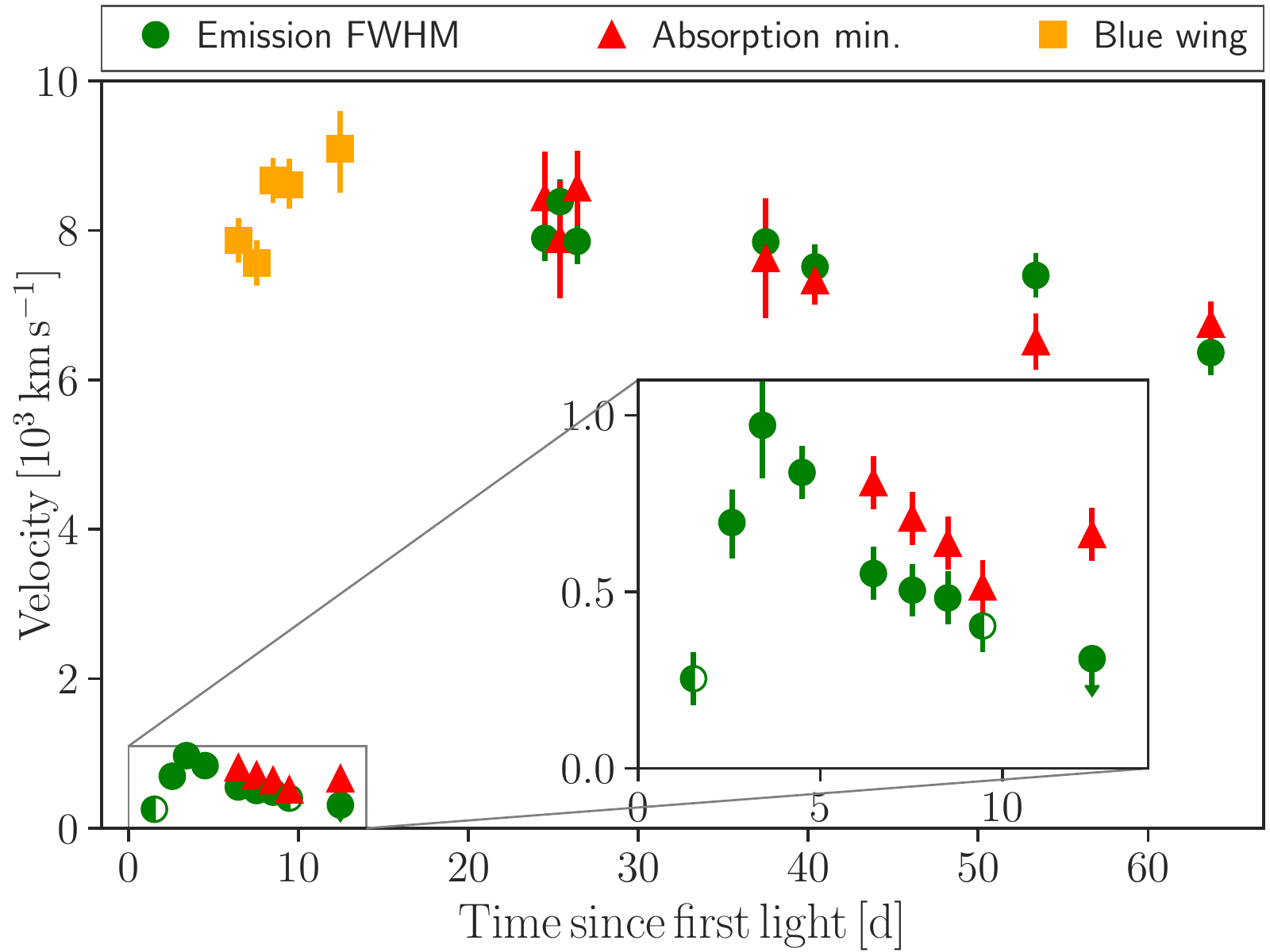}
    \caption{
    \textit{Top panel:} Details of the Keck I+LRIS spectrum of \sn{} obtained on 2020 July 25.26 ($\sim9.5$\,days after first light), showing an example of the modeling of the \Ha\ and \Hb\ line profiles. Starting $\sim5$\,days after the time of first light, a broad absorption component is present, which can be modeled with a boxy profile having sigmoidal boundaries. The narrow emission and absorption profiles are modeled with single Lorentzian functions. \textit{Bottom panel:} Evolution of the \Ha\ velocity components with time. The symbols refer to the velocities measured at the positions marked in the upper panel. We use half-hollow symbols to denote marginally resolved lines, and upper limits for unresolved lines. The inset presents a zoom-in of the early phases, when the narrow emission first shows an increase in velocity and then a decrease, which we interpret as a signature of complex CSM.}
    \label{fig: line_vel}
\end{figure}

At 6\,days after first light, a very broad, although shallow, absorption starts to appear on the blue side of \Ha{} (see Figure \ref{fig:Ha_ef}). This feature extends to $\sim 8000\,\kms$, while the emission component remains narrow. The broad component is clearly associated with the SN ejecta. After a few days, the absorption component develops a more pronounced dip. From the position of the minimum, we measure a velocity in close agreement with what is determined from the FWHM of the emission feature. The narrow \Ha{} now has the full appearance of a P~Cygni profile. The broad component is still present at these phases. This morphology, with a narrow P~Cygni profile superimposed on a broader absorption, was also observed in SN~1998S \citep{leonard2000,fassia01,shivvers15,Dessart2016}. At 14\,days after first light, the emission component disappears completely, initially leaving only the narrow absorption, and then a featureless continuum. At 20\,days a broad emission component starts to emerge with FWHM $\approx 8000\,\kms$, accompanied by a shallow absorption with a minimum at similar velocities, forming the classical P~Cygni profile of typical evolved SNe~II.

We now focus on the complex evolution of the \Ha{} profile by studying the velocity evolution of each component described above. At $t<5$\,days, in order to reproduce both the narrow core and the broad wings of the pure emission feature, we used two Lorentzian components. Beyond this phase and until $t<13$\,days, we used single Lorentzian profiles to reproduce the narrow emission and absorption, while the broad absorption is well reproduced by a flat, boxy profile, with boundaries defined by a sigmoidal function. In Figure \ref{fig: line_vel} we show an example of the modeling we performed during these phases. When the broad P~Cygni profile is fully formed ($t>20$\,days), we instead used only two Gaussians to reproduce the line profile. We then used the FWHM as a proxy for the velocity of the emission component and the minimum of the absorption component as a proxy for the bulk velocity of the narrow absorption component. We also kept track of the middle point of the red-most sigmoidal boundary of the boxy profile (orange squares in Figure~\ref{fig: line_vel}).

The evolution in time of the velocity of these features is shown in the bottom panel of Figure \ref{fig: line_vel}. At early phases, the narrow emission shows a clear increase in velocity, reaches a peak $\sim 4$ days after first light, and then starts to decrease. We remark that this is not a spectral resolution effect, as the lines are fully resolved in all of the spectra (apart from the first epoch with the SEDM). One possible conclusion could be that the ejecta inside the CSM are accelerating the inner material \citep{Moriya11}. However, the emission originates from the CSM in front of the shock (this is unshocked CSM). The shock then has yet to reach this part, so it could not be responsible for any acceleration at this phase.

The increasing velocity with time likely maps a velocity gradient of the CSM at larger radii from the explosion, hinting at complex CSM. Given that we see these narrow lines only during the early phases of evolution of \sn{}, it is fair to assume that this CSM was created by recent mass loss, and it is possible that the progenitor lost material having different velocities with time. In particular, we observe that the progenitor lost material with larger velocities at earlier times and then smaller velocities as it approached the explosion time. We point out, though, that this does not naturally reflect a variable mass-loss rate during this phase, as the velocity of the material is not necessarily linked to the amount of the material lost by the progenitor of \sn{}.

Finally, from Figure \ref{fig: line_vel}, one could also argue that the boxy profile extension evolves with time, as the orange square seems to increase in velocity from $\sim8000$ to $\sim9500\,\kms$. However, given the modeling we performed, this parameter does not directly link to a specific physical property of the explosion. Although this parameter could be seen as a proxy for the maximum velocity of the ejecta, the signal-to-noise ratio (S/N) of the spectrum, the depth of the absorption function, the shallowness of the transition from the continuum to the floor of the boxy profile, and the fit to the continuum level all play a role in inferring this value. Therefore, the observed increase in velocity is probably not as significant as the figure might suggest, as the uncertainties are also likely underestimated. However, further studies of other objects showing this extended absorption are encouraged, and possibly a more meaningful physical quantity could be inferred from a larger sample.

\subsection{Inferences on the Explosion's Environment} \label{SubSec:environmentmodels}
\subsubsection{CSM Properties at \texorpdfstring{$r\le 10^{15}\,\rm{cm}$}{r\textless 10\textasciicircum15 cm} from Early Light-curve Modeling} \label{SubSubSec:earlyLC}

A number of observational features suggest that the shock's radiation is breaking out of a compact shell of dense CSM extending out to a radius $R_w$. Specifically, these include the fast rise to maximum optical light over a timescale of $t_{r}\approx2$\,days, the bright and hot UV emission reaching a color temperature $T \geqslant 20,000$\,K (e.g., Sec.~\ref{phot_analysis} and Figure~\ref{fig:LC}), the rapid fading of shock-ionized spectral features, and the narrow P~Cygni profiles (Figure~\ref{fig:Ha_ef} and \ref{fig:spec_ev}; Sec.~\ref{SubSubSec:specmodel}) by $\sim 15$\,days after first light. We expect $R_w\approx 10^{15}\,\rm{cm}$, comparable to the inferred best-fitting blackbody radius when the shock-ionized spectral features become subdominant (e.g., Sec.~\ref{phot_analysis}). 

We employ the formalism by \cite{Chevalier11} (see \citealt{Waxman17} for a recent review) to model the onset of the emission that breaks out from the thick shell of CSM under the reasonable assumption $R_d<R_w$, where $R_d$ is the radius of the contact discontinuity at the diffusion time $t_d$ (i.e., the radius where the diffusion of radiation becomes important). This assumption is motivated by the persistence of shock-ionized spectral features well beyond the time of bolometric peak. Indeed, the low-velocity P~Cygni profiles of \Ha{} are detectable until at least 15\,days after first light. Under these circumstances, the escape of the radiation is delayed with respect to the onset of the explosion on a timescale $\sim t_d$, which is set by the time necessary for the radiation to reach an optical depth $\tau_w\approx c/v_{\rm sh}$, where $v_{\rm sh}$ is the shock velocity. Radiation is also released on the diffusion timescale, leading to a bolometric rise time $t_{\rm{rise}}\approx t_d$, which implies that the explosion started $\sim t_{\rm rise}$ (i.e., at most a few days) before the estimated time of first light (Table \ref{tab:params}). This result is consistent with the time of explosion estimated from the emergence of spectral features with $v\approx8000\,\kms$ at $t\approx8$ days since first light and the measured radius of the photosphere at this time, which implies that the time of first light is delayed from the time of explosion by at most 1--2 days.

Following \cite{Chevalier11} and using the solutions by \cite{Margutti1409ip}, we find that the observed $t_{\rm rise}$, radiated energy at breakout $E_{\rm rad}\approx 0.5\times 10^{49}\,\rm{erg}$, and breakout radius $R_{\rm bo}\approx 2\times 10^{14}\,\rm{cm}$ constrain the wind mass-loss rate to $\dot M \approx0.01 \,\rm{M_{\sun}\,yr^{-1}}$ for a wind velocity $v_w=200\,\rm{km\,s^{-1}}$, similar to the FWHM of the narrow hydrogen component in the first optical spectrum. For these parameters the wind-shell mass is $M_w\approx 0.02\,\rm{M_{\sun}}$ enclosed within $R_w\approx 10^{15}\,\rm{cm}$. The wind mass within the breakout radius is $M_w(\le R_{\rm bo})\approx 0.005\,\rm{M_{\sun}}$. In this model, after shock breakout, continued interaction with the wind material supports a luminosity $\gtrsim 10^{43}\,\rm{erg\,s^{-1}}$ for a few days, consistent with the observations.
We note that these parameter values should be treated as order-of-magnitude estimates given the likely complexity of the SN environment and some simplifying assumptions inherent to our analytical modeling approach. 

\subsubsection{CSM Properties at \texorpdfstring{$r\le 10^{15}\,\rm{cm}$}{r\textless 10\textasciicircum15 cm} from Spectral Modeling} \label{SubSubSec:specmodel}

\begin{figure*}[t]
\centering
\includegraphics[width=\textwidth]{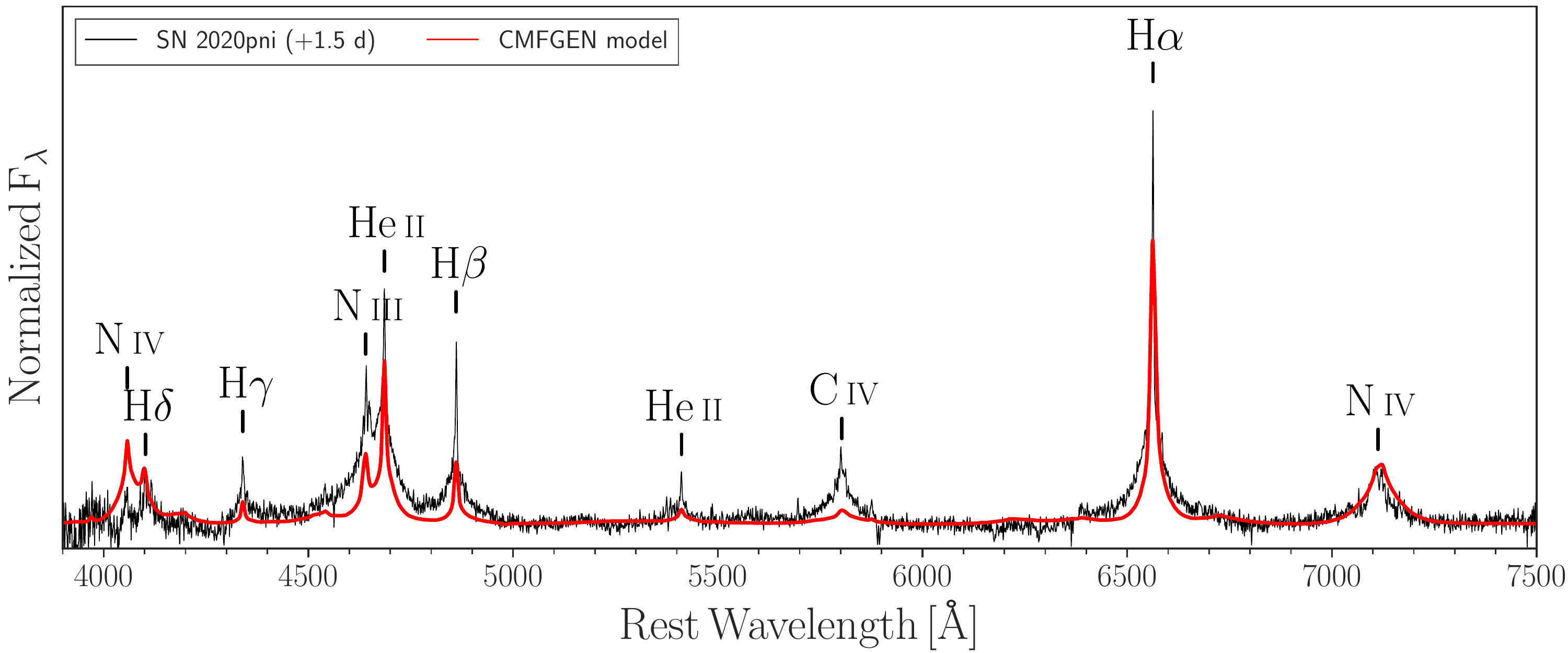} 
\caption{Best-fitting CMFGEN model (red) compared to the $+1.5$\,days optical spectrum of \sn{}. Assuming a wind velocity of $v_{w}=200\,\kms$, we obtained a mass-loss rate $\dot{M}=(3.5-5.3)\times10^{-3}$\msunyr.}
\label{fig:cmfgenmodel}
\end{figure*}

We employ the radiative transfer code \mbox{CMFGEN} \citep{Hillier96}, with the implementation of \cite{Groh14},\footnote{See also \cite{Boian18}, \cite{Boian19}, and \cite{Boian20} for more details.} to study the properties of the material surrounding the progenitor of \sn{} a few days after the time of first light. The photons produced by the interaction between the SN ejecta and the progenitor wind diffuse out through the extended CSM around the progenitor. Subsequently, the material heats up, achieves radiative equilibrium, and reemits according to its composition, velocity, temperature, and density structure. For simplicity, we assume a spherical and stationary wind,\footnote{Although the observations indicate a wind with stratified velocity (see Sec.~\ref{sec: line_ev}) these assumptions concern the volume of the CSM swept by the ejecta at the time of the modeled spectrum. Considering that we use the spectrum taken $\sim1.5$\,days after first light, we can assume the inner wind to be stationary without loss of generality.}, in non-LTE. 
We also assume that no energy is generated in the progenitor wind, that time-dependent effects are negligible, and that the medium is not clumpy. These assumptions and caveats affect the inferred observables, but we expect our modeling to still provide realistic quantities during the early phases, when the photosphere is located in the progenitor wind and the SN shock front is located at high optical depths. 

We fit our first spectrum, obtained with Keck~II+DEIMOS $\sim 1.5$\,days after first light, which also exhibits the high S/N necessary for this analysis. With the CMFGEN modeling we are able to constrain the progenitor mass loss $\dot{M}$, its chemical surface abundances, as well as the inner boundary\footnote{The inner boundary roughly corresponds to where the source of ionization photons is in the models (actually the steep density increase) at the time of the observations. This is not to be mistaken for the inner boundary of the actual CSM, as we cannot infer anything for $R<R_{\rm in}$.} of the CSM $R_{\rm in}$, and the bolometric luminosity of the event $L_{\rm SN}$. Figure \ref{fig:cmfgenmodel} shows the comparison between the best-fit model and the observations of \sn{} at $\sim 1.5$\,days. We find excellent agreement between the observed and modeled spectral morphologies, with strong \ion{He}{ii}, \ion{H}{i}, \ion{N}{iii}, and \ion{N}{iv} features.

Our modeling suggests $L_{\rm SN}=(2.3-3.5)\times10^{43}$\,erg\,s$^{-1}$, $\dot{M}=(3.5-5.3)\times10^{-3}$\,\msunyr\ (assuming a wind velocity $v_{\rm wind}=200\,\kms$), $R_{\rm in}=2.5\times10^{14}\,\rm{cm}$, and a flux temperature at a Rosseland optical depth of 10 of $T_{\star}=27,500-29,400\,\rm{K}$. We obtain a progenitor helium surface mass fraction of $Y\approx0.30$--0.40, and we constrain the CNO surface abundances to C$_{\rm sur}=2.6\times10^{-4}$, N$_{\rm sur}=8.2\times10^{-3}$, and O$_{\rm sur} \lesssim 1.3\times10^{-4}$, with an estimated 3$\sigma$ uncertainty of 50\%.

\begin{figure*}[t]
\centering
\includegraphics[width=\textwidth]{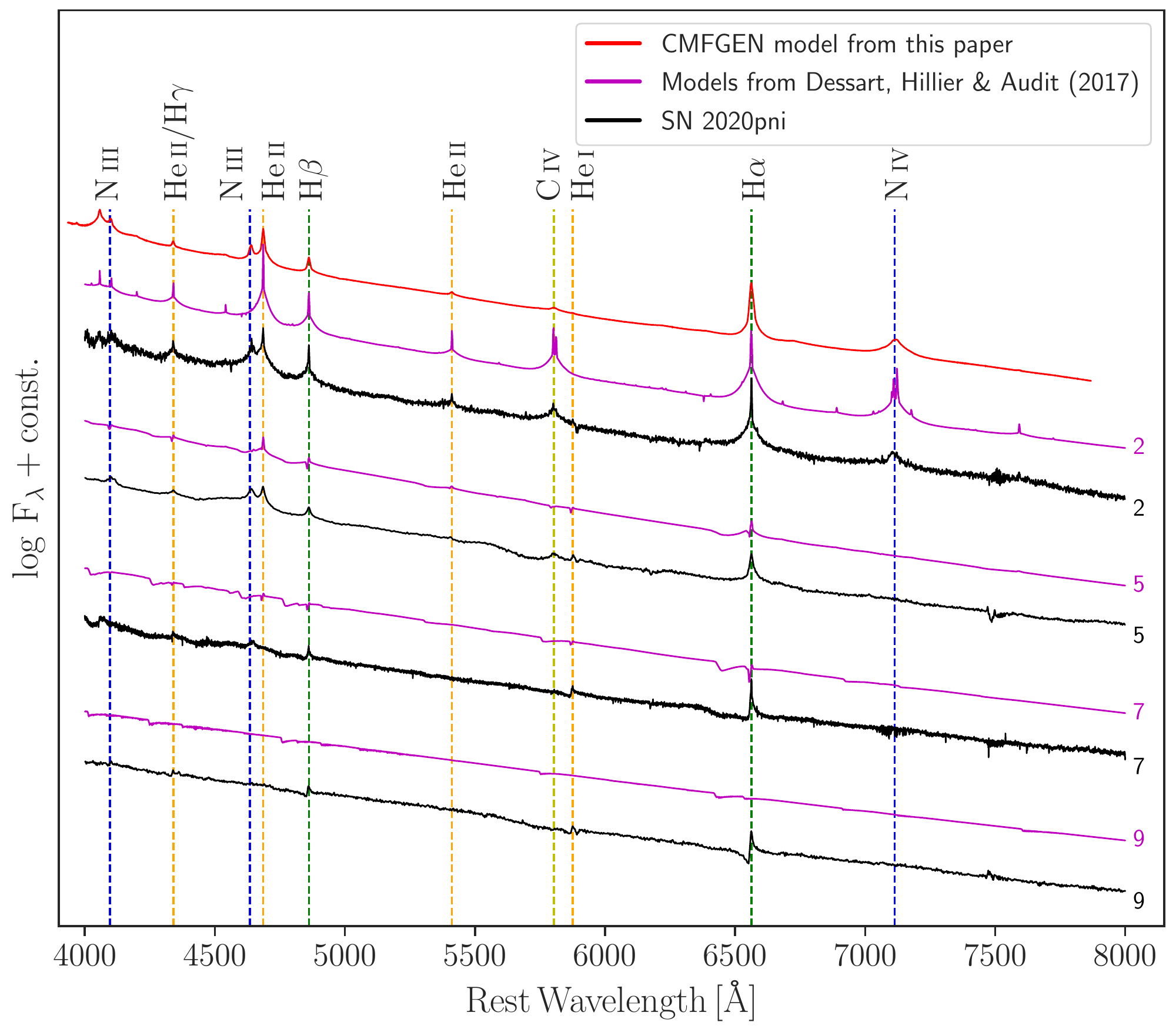} 
\caption{Comparison of the early-time spectra of \sn{} (black) with model r1w6 from \cite{Dessart2017} (magenta). We also included our CMFGEN modeling for comparison (red). The phases, from first light, are reported on the right. The models from \cite{Dessart2017} have been evolved from the progenitor model m15mlt3 \citep{Dessart2013}, which corresponds to a star with $R_\star=501\,R_{\odot}$, $M_{\rm ej}=12.52$\,\Msun, $E_k=1.35\times10^{51}$\,erg, and $\dot{M}=10^{-2}$\,\Msun\,yr${^{-1}}$. These results are consistent with those from our CMFGEN modeling. The main spectral features are labeled. Although the initial spectrum is well matched by the model, \sn{} shows long-lasting line emission, in particular the H lines, persisting for more than 10\,days (see Figure~\ref{fig:Ha_ef}). We interpret this discrepancy as the consequence of a more radially extended, thick CSM with respect to the one assumed by \cite{Dessart2017} in their grid of models. }
\label{fig:Dessart}
\end{figure*}

\cite{Dessart2017} computed a grid of models presenting the spectroscopic outcome of early-time interaction of the SN ejecta with material near the star's surface. They used radiation hydrodynamics and radiative transfer to reproduce the explosion of RSG stars, embedded in different dense material, in contrast to our approach, where we adopted a static wind configuration. We did not solve for the time-dependent radiation hydrodynamics of the CSM, as assuming the radiative equilibrium is still a reasonable approximation at very early phases, when the photosphere forms in the CSM. We compare the early-time spectra of \sn{} with the grid of models from \cite{Dessart2017} and find a good match with their r1w6 model. This was evolved from the progenitor model m15mlt3 from \cite{Dessart2013}, which corresponds to a star with $R_\star=501\,R_{\odot}$, $M_{\rm ej}=12.52$\,\Msun, $E_ k=1.35\times10^{51}$\,erg, and $\dot{M}=10^{-2}$\,\msunyr. We show a comparison of the early-time spectra of \sn{} with the model from \cite{Dessart2017} in Figure \ref{fig:Dessart}, as well as our CMFGEN model. All of the main features shown by \sn{} in its $\sim 2$\,day spectrum are also present in the r1w6 model by \cite{Dessart2017}, including the strong \ion{He}{ii} lines, although the model seems to show less prominent \ion{N}{iii} emission. For reference, \cite{Dessart2017} adopted a nitrogen surface abundance of N$_{\rm sur}\approx3\times10^{-3}$. Their first synthetic spectrum is remarkably similar to the one obtained with our CMFGEN modeling. The \ion{C}{iv} \lam5803 line, which is underestimated by our modeling, looks particularly prominent in the model by \cite{Dessart2017}, although in this case it is even stronger than the line shown by \sn{}.

\subsubsection{Properties of the Larger-scale \texorpdfstring{$r> 10^{15}\,\rm{cm}$}{r\textgreater 10\textasciicircum15 cm} Environment from Radio Observations} \label{SubSubSec:RadioModeling}

\begin{figure}[t]
\centering
\includegraphics[width=\columnwidth]{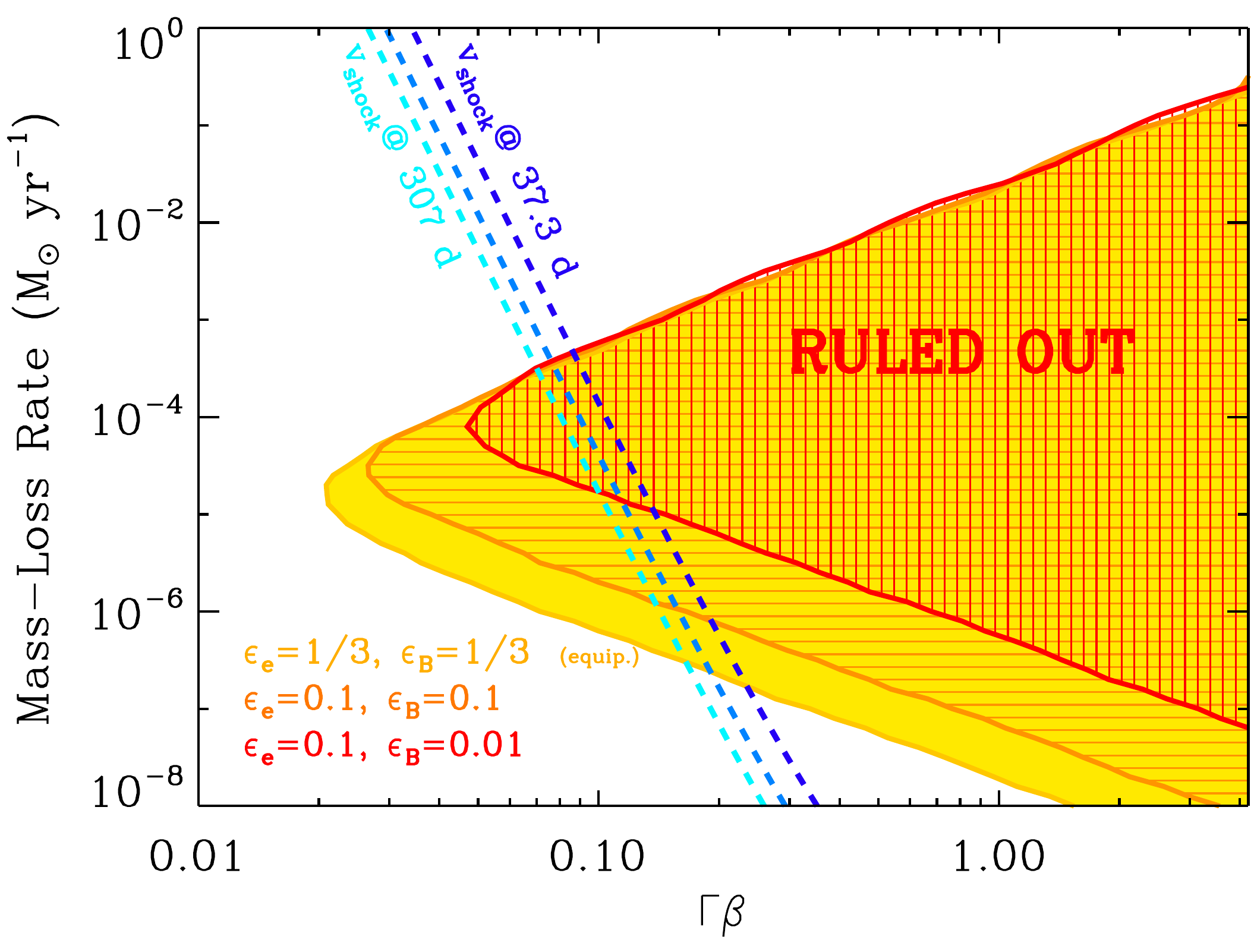} 
\caption{Region of the mass-loss rate vs. shock velocity ($\Gamma\beta$) parameter space that is ruled out by the radio nondetection (Sec.~\ref{SubSec:RadioData}) for different assumptions of shock microphysical parameters $\epsilon_B$ and $\epsilon_e$. The $\dot M$ values on the ordinate are for an assumed wind velocity of $200\,\rm{km\,s^{-1}}$. The blue dashed line is the expected shock velocity at the time of the radio observations at $\delta t=37.3$ days, $\delta t=129$ days, and $\delta t=307$ days for a massive-star explosion with $E_k\approx10^{51}\,\rm{erg}$ and ejecta mass of roughly a few $M_{\sun}$. }
\label{fig:radiodensity}
\end{figure}

We infer the density properties of the larger-scale environment at distances of (1--3) $\times 10^{16}\,\rm{cm}$ using the radio nondetection in Sec.~\ref{SubSec:RadioData}. In the context of synchrotron emission from the explosion's forward shock, and self-consistently accounting for both synchrotron self-absorption (SSA) and free-free absorption (FFA) \citep[e.g.,][]{Chevalier98,Weiler02}, the radio nondetections at $\delta t=37.3$--307 days enable constraints on the $\dot M$ vs.\ $v_{\rm{shock}}$ parameter space shown in Figure \ref{fig:radiodensity}. We followed the prescriptions from \citet{Chevalier98} to compute the SSA emission as a function of the radio-spectrum observables \citep[see also the equations reported by][]{Terreran2019} and we accounted for external FFA using the formalism for the optical depth to free-free radiation by \cite{Weiler02}. The shock velocity is self-consistently calculated using the self-similar solutions by \cite{Chevalier82}.
For this calculation, we have assumed a wind-like density profile of the CSM ($\rho_{\rm{CSM}}\propto r^{-2}$) and a plasma temperature of $\sim 10^{4}$\,K. We describe this process in further detail in Appendix \ref{app:radio}. We find that for a typical shock velocity of $\sim 0.1 c$, the lack of detectable radio emission is consistent with either a low-density medium, with density corresponding to $\dot M<10^{-5}$--$10^{-6}\,\rm{M_{\sun}\,yr^{-1}}$, or a higher-density medium, with $\dot M>5 \times 10^{-4}\,\rm{M_{\sun}\,yr^{-1}}$ that would absorb the emission (Figure~\ref{fig:radiodensity}). These $\dot M$ values reported are for a wind velocity $v_w=200\,\rm{km\,s^{-1}}$. The range of allowed mass-loss rates in the lower-density case is sensitive to the choice of shock microphysical parameter values $\epsilon_e$ and $\epsilon_B$, which represent the fraction of post-shock thermal energy in relativistic electrons and magnetic fields, respectively. 

\begin{figure}[t]
\centering
\includegraphics[width=\columnwidth]{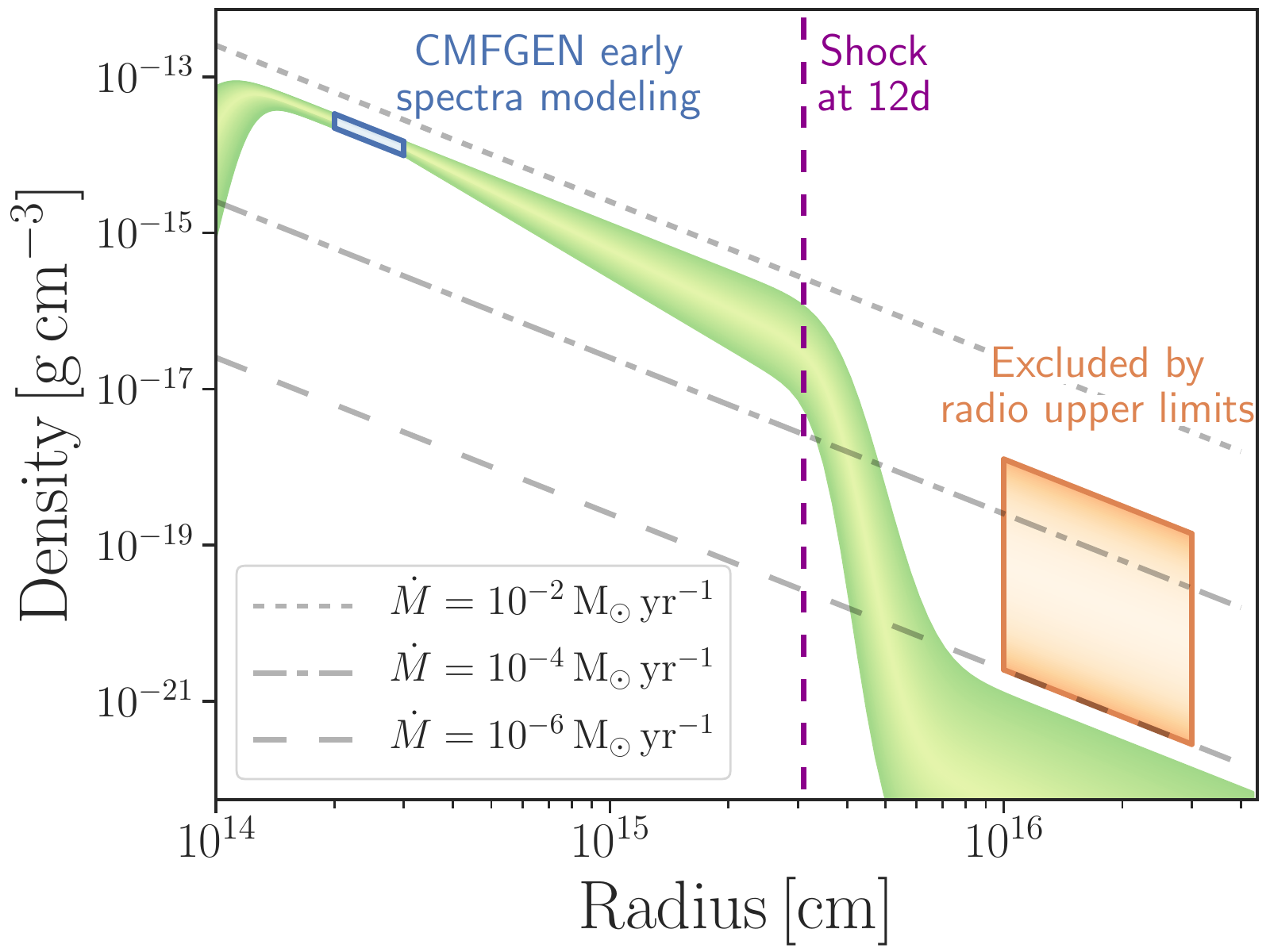} 
\caption{Proposed CSM configuration around the progenitor star of \sn{}. The dashed diagonal lines represent constant mass-loss rates as described in the legend, assuming a wind velocity of 200~$\kms$. The blue box is inferred from the early optical spectrum modeling. The orange box is the region excluded by the radio upper limits. The dashed magenta line marks the position of the shock at 12~days after first light (assuming a typical shock velocity of $\sim 0.1 c$). The narrow lines from shock ionization disappear after this epoch; therefore, we can assume that the confined CSM extended to a radius of $\sim$ (2--4) $\times10^{15}$~cm.}
\label{fig:CSM_config}
\end{figure}

In Figure \ref{fig:CSM_config} we summarize our inferences concerning the CSM that surrounded the progenitor of \sn{} at the time of explosion, similar to what \cite{Yaron17} did for SN\,2013fs. We marked in blue the densities we inferred from the CMFGEN modeling of the Keck~II+DEIMOS optical spectrum taken $\sim 1.5$\,days after first light (see Sec.~\ref{SubSubSec:specmodel}). The radio upper limits from  the previous paragraph translate to an excluded region at a higher distance, marked in orange in the figure. The shock-ionization features are present in the spectra until $\sim$12~d after first light. The position of the shock at this phase (assuming a typical velocity of $0.1 c$) is marked in the figure with a vertical dashed, magenta line. Considering the lack of narrow features after this epoch, it is fair to assume that the CSM was less dense beyond a radius of (2--4) $\times10^{15}$~cm.
The radio analysis suggested that both a high-density configuration and a low-density configuration were possible; however, the lack of narrow lines at later phases disfavors the high-density scenario. In Figure \ref{fig:CSM_config} we suggest a possible configuration of the CSM.

\section{Discussion} \label{sec:discussion}

\subsection{The Population of Shock-ionization Events} \label{sec: sample}

\begin{figure*}[t]
\centering
\includegraphics[width=\textwidth]{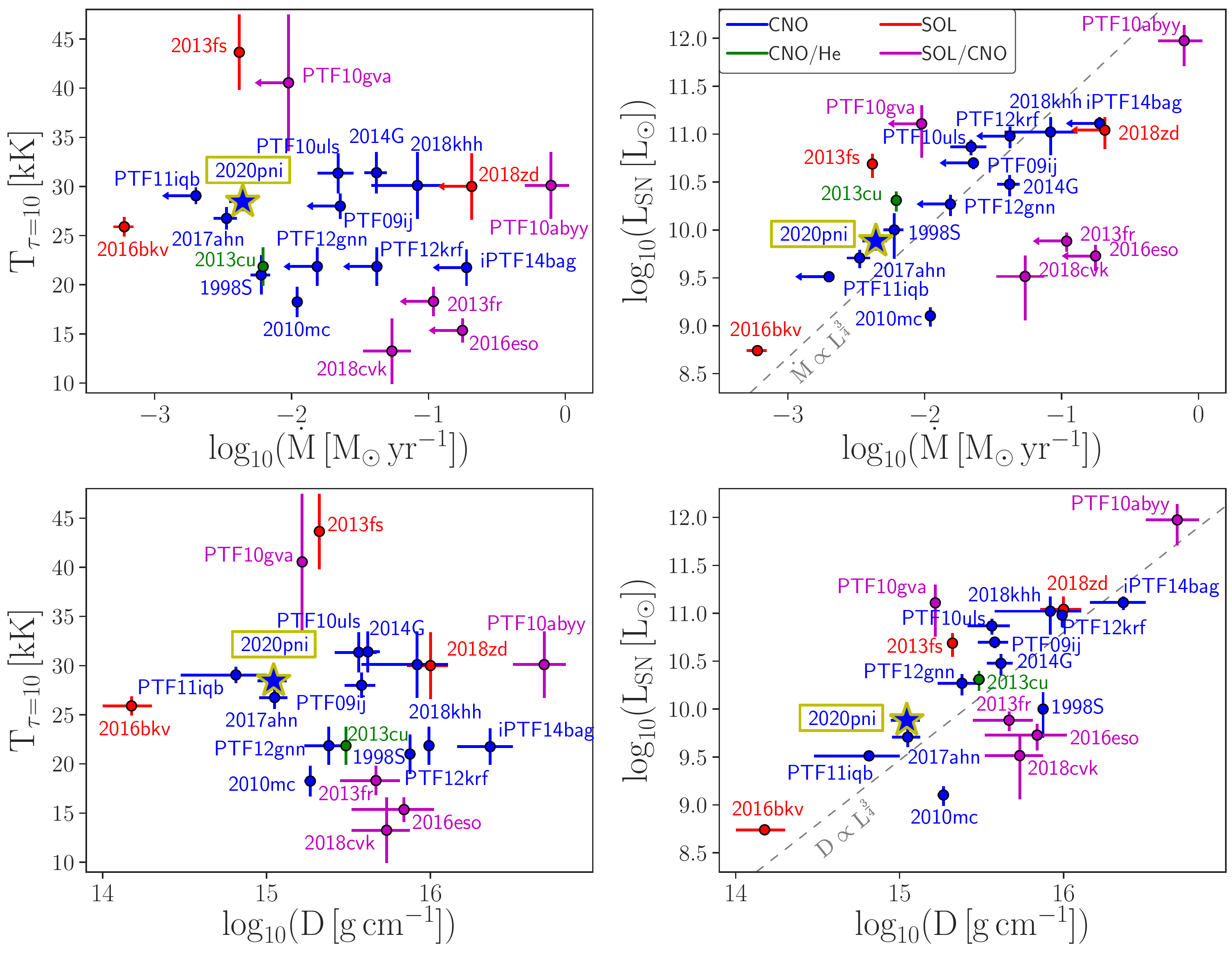} 
\caption{Adapted from Figure 8 of \cite{Boian20}; temperature, luminosity, mass-loss rate, and density relations from their sample. We also add SN 2017ahn \citep{Tartaglia21}, which was not included in the \cite{Boian20} sample, and SN 2020pni. The latter is highlighted with a star symbol. The surface abundances of the progenitor of each object are color-coded following the legend in the top right panel \citep[solar-like in red, CNO-processed in blue, intermediate between these two in magenta, and He-rich in green; see][for more details]{Boian20}. All of the values are derived from CMFGEN modeling of the optical spectra of the presented SNe.}
\label{fig:boian_comp}
\end{figure*}

In this section we compare the inferred physical properties of \sn{} with the growing population of flash-spectroscopy events. For this purpose we use the sample by \cite{Boian20}, as they performed spectral modeling using the same code, CMFGEN. \cite{Boian20} used a grid of models for their analysis and three different surface abundance scenarios: solar-like, CNO-processed, and He-rich \citep[see][for more details]{Boian20}. In Figure \ref{fig:boian_comp} we show the inferred relations between the mass-loss rate $\dot{M}$, density factor $D\equiv{\dot{M}}/({4\pi v_{\rm wind}}$), temperature $T$ at the CSM inner boundary (where the optical depth to electron scattering is $\tau\approx10$), and SN luminosity $L_{\rm SN}$. The mass-loss rate estimates rely on a measure of the wind velocity, which is not always possible to obtain, owing to the resolution of the classification spectra. Plotting the density factor instead of the mass loss has the advantage of not relying on an assumed wind velocity; however, it is less trivial to link this quantity with the general characteristics of the progenitor star. The electron temperature can be used as a proxy for the ionization level of the CSM, with higher temperatures indicating a higher ionization. \sn{} sits right in the middle of the distribution of all parameters. Indeed, the initial spectrum did not show features related to highly ionized ions such as \ion{O}{iv}, \ion{O}{v}, or \ion{N}{v} the way SN 2013fs did (Figure~\ref{fig:early_spec}). However, the temperature (and therefore the ionization level) is strongly influenced by the epoch at which the classification spectrum was acquired. SN 2013fs and PTF10gva are among the objects with the earliest spectra ($t<1$~day), while the first spectrum of \sn{} was acquired later.

In flash-ionization events, the ejecta lose kinetic energy as they are slowed down by the CSM. A denser medium allows for a more efficient conversion of kinetic energy into radiation, which can power a more luminous continuum and contribute to a more luminous SN at early phases. This energy conversion should follow $D\propto L^{3/4}$, which is in rough agreement with what is shown in the right panel of Figure \ref{fig:boian_comp}. Also, in this case \sn{} sits roughly in the middle of the distribution, with mass loss from the progenitor and therefore density parameter slightly lower than the average.

\begin{figure*}[t]
\centering
\includegraphics[width=\textwidth]{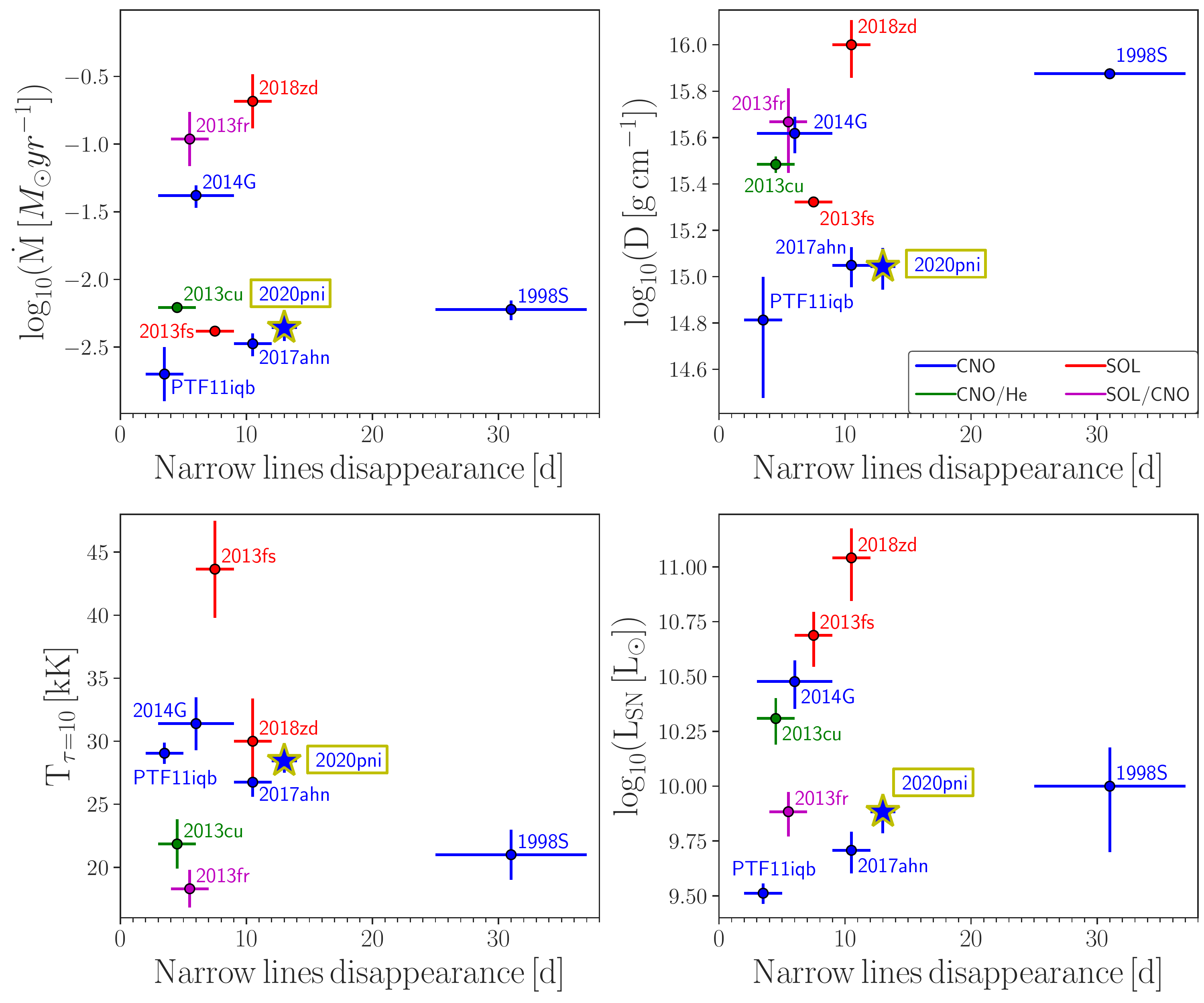} 
\caption{Subsample of the objects studied by \cite{Boian20} for which the extensive spectral coverage allowed for an estimate of the duration of the flash-ionization lines. We compare this quantity with the temperature, luminosity, mass-loss rate, and density relations from this sample derived from CMFGEN modeling. The surface abundances of the progenitor of each object are color-coded following the legend in the top right panel \citep[solar-like in red, CNO-processed in blue, intermediate between these two in magenta, and He-rich in green; see][for more details]{Boian20}.}
\label{fig:persistence_comp}
\end{figure*}

One conclusion by \cite{Boian20} was that the overall mass-loss estimates from the spectral modeling do not significantly differ from those inferred for SNe~II (from optical observations), showing narrow lines for hundreds of days. This suggests that the difference between these more extreme interacting SNe and the shock-ionization events is not to be found in the density of the CSM surrounding the progenitor stars. \cite{Boian20} suggest, therefore, that the radial extension of the thick CSM could play a major role in shaping the evolution of the SN. 

The persistence of the narrow lines in time $t_{\rm line}$ (i.e., for how long the shock-ionization lines are visible in the spectra) is an observable that was not taken into consideration by the analysis of \cite{Boian20}. The hydrogen recombination timescale (for pure H composition) can be approximated by $t_{\rm{rec}}=(\alpha n)^{-1}$, where $\alpha\approx (2-4)\times10^{-13}$\,cm$^3$\,s$^{-1}$ for $T=10^4$\,K \citep{Osterbrock2006}. Given the previously inferred density of $\sim 1.5\times10^8$\,particles\,cm$^{-3}$, we find $t_{\rm{rec}}\approx2.2\times10^4$\,s, or 0.26\,days. This is considerably smaller than $t_{\rm line}$, implying the need for a source of ionizing photons that is active well beyond the time of shock breakout. Therefore, the CSM is likely kept ionized by the prolonged interaction of the SN ejecta with the inner boundary of the CSM. We can thus link the persistence of the narrow shock-ionization lines with some physical properties of the explosion itself: (i) an explosion that launches faster shocks would have the ejecta ram through the thick CSM earlier, and (ii) a radially less extended CSM would be engulfed at earlier times by the ejecta, and therefore the narrow lines would disappear sooner. This second point is particularly important, as a less extended CSM would have been created by the progenitor star at a time closer to the explosion. Hence, the persistence of these lines could be linked to the mass-loss history right before explosion. 

We estimate the persistence of the narrow lines for all of the objects from the \cite{Boian20} sample that also had a published spectroscopic sequence beyond the classification spectrum. We identified the last spectrum showing narrow lines from the ionized CSM and the first spectrum in which the lines were absent, taking the midway point as the time of disappearance. For the majority of the targets, especially those with long-lived narrow lines, hydrogen is usually the only species left at the time of narrow-line disappearance, while the He and CNO lines disappear at earlier phases. These measurements are presented in Figure \ref{fig:persistence_comp}, where we look for correlations with the other physical properties studied above. The sample size does not allow us to come to any definitive conclusion, but at this time we do not see any clear correlations among the plotted quantities. However, an important conclusion to take is that the persistence of the narrow lines does not seem to depend on the composition of the CSM, considering that no clear distinction is evident among solar-like, CNO-processed, and He-rich events. Assuming that the CSM composition could be used as a proxy for the composition of the progenitor, this suggests that the composition probably does not play a major role in shaping the CSM.

If we refer to point (ii) mentioned above, this conclusion could lead us to suggest that any physical mechanism responsible for the observed late-time mass loss must operate under different physical conditions and stellar progenitors. We caution that the sample studied here is quite limited. A significantly larger sample size than what is currently available might reveal correlations among the parameters of Figures \ref{fig:boian_comp} and \ref{fig:persistence_comp}, and has the potential to constrain the nature of the mass-loss mechanism at work.

\subsection{The Progenitor of SN~2020pni}
So far we have shown that \sn{} was a particularly luminous (see Figure~\ref{fig:bol_LCs}) hydrogen-rich SN. We found that the CSM surrounding the progenitor star was He-rich, suggesting that the progenitor had shed a large part of its envelope at the time of explosion, or alternatively that mixing was particularly high in the progenitor star. Assuming that the progenitor was a single star, the abundance of CN-processed material at the surface of the star, and therefore in the CSM immediately surrounding it, tends to increase with the ZAMS mass of the progenitor \citep{Ekstrom2012}. Therefore, the strong nitrogen and carbon lines observed at early phases would favor a stellar mass in the higher end of the progenitors of SNe~II. A comparison with stellar evolution models computed with the Geneva code \citep{Ekstrom2012} suggests an RSG or a yellow hypergiant progenitor star for \sn{}, with a ZAMS mass between 15 and 25\,\Msun\ \citep{Groh13}. These studies have also shown that luminous blue variables (LBVs) could be the potential progenitors of some of the more H-depleted SNe~II. Both direct observations and models of stellar evolution revealed that LBVs are enriched in helium and nitrogen, while being depleted in carbon and oxygen \citep[e.g.,][]{Meynet94,Smith94,Crowther97,Najarro97}, just as is shown by \sn{}. In addition, in Section~\ref{sec: line_ev}, we demonstrated that the CSM surrounding the progenitor of \sn{} was not uniform. This is something that is also often observed in LBV nebulae \citep[e.g.,][]{Smith06,Smith2014}. Considering the further evolution of \sn{} as a relatively normal SN~II, we favor an RSG as its most probable progenitor. However, the fact that some characteristics of the CSM resemble LBV-like winds is remarkable. This once again highlights the lack of a full understanding of the mass-loss processes in RSG stars, especially during the last phases of their lives.

On a final note, binarity could have played a major role in shaping the CSM surrounding the progenitor of \sn{} and the other shock-ionization events. Considering the high percentage of progenitors of SNe~II that are expected to be in binary systems \citep[e.g.,][]{Zapartas2019}, it is possible that the surface composition of the progenitor star could have been heavily modified by the interaction of a stellar companion or even by a merger event. However, one key aspect that emerges from Section \ref{sec: sample} is that the mass loss observed in shock-ionization events appears to have been sustained for a very brief period of time (years), when compared to the whole lives of the stars themselves (millions of years). This necessarily means that the physical mechanism responsible for this mass loss is somehow linked to the actual end of the star's life. In other words, the mechanism ``knows'' that the end is coming. Interaction with a binary companion would not know about the advanced stage of evolution of the companion star.
Considering that 30\% of core-collapse SNe observed within 2 days since explosion show shock-ionization features \citep{Bruch21}, the chances that so many objects exhibit enhanced mass loss induced by the binary interaction with a companion right before the death of the progenitor star are remarkably low. Therefore, although we cannot exclude the presence of a binary companion in the system of these stars, we disfavor the idea that the physical mechanism responsible for the appearance of shock-ionization features in young core-collapse SNe is directly related to the presence of a companion star.

\section{Conclusions} \label{sec:conclusion}

In this work we presented the multiwavelength evolution of the Type II \sn{}, which exploded in the host galaxy UGC 09684. The object was discovered by ALeRCE in the ZTF data stream only a few hours after explosion, and we promptly activated a multiwavelength radio through X-ray follow-up campaign. Our first optical spectrum, obtained $\sim 1.5$~days after our estimated explosion epoch, highlighted the presence of flash-ionization features of \ion{He}{ii}, \ion{N}{iii}, \ion{N}{iv}, and \ion{C}{iv}, with a partially resolved FWHM of 200--250\,km\,s$^{-1}$. We interpret these features to have originated from a dense, confined shell of material, likely ejected by the progenitor of \sn{} in the last year before explosion. From the modeling of the first spectrum using the non-LTE radiative transfer code CMFGEN, we inferred a mass-loss rate of the progenitor of $\dot{M}=(3.5-5.3)\times10^{-3}$\,\msunyr. This is in agreement with the constraints obtained from the radio upper-limit analysis, which allows for a thick absorbing medium produced by a mass loss of $\dot M>5 \times 10^{-4}\,\rm{M_{\sun}\,yr^{-1}}$.

We then compared the inferred physical properties with those of other shock-ionization SNe, in particular with the sample presented by \cite{Boian20}. \sn{} displays characteristics that are typical of this class of objects, and no correlation between the shown physical quantities appears evident. We further investigated the persistence of the narrow lines in these objects, as this parameter could be linked to the physical extent of the CSM at the time of explosion and the timing of the mass-loss episodes responsible for the creation of this thick CSM. The sample of objects where this timescale can be measured is smaller, owing to the need for further spectroscopic follow-up observations beyond the classification spectrum in order to assess the actual persistence of the narrow lines; the comparison does not highlight any strong correlation either. This leads to a possible conclusion that the mass loss responsible for the creation of the nearby CSM is not linked to any obvious characteristics of the progenitor star. The mechanism inducing this mass loss not only has to be common to a relatively wide range of progenitor stars (with different surface compositions and size) but also needs to be linked to processes occurring toward the end of the life of a star in order to explain the timing of the mass loss. The core convection that occurs during the late-stage nuclear burning could be a viable mechanism to transmit energy, through gravity waves, from the core of the progenitor to the envelope. \cite{Wu2021} estimated that up to $10^{46}-10^{47}$\,erg could be transmitted to the outer envelope during oxygen and neon burning, and this should happen 0.1--10\,yr before core collapse. The amount of energy transferred is even higher for low-mass ($< 12$\,\Msun) and high-mass ($>30$\,\Msun) stars. The growing number of objects discovered within days of explosion should provide more evidence for these types of phenomena, resulting in a better understanding of the final phases of massive stars.

\section{Acknowledgements} \label{Sec:ack}

W.J.-G. is supported by the National Science Foundation (NSF) Graduate Research Fellowship Program under grant DGE-1842165 and the Data Science Initiative Fellowship from Northwestern University.
M.R.S. is supported by the NSF Graduate Research Fellowship Program Under grant 1842400.
R.M. acknowledges support by the NSF under grants AST-1909796 and AST-1944985. She is a CIFAR Azrieli Global Scholar in the Gravity \& the Extreme Universe Program (2019) and an Alfred P. Sloan Fellow in Physics (2019). Her team at Northwestern University is partially funded by the Heising-Simons Foundation under grant 2018-0911 (PI Margutti). The Northwestern team is partially supported by National Aeronautics and Space Administration (NASA) grant 80NSSC20K1575.
Support for D.O.J. was provided by NASA through Hubble Fellowship
grant HF2-51462.001 awarded by the Space Telescope Science Institute (STScI), which is operated by the Association of Universities for Research in Astronomy, Inc., for NASA, under contract NAS5-26555.
The UCSC team is supported in part by NASA grant 80NSSC20K0953, NSF grant AST-1815935, the Gordon \& Betty Moore Foundation, the Heising-Simons Foundation, and by a fellowship from the David and Lucile Packard Foundation to R.J.F.
C.G. is supported by a VILLUM FONDEN Young Investigator Grant (project \#25501).
H.P. is indebted to the Danish National Research Foundation (DNRF132) and the Hong Kong government (GRF grant HKU27305119) for support.
K.E. was supported by an NSF graduate research fellowship.
Y.Z. is supported by the CHE Israel Excellence Fellowship.
A.V.F.'s group at U.C. Berkeley has been supported by the Christopher R. Redlich Fund, the Miller Institute for Basic Research in Science (where A.V.F. is a Senior Miller Fellow), and many individual donors.

The Pan-STARRS1 Surveys (PS1) and the PS1 public science archive have been made possible through contributions by the Institute for Astronomy, the University of Hawaii, the Pan-STARRS Project Office, the Max-Planck Society and its participating institutes, the Max Planck Institute for Astronomy, Heidelberg and the Max Planck Institute for Extraterrestrial Physics, Garching, The Johns Hopkins University, Durham University, the University of Edinburgh, the Queen's University Belfast, the Harvard-Smithsonian Center for Astrophysics, the Las Cumbres Observatory Global Telescope Network Incorporated, the National Central University of Taiwan, STScI, NASA under grant NNX08AR22G issued through the Planetary Science Division of the NASA Science Mission Directorate, NSF grant AST-1238877, the University of Maryland, Eotvos Lorand University (ELTE), the Los Alamos National Laboratory, and the Gordon and Betty Moore Foundation.

Some of the data presented herein were obtained at the W. M. Keck Observatory, which is operated as a scientific partnership among the California Institute of Technology, the University of California, and NASA. The Observatory was made possible by the generous financial support of the W. M. Keck Foundation. The authors wish to recognize and acknowledge the very significant cultural role and reverence that the summit of Maunakea has always had within the indigenous Hawaiian community. We are most fortunate to have the opportunity to conduct observations from this mountain.
W. M. Keck Observatory and MMT Observatory access was supported by Northwestern University and the Center for Interdisciplinary Exploration and Research in Astrophysics (CIERA).

A major upgrade of the Kast spectrograph on the Shane 3~m telescope at Lick Observatory was made possible through generous gifts from the Heising-Simons Foundation as well as William and Marina Kast. Research at Lick Observatory is partially supported by a generous gift from Google. %
Based in part on observations obtained with the Samuel Oschin 48-inch Telescope at the Palomar Observatory as part of the Zwicky Transient Facility project. ZTF is supported by the NSF under grant AST-1440341 and a collaboration including Caltech, IPAC, the Weizmann Institute for Science, the Oskar Klein Center at Stockholm University, the University of Maryland, the University of Washington, Deutsches Elektronen-Synchrotron and Humboldt University, Los Alamos National Laboratories, the TANGO Consortium of Taiwan, the University of Wisconsin at Milwaukee, and the Lawrence Berkeley National Laboratory. Operations are conducted by the Caltech Optical Observatories (COO), the Infrared Processing and Analysis Center (IPAC), and the University of Washington (UW).

We acknowledge the use of public data from the {\it Neil Gehrels Swift Observatory} data archive.
Parts of this research were supported by the Australian Research Council Centre of Excellence for All Sky Astrophysics in 3 Dimensions (ASTRO 3D), through project \#CE170100013.
%
The National Radio Astronomy Observatory is a facility of the NSF operated under cooperative agreement by Associated Universities, Inc.

\facilities{Gemini, Keck, LCO, MMT, NOT, Pan-STARRS, Lick Shane 3\,m, Swift, VLA, ZTF}
\software{
APLpy \citep[v1.1.1;][]{Robitaille2012},
Astropy \citep[v2.0.1;][]{AstropyCollaboration2013,AstropyCollaboration2018},
CASA \citep[v5.4.1;][]{McMullin2007},
DAOPHOT \citep[v2.14.1;][]{Stetson1987},
HEAsoft (v6.22; HEASARC 2014),
IRAF \citep[v2.16;][]{Tody1986,Tody1993},
NumPy \citep[v1.13.1;][]{Oliphant2016},
Matplotlib \citep[v2.0.2;][]{Hunter2007},
PypeIt \citep[v1.0.6;][]{Prochaska2020},
SciPy \citep[v0.19.1;][]{Jones2001},
SNOoPY (Cappellaro, E. 2014; \url{http://sngroup.oapd.inaf.it/snoopy.html}),
Source Extractor \citep[v2.19.5;][]{Bertin1996},
}

\bibliographystyle{aasjournal} 
\bibliography{references}

\begin{thebibliography}{}
\expandafter\ifx\csname natexlab\endcsname\relax\def\natexlab#1{#1}\fi
\providecommand{\url}[1]{\href{#1}{#1}}
\providecommand{\dodoi}[1]{doi:~\href{http://doi.org/#1}{\nolinkurl{#1}}}
\providecommand{\doeprint}[1]{\href{http://ascl.net/#1}{\nolinkurl{http://ascl.net/#1}}}
\providecommand{\doarXiv}[1]{\href{https://arxiv.org/abs/#1}{\nolinkurl{https://arxiv.org/abs/#1}}}

\bibitem[{{Astropy Collaboration} {et~al.}(2013){Astropy Collaboration},
  {Robitaille}, {Tollerud}, {Greenfield}, {Droettboom}, {Bray}, {Aldcroft},
  {Davis}, {Ginsburg}, {Price-Whelan}, {Kerzendorf}, {Conley}, {Crighton},
  {Barbary}, {Muna}, {Ferguson}, {Grollier}, {Parikh}, {Nair}, {Unther},
  {Deil}, {Woillez}, {Conseil}, {Kramer}, {Turner}, {Singer}, {Fox}, {Weaver},
  {Zabalza}, {Edwards}, {Azalee Bostroem}, {Burke}, {Casey}, {Crawford},
  {Dencheva}, {Ely}, {Jenness}, {Labrie}, {Lim}, {Pierfederici}, {Pontzen},
  {Ptak}, {Refsdal}, {Servillat}, \& {Streicher}}]{AstropyCollaboration2013}
{Astropy Collaboration}, {Robitaille}, T.~P., {Tollerud}, E.~J., {et~al.} 2013,
  \aap, 558, A33, \dodoi{10.1051/0004-6361/201322068}

\bibitem[{{Astropy Collaboration} {et~al.}(2018){Astropy Collaboration},
  {Price-Whelan}, {Sip{\H o}cz}, {G{\"u}nther}, {Lim}, {Crawford}, {Conseil},
  {Shupe}, {Craig}, {Dencheva}, {Ginsburg}, {VanderPlas}, {Bradley},
  {P{\'e}rez-Su{\'a}rez}, {de Val-Borro}, {Aldcroft}, {Cruz}, {Robitaille},
  {Tollerud}, {Ardelean}, {Babej}, {Bach}, {Bachetti}, {Bakanov}, {Bamford},
  {Barentsen}, {Barmby}, {Baumbach}, {Berry}, {Biscani}, {Boquien}, {Bostroem},
  {Bouma}, {Brammer}, {Bray}, {Breytenbach}, {Buddelmeijer}, {Burke},
  {Calderone}, {Cano Rodr{\'{\i}}guez}, {Cara}, {Cardoso}, {Cheedella},
  {Copin}, {Corrales}, {Crichton}, {D'Avella}, {Deil}, {Depagne}, {Dietrich},
  {Donath}, {Droettboom}, {Earl}, {Erben}, {Fabbro}, {Ferreira}, {Finethy},
  {Fox}, {Garrison}, {Gibbons}, {Goldstein}, {Gommers}, {Greco}, {Greenfield},
  {Groener}, {Grollier}, {Hagen}, {Hirst}, {Homeier}, {Horton}, {Hosseinzadeh},
  {Hu}, {Hunkeler}, {Ivezi{\'c}}, {Jain}, {Jenness}, {Kanarek}, {Kendrew},
  {Kern}, {Kerzendorf}, {Khvalko}, {King}, {Kirkby}, {Kulkarni}, {Kumar},
  {Lee}, {Lenz}, {Littlefair}, {Ma}, {Macleod}, {Mastropietro}, {McCully},
  {Montagnac}, {Morris}, {Mueller}, {Mumford}, {Muna}, {Murphy}, {Nelson},
  {Nguyen}, {Ninan}, {N{\"o}the}, {Ogaz}, {Oh}, {Parejko}, {Parley}, {Pascual},
  {Patil}, {Patil}, {Plunkett}, {Prochaska}, {Rastogi}, {Reddy Janga},
  {Sabater}, {Sakurikar}, {Seifert}, {Sherbert}, {Sherwood-Taylor}, {Shih},
  {Sick}, {Silbiger}, {Singanamalla}, {Singer}, {Sladen}, {Sooley},
  {Sornarajah}, {Streicher}, {Teuben}, {Thomas}, {Tremblay}, {Turner},
  {Terr{\'o}n}, {van Kerkwijk}, {de la Vega}, {Watkins}, {Weaver}, {Whitmore},
  {Woillez}, {Zabalza}, \& {Astropy Contributors}}]{AstropyCollaboration2018}
{Astropy Collaboration}, {Price-Whelan}, A.~M., {Sip{\H o}cz}, B.~M., {et~al.}
  2018, \aj, 156, 123, \dodoi{10.3847/1538-3881/aabc4f}

\bibitem[{{Becker}(2015)}]{becker15}
{Becker}, A. 2015, {HOTPANTS: High Order Transform of PSF ANd Template
  Subtraction}, Astrophysics Source Code Library.
\newblock \doeprint{1504.004}

\bibitem[{{Bellm} {et~al.}(2019){Bellm}, {Kulkarni}, {Graham}, {Dekany},
  {Smith}, {Riddle}, {Masci}, {Helou}, {Prince}, {Adams}, {Barbarino},
  {Barlow}, {Bauer}, {Beck}, {Belicki}, {Biswas}, {Blagorodnova}, {Bodewits},
  {Bolin}, {Brinnel}, {Brooke}, {Bue}, {Bulla}, {Burruss}, {Cenko}, {Chang},
  {Connolly}, {Coughlin}, {Cromer}, {Cunningham}, {De}, {Delacroix}, {Desai},
  {Duev}, {Eadie}, {Farnham}, {Feeney}, {Feindt}, {Flynn}, {Franckowiak},
  {Frederick}, {Fremling}, {Gal-Yam}, {Gezari}, {Giomi}, {Goldstein},
  {Golkhou}, {Goobar}, {Groom}, {Hacopians}, {Hale}, {Henning}, {Ho}, {Hover},
  {Howell}, {Hung}, {Huppenkothen}, {Imel}, {Ip}, {Ivezi{\'c}}, {Jackson},
  {Jones}, {Juric}, {Kasliwal}, {Kaspi}, {Kaye}, {Kelley}, {Kowalski},
  {Kramer}, {Kupfer}, {Landry}, {Laher}, {Lee}, {Lin}, {Lin}, {Lunnan},
  {Giomi}, {Mahabal}, {Mao}, {Miller}, {Monkewitz}, {Murphy}, {Ngeow},
  {Nordin}, {Nugent}, {Ofek}, {Patterson}, {Penprase}, {Porter}, {Rauch},
  {Rebbapragada}, {Reiley}, {Rigault}, {Rodriguez}, {van Roestel}, {Rusholme},
  {van Santen}, {Schulze}, {Shupe}, {Singer}, {Soumagnac}, {Stein}, {Surace},
  {Sollerman}, {Szkody}, {Taddia}, {Terek}, {Van Sistine}, {van Velzen},
  {Vestrand}, {Walters}, {Ward}, {Ye}, {Yu}, {Yan}, \& {Zolkower}}]{bellm19}
{Bellm}, E.~C., {Kulkarni}, S.~R., {Graham}, M.~J., {et~al.} 2019, \pasp, 131,
  018002, \dodoi{10.1088/1538-3873/aaecbe}

\bibitem[{{Berger} {et~al.}(2002){Berger}, {Kulkarni}, \&
  {Chevalier}}]{Berger02}
{Berger}, E., {Kulkarni}, S.~R., \& {Chevalier}, R.~A. 2002, \apjl, 577, L5,
  \dodoi{10.1086/344045}

\bibitem[{{Bertin} \& {Arnouts}(1996)}]{Bertin1996}
{Bertin}, E., \& {Arnouts}, S. 1996, \aaps, 117, 393,
  \dodoi{10.1051/aas:1996164}

\bibitem[{{Blagorodnova} {et~al.}(2018){Blagorodnova}, {Neill}, {Walters},
  {Kulkarni}, {Fremling}, {Ben-Ami}, {Dekany}, {Fucik}, {Konidaris}, {Nash},
  {Ngeow}, {Ofek}, {O' Sullivan}, {Quimby}, {Ritter}, \&
  {Vyhmeister}}]{Blagorodnova2018}
{Blagorodnova}, N., {Neill}, J.~D., {Walters}, R., {et~al.} 2018, \pasp, 130,
  035003, \dodoi{10.1088/1538-3873/aaa53f}

\bibitem[{{Boian} \& {Groh}(2018)}]{Boian18}
{Boian}, I., \& {Groh}, J.~H. 2018, \aap, 617, A115,
  \dodoi{10.1051/0004-6361/201731794}

\bibitem[{{Boian} \& {Groh}(2019)}]{Boian19}
---. 2019, \aap, 621, A109, \dodoi{10.1051/0004-6361/201833779}

\bibitem[{{Boian} \& {Groh}(2020)}]{Boian20}
---. 2020, \mnras, 496, 1325, \dodoi{10.1093/mnras/staa1540}

\bibitem[{{Brown} {et~al.}(2014){Brown}, {Breeveld}, {Holland}, {Kuin}, \&
  {Pritchard}}]{Brown14}
{Brown}, P.~J., {Breeveld}, A.~A., {Holland}, S., {Kuin}, P., \& {Pritchard},
  T. 2014, \apss, 354, 89, \dodoi{10.1007/s10509-014-2059-8}

\bibitem[{{Brown} {et~al.}(2009){Brown}, {Holland}, {Immler}, {Milne},
  {Roming}, {Gehrels}, {Nousek}, {Panagia}, {Still}, \& {Vanden
  Berk}}]{Brown09}
{Brown}, P.~J., {Holland}, S.~T., {Immler}, S., {et~al.} 2009, \aj, 137, 4517,
  \dodoi{10.1088/0004-6256/137/5/4517}

\bibitem[{{Brown} {et~al.}(2010){Brown}, {Roming}, {Milne}, {Bufano},
  {Ciardullo}, {Elias-Rosa}, {Filippenko}, {Foley}, {Gehrels}, {Gronwall},
  {Hicken}, {Holland}, {Hoversten}, {Immler}, {Kirshner}, {Li}, {Mazzali},
  {Phillips}, {Pritchard}, {Still}, {Turatto}, \& {Vanden Berk}}]{brown10}
{Brown}, P.~J., {Roming}, P. W.~A., {Milne}, P., {et~al.} 2010, \apj, 721,
  1608, \dodoi{10.1088/0004-637X/721/2/1608}

\bibitem[{{Brown} {et~al.}(2013){Brown}, {Baliber}, {Bianco}, {Bowman},
  {Burleson}, {Conway}, {Crellin}, {Depagne}, {De Vera}, {Dilday}, {Dragomir},
  {Dubberley}, {Eastman}, {Elphick}, {Falarski}, {Foale}, {Ford}, {Fulton},
  {Garza}, {Gomez}, {Graham}, {Greene}, {Haldeman}, {Hawkins}, {Haworth},
  {Haynes}, {Hidas}, {Hjelstrom}, {Howell}, {Hygelund}, {Lister}, {Lobdill},
  {Martinez}, {Mullins}, {Norbury}, {Parrent}, {Paulson}, {Petry}, {Pickles},
  {Posner}, {Rosing}, {Ross}, {Sand}, {Saunders}, {Shobbrook}, {Shporer},
  {Street}, {Thomas}, {Tsapras}, {Tufts}, {Valenti}, {Vander Horst}, {Walker},
  {White}, \& {Willis}}]{Brown13}
{Brown}, T.~M., {Baliber}, N., {Bianco}, F.~B., {et~al.} 2013, \pasp, 125,
  1031, \dodoi{10.1086/673168}

\bibitem[{{Bruch} {et~al.}(2020){Bruch}, {Schulze}, \& {Gal-Yam}}]{Bruch2020b}
{Bruch}, R., {Schulze}, S., \& {Gal-Yam}, A. 2020, Transient Name Server
  Classification Report, 2020-2170, 1

\bibitem[{{Bruch} {et~al.}(2021){Bruch}, {Gal-Yam}, {Schulze}, {Yaron}, {Yang},
  {Soumagnac}, {Rigault}, {Strotjohann}, {Ofek}, {Sollerman}, {Masci},
  {Barbarino}, {Ho}, {Fremling}, {Perley}, {Nordin}, {Cenko}, {Adams},
  {Adreoni}, {Bellm}, {Blagorodnova}, {Bulla}, {Burdge}, {De}, {Dhawan},
  {Drake}, {Duev}, {Dugas}, {Graham}, {Graham}, {Irani}, {Jencson},
  {Karamehmetoglu}, {Kasliwal}, {Kim}, {Kulkarni}, {Kupfer}, {Liang},
  {Mahabal}, {Miller}, {Prince}, {Riddle}, {Sharma}, {Smith}, {Taddia},
  {Taggart}, {Walters}, \& {Yan}}]{Bruch21}
{Bruch}, R.~J., {Gal-Yam}, A., {Schulze}, S., {et~al.} 2021, \apj, 912, 46,
  \dodoi{10.3847/1538-4357/abef05}

\bibitem[{{Bruzual} \& {Charlot}(2003)}]{Bruzual2003}
{Bruzual}, G., \& {Charlot}, S. 2003, \mnras, 344, 1000,
  \dodoi{10.1046/j.1365-8711.2003.06897.x}

\bibitem[{{Burrows} {et~al.}(2005){Burrows}, {Hill}, {Nousek}, {Kennea},
  {Wells}, {Osborne}, {Abbey}, {Beardmore}, {Mukerjee}, {Short}, {Chincarini},
  {Campana}, {Citterio}, {Moretti}, {Pagani}, {Tagliaferri}, {Giommi},
  {Capalbi}, {Tamburelli}, {Angelini}, {Cusumano}, {Br{\"a}uninger}, {Burkert},
  \& {Hartner}}]{burrows05}
{Burrows}, D.~N., {Hill}, J.~E., {Nousek}, J.~A., {et~al.} 2005, \ssr, 120,
  165, \dodoi{10.1007/s11214-005-5097-2}

\bibitem[{{Calzetti} {et~al.}(2000){Calzetti}, {Armus}, {Bohlin}, {Kinney},
  {Koornneef}, \& {Storchi-Bergmann}}]{Calzetti2000}
{Calzetti}, D., {Armus}, L., {Bohlin}, R.~C., {et~al.} 2000, \apj, 533, 682,
  \dodoi{10.1086/308692}

\bibitem[{{Chabrier}(2003)}]{Chabrier2003}
{Chabrier}, G. 2003, in IAU Symposium, Vol. 221, IAU Symposium, P67

\bibitem[{{Chambers} {et~al.}(2017){Chambers}, {Huber}, {Flewelling},
  {Magnier}, {Schultz}, {Lowe}, {Smartt}, {Smith}, {Tonry}, {Waters}, {Wright},
  \& {Young}}]{Chambers2017}
{Chambers}, K.~C., {Huber}, M.~E., {Flewelling}, H., {et~al.} 2017, Transient
  Name Server Discovery Report, 2017-324, 1

\bibitem[{{Chevalier}(1982)}]{Chevalier82}
{Chevalier}, R.~A. 1982, \apj, 258, 790, \dodoi{10.1086/160126}

\bibitem[{{Chevalier}(1998)}]{Chevalier98}
---. 1998, \apj, 499, 810, \dodoi{10.1086/305676}

\bibitem[{{Chevalier} \& {Fransson}(1994)}]{Chevalier1994}
{Chevalier}, R.~A., \& {Fransson}, C. 1994, \apj, 420, 268,
  \dodoi{10.1086/173557}

\bibitem[{{Chevalier} \& {Fransson}(2006)}]{Chevalier06}
---. 2006, \apj, 651, 381, \dodoi{10.1086/507606}

\bibitem[{{Chevalier} \& {Irwin}(2011)}]{Chevalier11}
{Chevalier}, R.~A., \& {Irwin}, C.~M. 2011, \apjl, 729, L6,
  \dodoi{10.1088/2041-8205/729/1/L6}

\bibitem[{{Conroy} {et~al.}(2009){Conroy}, {Gunn}, \& {White}}]{Conroy2009}
{Conroy}, C., {Gunn}, J.~E., \& {White}, M. 2009, \apj, 699, 486,
  \dodoi{10.1088/0004-637X/699/1/486}

\bibitem[{{Crowther}(1997)}]{Crowther97}
{Crowther}, P.~A. 1997, in Astronomical Society of the Pacific Conference
  Series, Vol. 120, Luminous Blue Variables: Massive Stars in Transition, ed.
  A.~{Nota} \& H.~{Lamers}, 51

\bibitem[{{de Jaeger} {et~al.}(2019){de Jaeger}, {Zheng}, {Stahl},
  {Filippenko}, {Brink}, {Bigley}, {Blanchard}, {Blanchard}, {Bradley},
  {Cargill}, {Casper}, {Cenko}, {Channa}, {Choi}, {Clubb}, {Cobb}, {Cohen}, {de
  Kouchkovsky}, {Ellison}, {Falcon}, {Fox}, {Fuller}, {Ganeshalingam}, {Gould},
  {Graham}, {Halevi}, {Hayakawa}, {Hestenes}, {Hyland}, {Jeffers}, {Joubert},
  {Kandrashoff}, {Kelly}, {Kim}, {Kim}, {Kumar}, {Leonard}, {Li}, {Lowe}, {Lu},
  {Mason}, {McAllister}, {Mauerhan}, {Modjaz}, {Molloy}, {Perley}, {Pina},
  {Poznanski}, {Ross}, {Shivvers}, {Silverman}, {Soler}, {Stegman}, {Taylor},
  {Tang}, {Wilkins}, {Wang}, {Wang}, {Yuk}, {Yunus}, \& {Zhang}}]{deJaeger2019}
{de Jaeger}, T., {Zheng}, W., {Stahl}, B.~E., {et~al.} 2019, \mnras, 490, 2799,
  \dodoi{10.1093/mnras/stz2714}

\bibitem[{{de Jager} {et~al.}(1988){de Jager}, {Nieuwenhuijzen}, \& {van der
  Hucht}}]{deJager1988}
{de Jager}, C., {Nieuwenhuijzen}, H., \& {van der Hucht}, K.~A. 1988, \aaps,
  72, 259

\bibitem[{{Dessart} {et~al.}(2016){Dessart}, {Hillier}, {Audit}, {Livne}, \&
  {Waldman}}]{Dessart2016}
{Dessart}, L., {Hillier}, D.~J., {Audit}, E., {Livne}, E., \& {Waldman}, R.
  2016, \mnras, 458, 2094, \dodoi{10.1093/mnras/stw336}

\bibitem[{{Dessart} {et~al.}(2013){Dessart}, {Hillier}, {Waldman}, \&
  {Livne}}]{Dessart2013}
{Dessart}, L., {Hillier}, D.~J., {Waldman}, R., \& {Livne}, E. 2013, \mnras,
  433, 1745, \dodoi{10.1093/mnras/stt861}

\bibitem[{{Dessart} {et~al.}(2017){Dessart}, {John Hillier}, \&
  {Audit}}]{Dessart2017}
{Dessart}, L., {John Hillier}, D., \& {Audit}, E. 2017, \aap, 605, A83,
  \dodoi{10.1051/0004-6361/201730942}

\bibitem[{{Dessart} {et~al.}(2010){Dessart}, {Livne}, \&
  {Waldman}}]{Dessart2010}
{Dessart}, L., {Livne}, E., \& {Waldman}, R. 2010, \mnras, 405, 2113,
  \dodoi{10.1111/j.1365-2966.2010.16626.x}

\bibitem[{{Ekstr{\"o}m} {et~al.}(2012){Ekstr{\"o}m}, {Georgy}, {Eggenberger},
  {Meynet}, {Mowlavi}, {Wyttenbach}, {Granada}, {Decressin}, {Hirschi},
  {Frischknecht}, {Charbonnel}, \& {Maeder}}]{Ekstrom2012}
{Ekstr{\"o}m}, S., {Georgy}, C., {Eggenberger}, P., {et~al.} 2012, \aap, 537,
  A146, \dodoi{10.1051/0004-6361/201117751}

\bibitem[{{Fassia} {et~al.}(2001){Fassia}, {Meikle}, {Chugai}, {Geballe},
  {Lundqvist}, {Walton}, {Pollacco}, {Veilleux}, {Wright}, {Pettini}, {Kerr},
  {Puchnarewicz}, {Puxley}, {Irwin}, {Packham}, {Smartt}, \&
  {Harmer}}]{fassia01}
{Fassia}, A., {Meikle}, W.~P.~S., {Chugai}, N., {et~al.} 2001, \mnras, 325,
  907, \dodoi{10.1046/j.1365-8711.2001.04282.x}

\bibitem[{{Filippenko}(1982)}]{filippenko82}
{Filippenko}, A.~V. 1982, \pasp, 94, 715, \dodoi{10.1086/131052}

\bibitem[{{Filippenko}(1997)}]{filippenko97}
---. 1997, \araa, 35, 309, \dodoi{10.1146/annurev.astro.35.1.309}

\bibitem[{{Foley} {et~al.}(2006){Foley}, {Silverman}, {Moore}, \&
  {Filippenko}}]{Foley2006}
{Foley}, R.~J., {Silverman}, J.~M., {Moore}, M., \& {Filippenko}, A.~V. 2006,
  Central Bureau Electronic Telegrams, 604, 1

\bibitem[{{Foley} {et~al.}(2007){Foley}, {Smith}, {Ganeshalingam}, {Li},
  {Chornock}, \& {Filippenko}}]{Foley2007}
{Foley}, R.~J., {Smith}, N., {Ganeshalingam}, M., {et~al.} 2007, \apjl, 657,
  L105, \dodoi{10.1086/513145}

\bibitem[{{Forster} {et~al.}(2020){Forster}, {Bauer}, {Galbany}, {Pignata},
  {Camacho}, {Silva-Farfan}, {Mourao}, {Arredondo}, {Cabrera-Vives},
  {Carrasco-Davis}, {Estevez}, {Huijse}, {Reyes}, {Reyes}, {Sanchez-Saez},
  {Valenzuela}, {Castillo}, {Ruz-Mieres}, {Rodriguez-Mancini}, {Catelan},
  {Eyheramendy}, \& {Graham}}]{Forster2020}
{Forster}, F., {Bauer}, F.~E., {Galbany}, L., {et~al.} 2020, Transient Name
  Server Discovery Report, 2020-2150, 1

\bibitem[{{Fuller}(2017)}]{Fuller2017}
{Fuller}, J. 2017, \mnras, 470, 1642, \dodoi{10.1093/mnras/stx1314}

\bibitem[{{Gal-Yam} {et~al.}(2014){Gal-Yam}, {Arcavi}, {Ofek}, {Ben-Ami},
  {Cenko}, {Kasliwal}, {Cao}, {Yaron}, {Tal}, {Silverman}, {Horesh}, {De Cia},
  {Taddia}, {Sollerman}, {Perley}, {Vreeswijk}, {Kulkarni}, {Nugent},
  {Filippenko}, \& {Wheeler}}]{Gal-Yam2014}
{Gal-Yam}, A., {Arcavi}, I., {Ofek}, E.~O., {et~al.} 2014, \nat, 509, 471,
  \dodoi{10.1038/nature13304}

\bibitem[{{Gangopadhyay} {et~al.}(2020){Gangopadhyay}, {Misra}, {Hiramatsu},
  {Wang}, {Hosseinzadeh}, {Wang}, {Valenti}, {Zhang}, {Howell}, {Arcavi},
  {Anupama}, {Burke}, {Dastidar}, {Itagaki}, {Kumar}, {Kumar}, {Li}, {McCully},
  {Mo}, {Pandey}, {Pellegrino}, {Sai}, {Sahu}, {Sanwal}, {Singh}, {Singh},
  {Zhang}, {Zhang}, \& {Zhang}}]{Gangopadhyay20}
{Gangopadhyay}, A., {Misra}, K., {Hiramatsu}, D., {et~al.} 2020, \apj, 889,
  170, \dodoi{10.3847/1538-4357/ab6328}

\bibitem[{{Gehrels} {et~al.}(2004){Gehrels}, {Chincarini}, {Giommi}, {Mason},
  {Nousek}, {Wells}, {White}, {Barthelmy}, {Burrows}, {Cominsky}, {Hurley},
  {Marshall}, {M{\'e}sz{\'a}ros}, {Roming}, {Angelini}, {Barbier}, {Belloni},
  {Campana}, {Caraveo}, {Chester}, {Citterio}, {Cline}, {Cropper}, {Cummings},
  {Dean}, {Feigelson}, {Fenimore}, {Frail}, {Fruchter}, {Garmire}, {Gendreau},
  {Ghisellini}, {Greiner}, {Hill}, {Hunsberger}, {Krimm}, {Kulkarni}, {Kumar},
  {Lebrun}, {Lloyd-Ronning}, {Markwardt}, {Mattson}, {Mushotzky}, {Norris},
  {Osborne}, {Paczynski}, {Palmer}, {Park}, {Parsons}, {Paul}, {Rees},
  {Reynolds}, {Rhoads}, {Sasseen}, {Schaefer}, {Short}, {Smale}, {Smith},
  {Stella}, {Tagliaferri}, {Takahashi}, {Tashiro}, {Townsley}, {Tueller},
  {Turner}, {Vietri}, {Voges}, {Ward}, {Willingale}, {Zerbi}, \&
  {Zhang}}]{Gehrels04}
{Gehrels}, N., {Chincarini}, G., {Giommi}, P., {et~al.} 2004, \apj, 611, 1005,
  \dodoi{10.1086/422091}

\bibitem[{{Graham} {et~al.}(2019){Graham}, {Kulkarni}, {Bellm}, {Adams},
  {Barbarino}, {Blagorodnova}, {Bodewits}, {Bolin}, {Brady}, {Cenko}, {Chang},
  {Coughlin}, {De}, {Eadie}, {Farnham}, {Feindt}, {Franckowiak}, {Fremling},
  {Gezari}, {Ghosh}, {Goldstein}, {Golkhou}, {Goobar}, {Ho}, {Huppenkothen},
  {Ivezi{\'c}}, {Jones}, {Juric}, {Kaplan}, {Kasliwal}, {Kelley}, {Kupfer},
  {Lee}, {Lin}, {Lunnan}, {Mahabal}, {Miller}, {Ngeow}, {Nugent}, {Ofek},
  {Prince}, {Rauch}, {van Roestel}, {Schulze}, {Singer}, {Sollerman}, {Taddia},
  {Yan}, {Ye}, {Yu}, {Barlow}, {Bauer}, {Beck}, {Belicki}, {Biswas}, {Brinnel},
  {Brooke}, {Bue}, {Bulla}, {Burruss}, {Connolly}, {Cromer}, {Cunningham},
  {Dekany}, {Delacroix}, {Desai}, {Duev}, {Feeney}, {Flynn}, {Frederick},
  {Gal-Yam}, {Giomi}, {Groom}, {Hacopians}, {Hale}, {Helou}, {Henning},
  {Hover}, {Hillenbrand}, {Howell}, {Hung}, {Imel}, {Ip}, {Jackson}, {Kaspi},
  {Kaye}, {Kowalski}, {Kramer}, {Kuhn}, {Landry}, {Laher}, {Mao}, {Masci},
  {Monkewitz}, {Murphy}, {Nordin}, {Patterson}, {Penprase}, {Porter},
  {Rebbapragada}, {Reiley}, {Riddle}, {Rigault}, {Rodriguez}, {Rusholme}, {van
  Santen}, {Shupe}, {Smith}, {Soumagnac}, {Stein}, {Surace}, {Szkody}, {Terek},
  {Van Sistine}, {van Velzen}, {Vestrand}, {Walters}, {Ward}, {Zhang}, \&
  {Zolkower}}]{graham19}
{Graham}, M.~J., {Kulkarni}, S.~R., {Bellm}, E.~C., {et~al.} 2019, \pasp, 131,
  078001, \dodoi{10.1088/1538-3873/ab006c}

\bibitem[{{Groh}(2014)}]{Groh14}
{Groh}, J.~H. 2014, \aap, 572, L11, \dodoi{10.1051/0004-6361/201424852}

\bibitem[{{Groh} {et~al.}(2013){Groh}, {Meynet}, {Georgy}, \&
  {Ekstr{\"o}m}}]{Groh13}
{Groh}, J.~H., {Meynet}, G., {Georgy}, C., \& {Ekstr{\"o}m}, S. 2013, \aap,
  558, A131, \dodoi{10.1051/0004-6361/201321906}

\bibitem[{{Guillochon} {et~al.}(2017){Guillochon}, {Parrent}, {Kelley}, \&
  {Margutti}}]{Guillochon2017}
{Guillochon}, J., {Parrent}, J., {Kelley}, L.~Z., \& {Margutti}, R. 2017, \apj,
  835, 64, \dodoi{10.3847/1538-4357/835/1/64}

\bibitem[{{Guti{\'e}rrez} {et~al.}(2017){Guti{\'e}rrez}, {Anderson}, {Hamuy},
  {Gonz{\'a}lez-Gaitan}, {Galbany}, {Dessart}, {Stritzinger}, {Phillips},
  {Morrell}, \& {Folatelli}}]{Gutierrez2017}
{Guti{\'e}rrez}, C.~P., {Anderson}, J.~P., {Hamuy}, M., {et~al.} 2017, \apj,
  850, 90, \dodoi{10.3847/1538-4357/aa8f42}

\bibitem[{{Hamuy} {et~al.}(2001){Hamuy}, {Pinto}, {Maza}, {Suntzeff},
  {Phillips}, {Eastman}, {Smith}, {Corbally}, {Burstein}, {Li}, {Ivanov},
  {Moro-Martin}, {Strolger}, {de Souza}, {dos Anjos}, {Green}, {Pickering},
  {Gonz{\'a}lez}, {Antezana}, {Wischnjewsky}, {Galaz}, {Roth}, {Persson}, \&
  {Schommer}}]{Hamuy2001}
{Hamuy}, M., {Pinto}, P.~A., {Maza}, J., {et~al.} 2001, \apj, 558, 615,
  \dodoi{10.1086/322450}

\bibitem[{{Heger} {et~al.}(1997){Heger}, {Jeannin}, {Langer}, \&
  {Baraffe}}]{Heger1997}
{Heger}, A., {Jeannin}, L., {Langer}, N., \& {Baraffe}, I. 1997, \aap, 327,
  224.
\newblock \doarXiv{astro-ph/9705097}

\bibitem[{{Hillier} \& {Miller}(1998)}]{Hillier96}
{Hillier}, D.~J., \& {Miller}, D.~L. 1998, \apj, 496, 407,
  \dodoi{10.1086/305350}

\bibitem[{{Hosseinzadeh} {et~al.}(2018){Hosseinzadeh}, {Valenti}, {McCully},
  {Howell}, {Arcavi}, {Jerkstrand}, {Guevel}, {Tartaglia}, {Rui}, {Mo}, {Wang},
  {Huang}, {Song}, {Zhang}, \& {Itagaki}}]{Hosseinzadeh2018}
{Hosseinzadeh}, G., {Valenti}, S., {McCully}, C., {et~al.} 2018, \apj, 861, 63,
  \dodoi{10.3847/1538-4357/aac5f6}

\bibitem[{Hunter(2007)}]{Hunter2007}
Hunter, J.~D. 2007, Computing in Science \& Engineering, 9, 90,
  \dodoi{10.1109/MCSE.2007.55}

\bibitem[{{Inserra} {et~al.}(2011){Inserra}, {Turatto}, {Pastorello},
  {Benetti}, {Cappellaro}, {Pumo}, {Zampieri}, {Agnoletto}, {Bufano},
  {Botticella}, {Della Valle}, {Elias Rosa}, {Iijima}, {Spiro}, \&
  {Valenti}}]{Inserra2011}
{Inserra}, C., {Turatto}, M., {Pastorello}, A., {et~al.} 2011, \mnras, 417,
  261, \dodoi{10.1111/j.1365-2966.2011.19128.x}

\bibitem[{{Jarrett} {et~al.}(2000){Jarrett}, {Chester}, {Cutri}, {Schneider},
  {Skrutskie}, \& {Huchra}}]{Jarrett2000}
{Jarrett}, T.~H., {Chester}, T., {Cutri}, R., {et~al.} 2000, \aj, 119, 2498,
  \dodoi{10.1086/301330}

\bibitem[{{Jerkstrand} {et~al.}(2017){Jerkstrand}, {Smartt}, {Inserra},
  {Nicholl}, {Chen}, {Kr{\"u}hler}, {Sollerman}, {Taubenberger}, {Gal-Yam},
  {Kankare}, {Maguire}, {Fraser}, {Valenti}, {Sullivan}, {Cartier}, \&
  {Young}}]{Jerkstrand2017}
{Jerkstrand}, A., {Smartt}, S.~J., {Inserra}, C., {et~al.} 2017, \apj, 835, 13,
  \dodoi{10.3847/1538-4357/835/1/13}

\bibitem[{{Jones} {et~al.}(2021){Jones}, {Foley}, {Narayan}, {Hjorth}, {Huber},
  {Aleo}, {Alexander}, {Angus}, {Auchettl}, {Baldassare}, {Bruun}, {Chambers},
  {Chatterjee}, {Coppejans}, {Coulter}, {DeMarchi}, {Dimitriadis}, {Drout},
  {Engel}, {French}, {Gagliano}, {Gall}, {Hung}, {Izzo}, {Jacobson-Gal{\'a}n},
  {Kilpatrick}, {Korhonen}, {Margutti}, {Raimundo}, {Ramirez-Ruiz}, {Rest},
  {Rojas-Bravo}, {Siebert}, {Smartt}, {Smith}, {Terreran}, {Wang}, {Wojtak},
  {Agnello}, {Ansari}, {Arendse}, {Baldeschi}, {Blanchard}, {Brethauer},
  {Bright}, {Brown}, {de Boer}, {Dodd}, {Fairlamb}, {Grillo}, {Hajela}, {Hede},
  {Kolborg}, {Law-Smith}, {Lin}, {Magnier}, {Malanchev}, {Matthews}, {Mockler},
  {Muthukrishna}, {Pan}, {Pfister}, {Ramanah}, {Rest}, {Sarangi},
  {Schr{\o}der}, {Stauffer}, {Stroh}, {Taggart}, {Tinyanont}, {Wainscoat}, \&
  {Young Supernova Experiment}}]{Jones2021}
{Jones}, D.~O., {Foley}, R.~J., {Narayan}, G., {et~al.} 2021, \apj, 908, 143,
  \dodoi{10.3847/1538-4357/abd7f5}

\bibitem[{{Jones} {et~al.}(2001){Jones}, {Oliphant}, {Peterson},
  {et~al.}}]{Jones2001}
{Jones}, E., {Oliphant}, T., {Peterson}, P., {et~al.} 2001, {SciPy}: Open
  source scientific tools for {Python}.
\newblock \url{http://www.scipy.org/}

\bibitem[{{Joubert} \& {Li}(2006)}]{Joubert2006}
{Joubert}, N., \& {Li}, W. 2006, Central Bureau Electronic Telegrams, 597, 1

\bibitem[{{Kaiser} {et~al.}(2002){Kaiser}, {Aussel}, {Burke}, {Boesgaard},
  {Chambers}, {Chun}, {Heasley}, {Hodapp}, {Hunt}, {Jedicke}, {Jewitt},
  {Kudritzki}, {Luppino}, {Maberry}, {Magnier}, {Monet}, {Onaka}, {Pickles},
  {Rhoads}, {Simon}, {Szalay}, {Szapudi}, {Tholen}, {Tonry}, {Waterson}, \&
  {Wick}}]{Kaiser2002}
{Kaiser}, N., {Aussel}, H., {Burke}, B.~E., {et~al.} 2002, in Society of
  Photo-Optical Instrumentation Engineers (SPIE) Conference Series, Vol. 4836,
  Survey and Other Telescope Technologies and Discoveries, ed. J.~A. {Tyson} \&
  S.~{Wolff}, 154--164, \dodoi{10.1117/12.457365}

\bibitem[{{Kalberla} {et~al.}(2005){Kalberla}, {Burton}, {Hartmann}, {Arnal},
  {Bajaja}, {Morras}, \& {P{\"o}ppel}}]{Kalberla05}
{Kalberla}, P.~M.~W., {Burton}, W.~B., {Hartmann}, D., {et~al.} 2005, \aap,
  440, 775, \dodoi{10.1051/0004-6361:20041864}

\bibitem[{{Kelly} \& {Kirshner}(2012)}]{Kelly2012}
{Kelly}, P.~L., \& {Kirshner}, R.~P. 2012, \apj, 759, 107,
  \dodoi{10.1088/0004-637X/759/2/107}

\bibitem[{{Khazov} {et~al.}(2016){Khazov}, {Yaron}, {Gal-Yam}, {Manulis},
  {Rubin}, {Kulkarni}, {Arcavi}, {Kasliwal}, {Ofek}, {Cao}, {Perley},
  {Sollerman}, {Horesh}, {Sullivan}, {Filippenko}, {Nugent}, {Howell}, {Cenko},
  {Silverman}, {Ebeling}, {Taddia}, {Johansson}, {Laher}, {Surace},
  {Rebbapragada}, {Wozniak}, \& {Matheson}}]{Khazov2016}
{Khazov}, D., {Yaron}, O., {Gal-Yam}, A., {et~al.} 2016, \apj, 818, 3,
  \dodoi{10.3847/0004-637X/818/1/3}

\bibitem[{{Kriek} {et~al.}(2009){Kriek}, {van Dokkum}, {Labb{\'e}}, {Franx},
  {Illingworth}, {Marchesini}, \& {Quadri}}]{Kriek2009}
{Kriek}, M., {van Dokkum}, P.~G., {Labb{\'e}}, I., {et~al.} 2009, \apj, 700,
  221, \dodoi{10.1088/0004-637X/700/1/221}

\bibitem[{{Kulkarni}(2018)}]{Kulkarni2018}
{Kulkarni}, S.~R. 2018, The Astronomer's Telegram, 11266, 1

\bibitem[{{Leonard} {et~al.}(2000){Leonard}, {Filippenko}, {Barth}, \&
  {Matheson}}]{leonard2000}
{Leonard}, D.~C., {Filippenko}, A.~V., {Barth}, A.~J., \& {Matheson}, T. 2000,
  \apj, 536, 239, \dodoi{10.1086/308910}

\bibitem[{{Li} \& {Gong}(1994)}]{Li1994}
{Li}, Y., \& {Gong}, Z.~G. 1994, \aap, 289, 449

\bibitem[{{Linial} {et~al.}(2021){Linial}, {Fuller}, \& {Sari}}]{Linial2021}
{Linial}, I., {Fuller}, J., \& {Sari}, R. 2021, \mnras, 501, 4266,
  \dodoi{10.1093/mnras/staa3969}

\bibitem[{{Lipunov} {et~al.}(2012){Lipunov}, {Shumkov}, {Denisenko},
  {Gorbovskoy}, {Brimacombe}, {Tomasella}, {Benetti}, {Cappellaro}, {Ochner},
  {Pastorello}, \& {Turatto}}]{Lipunov2012}
{Lipunov}, V., {Shumkov}, V., {Denisenko}, D., {et~al.} 2012, Central Bureau
  Electronic Telegrams, 3359, 1

\bibitem[{{Margutti} {et~al.}(2014){Margutti}, {Milisavljevic}, {Soderberg},
  {Chornock}, {Zauderer}, {Murase}, {Guidorzi}, {Sanders}, {Kuin}, {Fransson},
  {Levesque}, {Chandra}, {Berger}, {Bianco}, {Brown}, {Challis},
  {Chatzopoulos}, {Cheung}, {Choi}, {Chomiuk}, {Chugai}, {Contreras}, {Drout},
  {Fesen}, {Foley}, {Fong}, {Friedman}, {Gall}, {Gehrels}, {Hjorth}, {Hsiao},
  {Kirshner}, {Im}, {Leloudas}, {Lunnan}, {Marion}, {Martin}, {Morrell},
  {Neugent}, {Omodei}, {Phillips}, {Rest}, {Silverman}, {Strader},
  {Stritzinger}, {Szalai}, {Utterback}, {Vinko}, {Wheeler}, {Arnett},
  {Campana}, {Chevalier}, {Ginsburg}, {Kamble}, {Roming}, {Pritchard}, \&
  {Stringfellow}}]{Margutti1409ip}
{Margutti}, R., {Milisavljevic}, D., {Soderberg}, A.~M., {et~al.} 2014, \apj,
  780, 21, \dodoi{10.1088/0004-637X/780/1/21}

\bibitem[{{Mauerhan} \& {Smith}(2012)}]{Mauerhan2012}
{Mauerhan}, J., \& {Smith}, N. 2012, \mnras, 424, 2659,
  \dodoi{10.1111/j.1365-2966.2012.21325.x}

\bibitem[{{Mauron} \& {Josselin}(2011)}]{Mauron2011}
{Mauron}, N., \& {Josselin}, E. 2011, \aap, 526, A156,
  \dodoi{10.1051/0004-6361/201013993}

\bibitem[{{McLean} {et~al.}(2012){McLean}, {Steidel}, {Epps}, {Konidaris},
  {Matthews}, {Adkins}, {Aliado}, {Brims}, {Canfield}, {Cromer}, {Fucik},
  {Kulas}, {Mace}, {Magnone}, {Rodriguez}, {Rudie}, {Trainor}, {Wang}, {Weber},
  \& {Weiss}}]{mclean2012}
{McLean}, I.~S., {Steidel}, C.~C., {Epps}, H.~W., {et~al.} 2012, in Society of
  Photo-Optical Instrumentation Engineers (SPIE) Conference Series, Vol. 8446,
  Ground-based and Airborne Instrumentation for Astronomy IV, 84460J,
  \dodoi{10.1117/12.924794}

\bibitem[{{McMullin} {et~al.}(2007{\natexlab{a}}){McMullin}, {Waters},
  {Schiebel}, {Young}, \& {Golap}}]{McMullin07}
{McMullin}, J.~P., {Waters}, B., {Schiebel}, D., {Young}, W., \& {Golap}, K.
  2007{\natexlab{a}}, in Astronomical Society of the Pacific Conference Series,
  Vol. 376, Astronomical Data Analysis Software and Systems XVI, ed. R.~A.
  {Shaw}, F.~{Hill}, \& D.~J. {Bell}, 127

\bibitem[{{McMullin} {et~al.}(2007{\natexlab{b}}){McMullin}, {Waters},
  {Schiebel}, {Young}, \& {Golap}}]{McMullin2007}
{McMullin}, J.~P., {Waters}, B., {Schiebel}, D., {Young}, W., \& {Golap}, K.
  2007{\natexlab{b}}, in Astronomical Society of the Pacific Conference Series,
  Vol. 376, Astronomical Data Analysis Software and Systems XVI, ed. R.~A.
  {Shaw}, F.~{Hill}, \& D.~J. {Bell}, 127

\bibitem[{{Meynet} {et~al.}(1994){Meynet}, {Maeder}, {Schaller}, {Schaerer}, \&
  {Charbonnel}}]{Meynet94}
{Meynet}, G., {Maeder}, A., {Schaller}, G., {Schaerer}, D., \& {Charbonnel}, C.
  1994, \aaps, 103, 97

\bibitem[{{Miller} \& {Stone}(1993)}]{KAST}
{Miller}, J.~S., \& {Stone}, R. P.~S. 1993, LOTRM

\bibitem[{{Mohan} \& {Rafferty}(2015)}]{Mohan15}
{Mohan}, N., \& {Rafferty}, D. 2015, {PyBDSF: Python Blob Detection and Source
  Finder}.
\newblock \doeprint{1502.007}

\bibitem[{{Moriya} {et~al.}(2011){Moriya}, {Tominaga}, {Blinnikov}, {Baklanov},
  \& {Sorokina}}]{Moriya11}
{Moriya}, T., {Tominaga}, N., {Blinnikov}, S.~I., {Baklanov}, P.~V., \&
  {Sorokina}, E.~I. 2011, \mnras, 415, 199,
  \dodoi{10.1111/j.1365-2966.2011.18689.x}

\bibitem[{{Morozova} {et~al.}(2017){Morozova}, {Piro}, \&
  {Valenti}}]{Morozova2017}
{Morozova}, V., {Piro}, A.~L., \& {Valenti}, S. 2017, \apj, 838, 28,
  \dodoi{10.3847/1538-4357/aa6251}

\bibitem[{{Morozova} {et~al.}(2018){Morozova}, {Piro}, \&
  {Valenti}}]{Morozova2018}
---. 2018, \apj, 858, 15, \dodoi{10.3847/1538-4357/aab9a6}

\bibitem[{{Najarro} {et~al.}(1997){Najarro}, {Hillier}, \& {Stahl}}]{Najarro97}
{Najarro}, F., {Hillier}, D.~J., \& {Stahl}, O. 1997, \aap, 326, 1117

\bibitem[{{Nakaoka} {et~al.}(2018){Nakaoka}, {Kawabata}, {Maeda}, {Tanaka},
  {Yamanaka}, {Moriya}, {Tominaga}, {Morokuma}, {Takaki}, {Kawabata},
  {Kawahara}, {Itoh}, {Shiki}, {Mori}, {Hirochi}, {Abe}, {Uemura}, {Yoshida},
  {Akitaya}, {Moritani}, {Ueno}, {Urano}, {Isogai}, {Hanayama}, \&
  {Nagayama}}]{Nakaoka2018}
{Nakaoka}, T., {Kawabata}, K.~S., {Maeda}, K., {et~al.} 2018, \apj, 859, 78,
  \dodoi{10.3847/1538-4357/aabee7}

\bibitem[{{Ofek} {et~al.}(2013){Ofek}, {Sullivan}, {Cenko}, {Kasliwal},
  {Gal-Yam}, {Kulkarni}, {Arcavi}, {Bildsten}, {Bloom}, {Horesh}, {Howell},
  {Filippenko}, {Laher}, {Murray}, {Nakar}, {Nugent}, {Silverman}, {Shaviv},
  {Surace}, \& {Yaron}}]{Ofek2013}
{Ofek}, E.~O., {Sullivan}, M., {Cenko}, S.~B., {et~al.} 2013, \nat, 494, 65,
  \dodoi{10.1038/nature11877}

\bibitem[{{Oke} \& {Gunn}(1983)}]{Oke83}
{Oke}, J.~B., \& {Gunn}, J.~E. 1983, \apj, 266, 713, \dodoi{10.1086/160817}

\bibitem[{Oliphant(2006)}]{Oliphant2016}
Oliphant, T. 2006, A guide to NumPy, Vol.~1 (Trelgol Publishing USA)

\bibitem[{{Osterbrock} \& {Ferland}(2006)}]{Osterbrock2006}
{Osterbrock}, D.~E., \& {Ferland}, G.~J. 2006, {Astrophysics of gaseous nebulae
  and active galactic nuclei}

\bibitem[{{Poznanski} {et~al.}(2012){Poznanski}, {Prochaska}, \&
  {Bloom}}]{Poznanski12}
{Poznanski}, D., {Prochaska}, J.~X., \& {Bloom}, J.~S. 2012, \mnras, 426, 1465,
  \dodoi{10.1111/j.1365-2966.2012.21796.x}

\bibitem[{{Prieto} {et~al.}(2008){Prieto}, {Stanek}, \& {Beacom}}]{Prieto2008}
{Prieto}, J.~L., {Stanek}, K.~Z., \& {Beacom}, J.~F. 2008, \apj, 673, 999,
  \dodoi{10.1086/524654}

\bibitem[{{Prochaska} {et~al.}(2020){Prochaska}, {Hennawi}, {Westfall},
  {Cooke}, {Wang}, {Hsyu}, {Davies}, \& {Farina}}]{Prochaska2020}
{Prochaska}, J.~X., {Hennawi}, J.~F., {Westfall}, K.~B., {et~al.} 2020, arXiv
  e-prints, arXiv:2005.06505.
\newblock \doarXiv{2005.06505}

\bibitem[{{Quataert} \& {Shiode}(2012)}]{Quataert2012}
{Quataert}, E., \& {Shiode}, J. 2012, \mnras, 423, L92,
  \dodoi{10.1111/j.1745-3933.2012.01264.x}

\bibitem[{{Rest} {et~al.}(2005){Rest}, {Stubbs}, {Becker}, {Miknaitis},
  {Miceli}, {Covarrubias}, {Hawley}, {Smith}, {Suntzeff}, {Olsen}, {Prieto},
  {Hiriart}, {Welch}, {Cook}, {Nikolaev}, {Huber}, {Prochtor}, {Clocchiatti},
  {Minniti}, {Garg}, {Challis}, {Keller}, \& {Schmidt}}]{Rest05}
{Rest}, A., {Stubbs}, C., {Becker}, A.~C., {et~al.} 2005, \apj, 634, 1103,
  \dodoi{10.1086/497060}

\bibitem[{{Rigault} {et~al.}(2019){Rigault}, {Neill}, {Blagorodnova}, {Dugas},
  {Feeney}, {Walters}, {Brinnel}, {Copin}, {Fremling}, {Nordin}, \&
  {Sollerman}}]{Rigault2019}
{Rigault}, M., {Neill}, J.~D., {Blagorodnova}, N., {et~al.} 2019, \aap, 627,
  A115, \dodoi{10.1051/0004-6361/201935344}

\bibitem[{{Robitaille} \& {Bressert}(2012)}]{Robitaille2012}
{Robitaille}, T., \& {Bressert}, E. 2012, {APLpy: Astronomical Plotting Library
  in Python}, Astrophysics Source Code Library.
\newblock \doeprint{1208.017}

\bibitem[{{Roming} {et~al.}(2005){Roming}, {Kennedy}, {Mason}, {Nousek}, {Ahr},
  {Bingham}, {Broos}, {Carter}, {Hancock}, {Huckle}, {Hunsberger}, {Kawakami},
  {Killough}, {Koch}, {McLelland}, {Smith}, {Smith}, {Soto}, {Boyd},
  {Breeveld}, {Holland}, {Ivanushkina}, {Pryzby}, {Still}, \&
  {Stock}}]{Roming05}
{Roming}, P. W.~A., {Kennedy}, T.~E., {Mason}, K.~O., {et~al.} 2005, \ssr, 120,
  95, \dodoi{10.1007/s11214-005-5095-4}

\bibitem[{{Roming} {et~al.}(2012){Roming}, {Pritchard}, {Prieto}, {Kochanek},
  {Fryer}, {Davidson}, {Humphreys}, {Bayless}, {Beacom}, {Brown}, {Holland},
  {Immler}, {Kuin}, {Oates}, {Pogge}, {Pojmanski}, {Stoll}, {Shappee},
  {Stanek}, \& {Szczygiel}}]{Roming2012}
{Roming}, P.~W.~A., {Pritchard}, T.~A., {Prieto}, J.~L., {et~al.} 2012, \apj,
  751, 92, \dodoi{10.1088/0004-637X/751/2/92}

\bibitem[{{Salpeter}(1955)}]{Salpeter1955}
{Salpeter}, E.~E. 1955, \apj, 121, 161, \dodoi{10.1086/145971}

\bibitem[{{Schlafly} \& {Finkbeiner}(2011)}]{schlafly11}
{Schlafly}, E.~F., \& {Finkbeiner}, D.~P. 2011, \apj, 737, 103,
  \dodoi{10.1088/0004-637X/737/2/103}

\bibitem[{{Schlegel}(1990)}]{Schlegel1990}
{Schlegel}, E.~M. 1990, \mnras, 244, 269

\bibitem[{{Seibert} {et~al.}(2012){Seibert}, {Wyder}, {Neill}, {Madore},
  {Bianchi}, {Smith}, {Shiao}, {Schiminovich}, {Rich}, {Conrow}, {Martin}, \&
  {GALEX Catalog Team}}]{Seibert2012}
{Seibert}, M., {Wyder}, T., {Neill}, J., {et~al.} 2012, in American
  Astronomical Society Meeting Abstracts, Vol. 219, American Astronomical
  Society Meeting Abstracts \#219, 340.01

\bibitem[{{Sheinis} {et~al.}(2002){Sheinis}, {Bolte}, {Epps}, {Kibrick},
  {Miller}, {Radovan}, {Bigelow}, \& {Sutin}}]{Sheinis2002}
{Sheinis}, A.~I., {Bolte}, M., {Epps}, H.~W., {et~al.} 2002, \pasp, 114, 851,
  \dodoi{10.1086/341706}

\bibitem[{{Shiode} \& {Quataert}(2014)}]{Shiode2014}
{Shiode}, J.~H., \& {Quataert}, E. 2014, \apj, 780, 96,
  \dodoi{10.1088/0004-637X/780/1/96}

\bibitem[{{Shivvers} {et~al.}(2015){Shivvers}, {Groh}, {Mauerhan}, {Fox},
  {Leonard}, \& {Filippenko}}]{shivvers15}
{Shivvers}, I., {Groh}, J.~H., {Mauerhan}, J.~C., {et~al.} 2015, \apj, 806,
  213, \dodoi{10.1088/0004-637X/806/2/213}

\bibitem[{{Sironi} {et~al.}(2015){Sironi}, {Keshet}, \& {Lemoine}}]{Sironi15}
{Sironi}, L., {Keshet}, U., \& {Lemoine}, M. 2015, \ssr, 191, 519,
  \dodoi{10.1007/s11214-015-0181-8}

\bibitem[{{Skrutskie} {et~al.}(2006){Skrutskie}, {Cutri}, {Stiening},
  {Weinberg}, {Schneider}, {Carpenter}, {Beichman}, {Capps}, {Chester},
  {Elias}, {Huchra}, {Liebert}, {Lonsdale}, {Monet}, {Price}, {Seitzer},
  {Jarrett}, {Kirkpatrick}, {Gizis}, {Howard}, {Evans}, {Fowler}, {Fullmer},
  {Hurt}, {Light}, {Kopan}, {Marsh}, {McCallon}, {Tam}, {Van Dyk}, \&
  {Wheelock}}]{Skrutskie2006}
{Skrutskie}, M.~F., {Cutri}, R.~M., {Stiening}, R., {et~al.} 2006, \aj, 131,
  1163, \dodoi{10.1086/498708}

\bibitem[{{Smith} {et~al.}(1994){Smith}, {Crowther}, \& {Prinja}}]{Smith94}
{Smith}, L.~J., {Crowther}, P.~A., \& {Prinja}, R.~K. 1994, \aap, 281, 833

\bibitem[{{Smith}(2006)}]{Smith06}
{Smith}, N. 2006, \apj, 644, 1151, \dodoi{10.1086/503766}

\bibitem[{{Smith}(2014)}]{Smith2014}
---. 2014, \araa, 52, 487, \dodoi{10.1146/annurev-astro-081913-040025}

\bibitem[{{Soker}(2021)}]{Soker2021}
{Soker}, N. 2021, \apj, 906, 1, \dodoi{10.3847/1538-4357/abca8f}

\bibitem[{{Stetson}(1987)}]{Stetson1987}
{Stetson}, P.~B. 1987, \pasp, 99, 191, \dodoi{10.1086/131977}

\bibitem[{{Tartaglia} {et~al.}(2021){Tartaglia}, {Sand}, {Groh}, {Valenti},
  {Wyatt}, {Bostroem}, {Brown}, {Yang}, {Burke}, {Chen}, {Davis},
  {F{\"o}rster}, {Galbany}, {Haislip}, {Hiramatsu}, {Hosseinzadeh}, {Howell},
  {Hsiao}, {Jha}, {Kouprianov}, {Kuncarayakti}, {Lyman}, {McCully}, {Phillips},
  {Rau}, {Reichart}, {Shahbandeh}, \& {Strader}}]{Tartaglia21}
{Tartaglia}, L., {Sand}, D.~J., {Groh}, J.~H., {et~al.} 2021, \apj, 907, 52,
  \dodoi{10.3847/1538-4357/abca8a}

\bibitem[{{Terreran} {et~al.}(2016){Terreran}, {Jerkstrand}, {Benetti},
  {Smartt}, {Ochner}, {Tomasella}, {Howell}, {Morales-Garoffolo},
  {Harutyunyan}, {Kankare}, {Arcavi}, {Cappellaro}, {Elias-Rosa},
  {Hosseinzadeh}, {Kangas}, {Pastorello}, {Tartaglia}, {Turatto}, {Valenti},
  {Wiggins}, \& {Yuan}}]{terreran16}
{Terreran}, G., {Jerkstrand}, A., {Benetti}, S., {et~al.} 2016, \mnras, 462,
  137, \dodoi{10.1093/mnras/stw1591}

\bibitem[{{Terreran} {et~al.}(2019){Terreran}, {Margutti}, {Bersier},
  {Brimacombe}, {Caprioli}, {Challis}, {Chornock}, {Coppejans}, {Dong},
  {Guidorzi}, {Hurley}, {Kirshner}, {Migliori}, {Milisavljevic}, {Palmer},
  {Prieto}, {Tomasella}, {Marchant}, {Pastorello}, {Shappee}, {Stanek},
  {Stritzinger}, {Benetti}, {Chen}, {DeMarchi}, {Elias-Rosa}, {Gall},
  {Harmanen}, \& {Mattila}}]{Terreran2019}
{Terreran}, G., {Margutti}, R., {Bersier}, D., {et~al.} 2019, \apj, 883, 147,
  \dodoi{10.3847/1538-4357/ab3e37}

\bibitem[{{Tody}(1986)}]{Tody1986}
{Tody}, D. 1986, in Society of Photo-Optical Instrumentation Engineers (SPIE)
  Conference Series, Vol. 627, \procspie, ed. D.~L. {Crawford}, 733,
  \dodoi{10.1117/12.968154}

\bibitem[{{Tody}(1993)}]{Tody1993}
{Tody}, D. 1993, in Astronomical Society of the Pacific Conference Series,
  Vol.~52, Astronomical Data Analysis Software and Systems II, ed. R.~J.
  {Hanisch}, R.~J.~V. {Brissenden}, \& J.~{Barnes}, 173

\bibitem[{{Tomasella} {et~al.}(2012){Tomasella}, {Benetti}, {Cappellaro},
  {Ochner}, {Pastorello}, \& {Turatto}}]{Tomasella2012}
{Tomasella}, L., {Benetti}, S., {Cappellaro}, E., {et~al.} 2012, The
  Astronomer's Telegram, 4672, 1

\bibitem[{{Turatto} {et~al.}(2003){Turatto}, {Benetti}, \&
  {Cappellaro}}]{Turatto2003b}
{Turatto}, M., {Benetti}, S., \& {Cappellaro}, E. 2003, in From Twilight to
  Highlight: The Physics of Supernovae, ed. W.~{Hillebrandt} \&
  B.~{Leibundgut}, 200, \dodoi{10.1007/10828549_26}

\bibitem[{{Vacca} {et~al.}(2003){Vacca}, {Cushing}, \& {Rayner}}]{vacca03}
{Vacca}, W.~D., {Cushing}, M.~C., \& {Rayner}, J.~T. 2003, \pasp, 115, 389,
  \dodoi{10.1086/346193}

\bibitem[{{Valenti} {et~al.}(2014){Valenti}, {Sand}, {Pastorello}, {Graham},
  {Howell}, {Parrent}, {Tomasella}, {Ochner}, {Fraser}, {Benetti}, {Yuan},
  {Smartt}, {Maund}, {Arcavi}, {Gal-Yam}, {Inserra}, \& {Young}}]{Valenti2014}
{Valenti}, S., {Sand}, D., {Pastorello}, A., {et~al.} 2014, \mnras, 438, L101,
  \dodoi{10.1093/mnrasl/slt171}

\bibitem[{{Vink} {et~al.}(2001){Vink}, {de Koter}, \& {Lamers}}]{Vink2001}
{Vink}, J.~S., {de Koter}, A., \& {Lamers}, H.~J.~G.~L.~M. 2001, \aap, 369,
  574, \dodoi{10.1051/0004-6361:20010127}

\bibitem[{{Waxman} \& {Katz}(2017)}]{Waxman17}
{Waxman}, E., \& {Katz}, B. 2017, {Shock Breakout Theory} (Springer), 967,
  \dodoi{10.1007/978-3-319-21846-5_33}

\bibitem[{{Weaver} \& {Woosley}(1979)}]{Weaver1979}
{Weaver}, T.~A., \& {Woosley}, S.~E. 1979, in Bulletin of the American
  Astronomical Society, Vol.~11, 724

\bibitem[{{Weiler} {et~al.}(2002){Weiler}, {Panagia}, {Montes}, \&
  {Sramek}}]{Weiler02}
{Weiler}, K.~W., {Panagia}, N., {Montes}, M.~J., \& {Sramek}, R.~A. 2002,
  \araa, 40, 387, \dodoi{10.1146/annurev.astro.40.060401.093744}

\bibitem[{{Weiler} {et~al.}(1986){Weiler}, {Sramek}, {Panagia}, {van der
  Hulst}, \& {Salvati}}]{Weiler86}
{Weiler}, K.~W., {Sramek}, R.~A., {Panagia}, N., {van der Hulst}, J.~M., \&
  {Salvati}, M. 1986, \apj, 301, 790, \dodoi{10.1086/163944}

\bibitem[{{Woosley} \& {Heger}(2015)}]{Woosley2015}
{Woosley}, S.~E., \& {Heger}, A. 2015, \apj, 810, 34,
  \dodoi{10.1088/0004-637X/810/1/34}

\bibitem[{{Woosley} {et~al.}(2002){Woosley}, {Heger}, \&
  {Weaver}}]{Woosley2002}
{Woosley}, S.~E., {Heger}, A., \& {Weaver}, T.~A. 2002, Reviews of Modern
  Physics, 74, 1015, \dodoi{10.1103/RevModPhys.74.1015}

\bibitem[{{Wu} \& {Fuller}(2021)}]{Wu2021}
{Wu}, S., \& {Fuller}, J. 2021, \apj, 906, 3, \dodoi{10.3847/1538-4357/abc87c}

\bibitem[{{Yaron} \& {Gal-Yam}(2012)}]{Yaron12}
{Yaron}, O., \& {Gal-Yam}, A. 2012, \pasp, 124, 668, \dodoi{10.1086/666656}

\bibitem[{{Yaron} {et~al.}(2017{\natexlab{a}}){Yaron}, {Perley}, {Gal-Yam},
  {Groh}, {Horesh}, {Ofek}, {Kulkarni}, {Sollerman}, {Fransson}, {Rubin},
  {Szabo}, {Sapir}, {Taddia}, {Cenko}, {Valenti}, {Arcavi}, {Howell},
  {Kasliwal}, {Vreeswijk}, {Khazov}, {Fox}, {Cao}, {Gnat}, {Kelly}, {Nugent},
  {Filippenko}, {Laher}, {Wozniak}, {Lee}, {Rebbapragada}, {Maguire},
  {Sullivan}, \& {Soumagnac}}]{Yaron17}
{Yaron}, O., {Perley}, D.~A., {Gal-Yam}, A., {et~al.} 2017{\natexlab{a}},
  Nature Physics, 13, 510, \dodoi{10.1038/nphys4025}

\bibitem[{{Yaron} {et~al.}(2017{\natexlab{b}}){Yaron}, {Perley}, {Gal-Yam},
  {Groh}, {Horesh}, {Ofek}, {Kulkarni}, {Sollerman}, {Fransson}, {Rubin},
  {Szabo}, {Sapir}, {Taddia}, {Cenko}, {Valenti}, {Arcavi}, {Howell},
  {Kasliwal}, {Vreeswijk}, {Khazov}, {Fox}, {Cao}, {Gnat}, {Kelly}, {Nugent},
  {Filippenko}, {Laher}, {Wozniak}, {Lee}, {Rebbapragada}, {Maguire},
  {Sullivan}, \& {Soumagnac}}]{Yaron2017}
---. 2017{\natexlab{b}}, Nature Physics, 13, 510, \dodoi{10.1038/nphys4025}

\bibitem[{{Yoon} \& {Cantiello}(2010)}]{Yoon2010}
{Yoon}, S.-C., \& {Cantiello}, M. 2010, \apjl, 717, L62,
  \dodoi{10.1088/2041-8205/717/1/L62}

\bibitem[{{Zapartas} {et~al.}(2019){Zapartas}, {de Mink}, {Justham}, {Smith},
  {de Koter}, {Renzo}, {Arcavi}, {Farmer}, {G{\"o}tberg}, \&
  {Toonen}}]{Zapartas2019}
{Zapartas}, E., {de Mink}, S.~E., {Justham}, S., {et~al.} 2019, \aap, 631, A5,
  \dodoi{10.1051/0004-6361/201935854}

\bibitem[{{Zhang} {et~al.}(2020){Zhang}, {Wang}, {J{\'o}zsef}, {Zhai}, {Zhang},
  {Filippenko}, {Brink}, {Zheng}, {Wyrzykowski}, {Miko{\l}ajczyk}, {Huang},
  {Rui}, {Mo}, {Sai}, {Zhang}, {Wang}, {DerKacy}, {Baron}, {S{\'a}rneczky},
  {B{\'o}di}, {Cs{\"o}rnyei}, {Hanyecz}, {Ign{\'a}cz}, {Kalup}, {Kriskovics},
  {K{\"o}nyves-T{\'o}th}, {Ordasi}, {P{\'a}l}, {S{\'o}dor}, {Szak{\'a}ts},
  {Vida}, \& {Zsidi}}]{Zhang20}
{Zhang}, J., {Wang}, X., {J{\'o}zsef}, V., {et~al.} 2020, \mnras, 498, 84,
  \dodoi{10.1093/mnras/staa2273}

\end{thebibliography}

\clearpage
\appendix

\section{Circumstellar Density Constraints from Radio Observations}\label{app:radio}

As the SN blast wave expands, it collides with the surrounding CSM and forms a shock front. The accelerated electrons within the forward-shock region produce synchrotron emission, which is attenuated by synchrotron self-absorption (SSA) and by external free-free absorption. The SSA spectrum can be approximated by a smoothed broken power law of the form
\begin{equation}\label{eqn:SSA_spectrum}
F=F_p\left(\left(\frac{\nu}{\nu_b}\right)^{-\beta/s}+\left(\frac{\nu}{\nu_b}\right)^{-\alpha/s}\right)^{-s},
\end{equation}
where $F_p$ is the peak flux at the SSA frequency ($\nu_b$), $\beta$ is the optically thick spectral index, $\alpha$ is the optically thin spectral index, and $s$ is a smoothing factor. For a fiducial SSA spectrum, $\beta=2.5$, $\alpha=-1$, and $s=1$.

This emission is attenuated by free-free absorption by the unshocked material in front of the blast wave. We follow \citet{Weiler86} (Equations (3)--(16)) to determine the optical depth of this material. Assuming that the progenitor star had a constant mass-loss rate in the years prior to explosion, its wind would form CSM with a density profile $\rho\propto r^{-2}$ (a wind-profile medium). The amount of material in front of the shock (and hence the optical depth) can thus be determined if the time-dependent radius of the forward shock is known. \citet{Chevalier98} and \citet{Chevalier06} show that for a progenitor star having a density profile $\rho\propto r^{-n}$ (where $r$ is the stellar radius) and a wind-profile CSM, the SN blast wave will expand according to $r_{\rm{sh}}\propto t^{(n-3)/(n-2)}$, which means that the shock velocity is given by
\begin{equation}\label{eqn:shock_velocity}
v_{\rm{sh}}=\left(\frac{n-3}{n-2}\right)\left(\frac{r_{\rm{sh}}}{t}\right),
\end{equation}
where $r_{\rm{sh}}$ and $v_{\rm{sh}}$ are, respectively, the shock radius and velocity at time $t$ post-explosion and appropriate progenitors (e.g., Wolf-Rayet stars) have $n\approx10$. If the peak flux and frequency of the SSA are measured, then \citet{Chevalier06} show that the shock radius is given by
\begin{equation}\label{eqn:shock_radius}
r_{\rm sh}=4.0\times10^{14}\left(\frac{\epsilon_{e}}{\epsilon_{B}}\right)^{-1/19}\left(\frac{f}{0.5}\right)^{-1/19}\left(\frac{F_p}{\rm{mJy}}\right)^{9/19}\left(\frac{D}{\rm{Mpc}}\right)^{18/19}\left(\frac{\nu_p}{5\,\rm{GHz}}\right)^{-1}\,\rm{cm},
\end{equation}
where $D$ is the angular distance of the SN, $\epsilon_{e}$ is the post-shock relativistic electron energy density, $\epsilon_{B}$ is the post-shock magnetic energy density, and $f$ is the filling factor such that the emitting volume of the blast wave is $f\times 4\pi r_{\rm sh}^3/3$. The parameters $f$ and $\epsilon_{B}$ are poorly constrained, but the shock radius is not strongly dependent on these parameters as $r_{\rm{sh}}\propto\alpha^{-1/19}$ and $r_{\rm{sh}}\propto (f/0.5)^{-1/19}$. A fiducial value of $f=0.5$ is used in the literature \citep[e.g.,][]{Chevalier06}, and for relativistic shocks $\epsilon_{e}\approx0.1$ \citep{Sironi15}.     

Following \citet{Weiler86}, for a blast wave expanding at this velocity (Eq. \ref{eqn:shock_velocity}, with $n=10$) in an ionized material with solar composition, the optical depth of the CSM is
\begin{equation}\label{eqn:tau_freefree}
    \tau_{\rm{ff}} = 2.1\times10^{6}\left(\frac{\nu}{\rm{GHz}}\right)^{2.1}\left(\frac{T_g}{10^4\,\rm{K}}\right)^{-1.36} \left(\frac{\dot{M}}{\rm{M_{\odot}\,yr^{-1}}}\right)^{2} \left(\frac{v_{\rm{sh}}}{0.1\rm{c}}\right)^{-3} \left(\frac{t}{7\,\rm{days}}\right)^{-3} \left(\frac{v_{\rm{w}}}{200\,\rm{km/s}}\right)^{-2},
\end{equation}
where $T_g$ is the plasma temperature normalized to $10^4$ K, $\nu$ is the frequency of the emission, $\dot{M}$ is the mass-loss rate of the progenitor star, $v_{\rm sh}$ is the velocity of the shock normalized to $0.1 c$, and $t$ is the time since explosion in days \citep{Weiler02}.

The radio emission spectrum at a given time can thus be calculated using Equations \ref{eqn:SSA_spectrum} and \ref{eqn:tau_freefree}. Specifically, we used a progenitor wind velocity of 200\,km\,s$^{-1}$ (as derived from our spectroscopic observations) and assumed a plasma temperature of $10^4$ K (which is typical of photoionized gas), $\beta=2.5$, $\alpha=-1$, $s=1$, $n=10$, and $f=0.5$. We then formed a grid of $\dot{M}$, $\nu_b$, $F_p$, $\epsilon_{e}$, and $\epsilon_{B}$ values covering the range appropriate to SNe. In particular, we used $\epsilon_{e}\approx0.1$, $\epsilon_{B}=0.1$ and $0.01$ (which are representative values derived from modeling of long-duration gamma-ray bursts), and $\epsilon_{e}=\epsilon_{B}=1/3$ (which assumes equipartition of the post-shock energy density between the relativistic electrons and magnetic fields). For each set of these parameters, we calculated the shock velocity (Eq. \ref{eqn:shock_radius} and $\Gamma=1/\sqrt{1-\beta/{\rm c}}$, where $\beta$ is the shock velocity in units of the speed of light $c$) and radio spectrum at the time of each radio observation (i.e., at 37.3, 129, and 307 days post-explosion). The radio upper limits rule out some of these spectra, and hence also the corresponding combinations of the progenitor mass-loss rate and shock velocity. The shaded regions of Figure \ref{fig:radiodensity} show the parameter space that we ruled out for \sn{} in this manner.

\citet{Berger02} used the same self-similar solution for the shock radius ($r_{\rm{sh}}\propto t^{(n-3)/(n-2)}$; \citealt{Chevalier98}) to derive the shock velocity for a Wolf-Rayet progenitor in a wind-profile environment as
\begin{equation}\label{eqn:vshock_with_normalization}
v_{\rm{sh}}=0.06\left(\frac{E_{k}}{10^{51}\,\rm{erg}}\right)^{0.44}\left(\frac{M_{\rm{ej}}}{\rm{M_{\odot}}}\right)^{-0.32}\left(\frac{\dot{M}}{\rm{M_{\odot}\,yr^{-1}}}\right)^{-0.12}\left(\frac{v_{\rm{w}}}{200\,\rm{km\,s^{-1}}}\right)^{0.12}\left(\frac{t}{7\,\rm{days}}\right)^{-0.12},
\end{equation}
where $v_{\rm{sh}}$ is in units of the speed of light, $E_{k}$ is the kinetic energy of the blast wave, and $M_{\rm{ej}}$ is the ejecta mass. In Figure \ref{fig:radiodensity} the dashed blue lines showing representative values for the progenitor mass-loss rate and shock velocity are derived using Equation \ref{eqn:vshock_with_normalization} for $E_{k}\approx10^{51}\,$erg and $M_{\rm{ej}}\approx1$.

\clearpage

\section{Follow-up Observations of \sn{}}\label{app:data}

\renewcommand\thetable{A\arabic{table}} 
\setcounter{table}{0}

In Table \ref{tab: griz} we report the ground-based optical photometry of 2020pni, while in Tables \ref{tab: UBV} and \ref{tab: SDA} we report the photometry quired with UVOT, on board the Neil Gehrels Swift Observatory. The only epoch of NIR photometry is reported in Table \ref{tab: NIR}. In Table \ref{tab: instr2}, we provide a summary of the telescopes and instruments employed in the spectroscopic follow-up of SN 2020pni, while the complete log of the spectroscopic observations is shown in Table \ref{fig: spec_log}. The radio observations measurements are reported in Table \ref{tbl:radio}.

\begin{table*}[h]
\centering
\renewcommand{\thefootnote}{\fnsymbol{footnote}}\caption{\textit{griz} Photometry.}
\begin{tabular}{>{\centering}m{1.2cm}>{\centering}m{1.8cm}>{\centering}m{1.8cm}>{\centering}m{1.8cm}>{\centering}m{1.8cm}>{\centering}m{1.8cm}}
\hline 
MJD & \textit{g} & \textit{r} & \textit{i} & \textit{z} & Instrument\tabularnewline 
 & (mag) & (mag) & (mag) & (mag) & \tabularnewline 
\hline 
59,042.30 &   $>$21.28   &   $>$21.42   &      $-$     &      $-$     & YSE/PS1 \tabularnewline
59,045.20 &   $>$20.82   &      $-$     &      $-$     &      $-$     & ZTF \tabularnewline
59,046.21 & 19.62 (0.13) & 19.80 (0.15) &      $-$     &      $-$     & ZTF \tabularnewline
59,047.26 & 17.66 (0.05) & 17.98 (0.06) &      $-$     &      $-$     & ZTF \tabularnewline
59,047.27 & 17.68 (0.02) & 17.93 (0.03) &      $-$     &      $-$     & YSE/PS1 \tabularnewline
59,048.26 & 16.79 (0.04) & 17.07 (0.05) &      $-$     &      $-$     & ZTF \tabularnewline
59,049.20 & 16.43 (0.04) & 16.66 (0.04) &      $-$     &      $-$     & ZTF \tabularnewline
59,050.29 &      $-$     & 16.50 (0.03) &      $-$     &      $-$     & ZTF \tabularnewline
59,051.18 & 16.24 (0.03) & 16.43 (0.03) &      $-$     &      $-$     & ZTF \tabularnewline
59,051.28 & 16.20 (0.01) &      $-$     & 16.59 (0.01) &      $-$     & YSE/PS1 \tabularnewline
59,052.31 & 16.24 (0.03) & 16.39 (0.03) &      $-$     &      $-$     & ZTF \tabularnewline
59,053.21 & 16.25 (0.03) &      $-$     &      $-$     &      $-$     & ZTF \tabularnewline
59,054.25 & 16.26 (0.04) & 16.32 (0.04) &      $-$     &      $-$     & ZTF \tabularnewline
59,055.20 &      $-$     & 16.31 (0.04) &      $-$     &      $-$     & ZTF \tabularnewline
59,057.18 &      $-$     & 16.34 (0.04) &      $-$     &      $-$     & ZTF \tabularnewline
59,058.30 & 16.28 (0.01) &      $-$     & 16.42 (0.01) &      $-$     & YSE/PS1 \tabularnewline
59,062.20 & 16.48 (0.03) & 16.37 (0.03) &      $-$     &      $-$     & ZTF \tabularnewline
59,063.24 & 16.55 (0.03) & 16.40 (0.05) &      $-$     &      $-$     & ZTF \tabularnewline
59,064.28 &      $-$     & 16.40 (0.04) &      $-$     &      $-$     & ZTF \tabularnewline
59,064.29 &      $-$     & 16.43 (0.02) &      $-$     & 16.50 (0.01) & YSE/PS1 \tabularnewline
59,065.21 & 16.66 (0.04) & 16.45 (0.04) &      $-$     &      $-$     & ZTF \tabularnewline
59,066.27 &      $-$     & 16.50 (0.02) & 16.51 (0.01) &      $-$     & YSE/PS1 \tabularnewline
59,067.20 & 16.78 (0.04) & 16.55 (0.04) &      $-$     &      $-$     & ZTF \tabularnewline
59,070.22 & 16.96 (0.05) & 16.70 (0.05) &      $-$     &      $-$     & ZTF \tabularnewline
59,070.29 &      $-$     & 16.71 (0.02) & 16.75 (0.02) &      $-$     & YSE/PS1 \tabularnewline
59,072.27 & 17.08 (0.02) &      $-$     &      $-$     & 16.82 (0.02) & YSE/PS1 \tabularnewline
59,074.28 & 17.18 (0.02) &      $-$     & 16.93 (0.01) &      $-$     & YSE/PS1 \tabularnewline
59,078.26 & 17.34 (0.02) &      $-$     & 17.01 (0.01) &      $-$     & YSE/PS1 \tabularnewline
59,081.25 & 17.46 (0.03) & 17.07 (0.02) &      $-$     &      $-$     & YSE/PS1 \tabularnewline
59,093.16 & 17.85 (0.07) & 17.13 (0.04) &      $-$     &      $-$     & ZTF \tabularnewline
59,093.25 &      $-$     & 17.18 (0.02) &      $-$     & 17.21 (0.03) & YSE/PS1 \tabularnewline
59,094.18 & 17.92 (0.12) & 17.21 (0.04) &      $-$     &      $-$     & ZTF \tabularnewline
59,095.25 &      $-$     & 17.21 (0.01) &      $-$     &      $-$     & YSE/PS1 \tabularnewline
59,097.24 &      $-$     & 17.23 (0.01) &      $-$     & 17.24 (0.02) & YSE/PS1 \tabularnewline
59,100.24 &      $-$     & 17.25 (0.02) &      $-$     & 17.30 (0.02) & YSE/PS1 \tabularnewline
59,104.24 & 18.06 (0.03) &      $-$     &      $-$     & 17.38 (0.02) & YSE/PS1 \tabularnewline
\hline 
\label{tab: griz}
\end{tabular}
\end{table*}

\begin{table*}
\centering
\renewcommand{\thefootnote}{\fnsymbol{footnote}}\caption{\textit{UBV} Photometry.}
\begin{tabular}{>{\centering}m{1.2cm}>{\centering}m{1.8cm}>{\centering}m{1.8cm}>{\centering}m{1.8cm}>{\centering}m{1.8cm}}
\hline 
MJD & \textit{U} & \textit{B} & \textit{V} & Instrument\tabularnewline 
 & (mag) & (mag) & (mag) & \tabularnewline 
\hline 
59,046.79 & 18.21 (0.11) & 18.40 (0.19) & 18.62 (0.40) & UVOT\tabularnewline
59,048.29 & 16.50 (0.09) & 16.79 (0.09) & 17.17 (0.15) & UVOT\tabularnewline
59,050.01 & 16.17 (0.08) & 16.45 (0.08) & 16.52 (0.11) & UVOT\tabularnewline
59,050.47 & 16.15 (0.08) & 16.27 (0.08) & 16.61 (0.12) & UVOT\tabularnewline
59,052.80 & 16.19 (0.08) & 16.27 (0.08) & 16.45 (0.14) & UVOT\tabularnewline
59,055.92 & 16.52 (0.08) & 16.40 (0.08) & 16.36 (0.10) & UVOT\tabularnewline
59,056.85 & 16.58 (0.08) & 16.44 (0.08) & 16.38 (0.10) & UVOT\tabularnewline
59,060.63 & 16.91 (0.08) & 16.46 (0.08) & 16.35 (0.10) & UVOT\tabularnewline
59,062.95 & 17.20 (0.09) & 16.89 (0.10) & 16.62 (0.12) & UVOT\tabularnewline
59,064.22 & 17.25 (0.09) & 16.94 (0.12) & 16.48 (0.10) & UVOT\tabularnewline
59,066.20 & 17.74 (0.10) & 16.98 (0.10) & 16.67 (0.11) & UVOT\tabularnewline
59,069.92 & 18.00 (0.11) & 17.15 (0.11) & 16.97 (0.13) & UVOT\tabularnewline
59,072.21 & 18.34 (0.12) & 17.26 (0.11) & 16.88 (0.12) & UVOT\tabularnewline
59,076.20 & 18.69 (0.15) & 17.58 (0.12) & 17.25 (0.15) & UVOT\tabularnewline
59,077.83 & 19.05 (0.18) & 17.68 (0.13) & 17.28 (0.15) & UVOT\tabularnewline
59,087.99 & $>$19.84 & 18.27 (0.22) & 17.69 (0.25) & UVOT\tabularnewline
59,090.19 & $>$19.65 & 18.31 (0.25) & 17.55 (0.26) & UVOT\tabularnewline
59,093.47 & $>$19.91 & 18.62 (0.27) & 17.47 (0.20) & UVOT\tabularnewline
59,102.66 & $>$19.61 & 18.80 (0.38) & 18.36 (0.50) & UVOT\tabularnewline
59,150.82 & $>$20.34 & $>$19.37 & $>$18.72 & UVOT\tabularnewline
\hline 
\label{tab: UBV}
\end{tabular}
\end{table*}

\begin{table*}
\centering
\renewcommand{\thefootnote}{\fnsymbol{footnote}}\caption{UV Photometry.}
\begin{tabular}{>{\centering}m{1.2cm}>{\centering}m{1.8cm}>{\centering}m{1.8cm}>{\centering}m{1.8cm}>{\centering}m{1.8cm}}
\hline 
MJD & \textit{UV-W2} & \textit{UV-M2} & \textit{UV-W1} & Instrument\tabularnewline 
 & (mag) & (mag) & (mag) & \tabularnewline 
\hline 
59,046.79  & 18.06 (0.11) & 18.30 (0.09) & 18.26 (0.09) & UVOT\tabularnewline 
59,048.28  & 16.42 (0.11) & 16.53 (0.10) & 16.52 (0.10) & UVOT\tabularnewline 
59,050.01  & 16.23 (0.13) & 16.35 (0.10) & 16.25 (0.09) & UVOT\tabularnewline 
59,050.47  & 16.24 (0.12) & 16.32 (0.10) & 16.21 (0.09) & UVOT\tabularnewline 
59,052.80  & 16.94 (0.11) & 16.90 (0.08) & 16.54 (0.09) & UVOT\tabularnewline 
59,055.92  & 17.79 (0.10) & 17.50 (0.09) & 17.11 (0.08) & UVOT\tabularnewline 
59,056.85  & 17.96 (0.10) & 17.67 (0.09) & 17.31 (0.08) & UVOT\tabularnewline 
59,060.64  & 18.78 (0.10) & 18.66 (0.09) & 18.01 (0.09) & UVOT\tabularnewline 
59,062.95  & 19.24 (0.12) & 19.07 (0.10) & 18.41 (0.10) & UVOT\tabularnewline 
59,064.22  & 19.78 (0.14) & 19.53 (0.12) & 18.63 (0.11) & UVOT\tabularnewline 
59,066.21  & 20.64 (0.22) & 20.04 (0.15) & 19.27 (0.13) & UVOT\tabularnewline 
59,069.92  & 20.85 (0.26) & 20.60 (0.21) & 19.53 (0.16) & UVOT\tabularnewline 
59,072.22  & 21.48 (0.40) & 21.08 (0.29) & 19.99 (0.19) & UVOT\tabularnewline 
59,076.20  & 21.95 (0.59) & 21.66 (0.45) & 20.90 (0.38) & UVOT\tabularnewline 
59,077.83  & 21.75 (0.50) & 21.92 (0.55) & 20.87 (0.36) & UVOT\tabularnewline 
59,087.99  & $>$21.30 & $>$21.04 & $>$20.77 & UVOT\tabularnewline 
59,090.19  & $>$21.24 & $>$21.14 & $>$20.50 & UVOT\tabularnewline 
59,093.47  & $>$21.27 & $>$21.48 & $>$20.86 & UVOT\tabularnewline 
59,102.66  & $>$21.12 & $>$21.20 & $>$20.49 & UVOT\tabularnewline 
59,150.82  & $>$21.63 & $>$21.64 & $>$21.15 & UVOT\tabularnewline 
\hline 
\label{tab: SDA}
\end{tabular}
\end{table*}

\begin{table*}
\centering
\renewcommand{\thefootnote}{\fnsymbol{footnote}}\caption{NIR Photometry.}
\begin{tabular}{>{\centering}m{1.2cm}>{\centering}m{1.8cm}>{\centering}m{1.8cm}>{\centering}m{1.8cm}>{\centering}m{1.8cm}}
\hline 
MJD & \textit{J} & \textit{H} & \textit{K} & Instrument\tabularnewline 
 & (mag) & (mag) & (mag) & \tabularnewline 
\hline 
59,058.25 & 15.94 (0.12) & 15.89 (0.11) & 15.98 (0.16) & MOSFIRE \tabularnewline 
\hline 
\label{tab: NIR}
\end{tabular}
\end{table*}

\setlength{\tabcolsep}{0pt}
\begin{table*}
 \centering
 \caption{Telescopes, Instruments, and Configurations Used for Spectroscopy of \sn{}.}
 \begin{tabular}{lccccr<{\hspace{-\tabcolsep}}>{\hspace{-\tabcolsep}\,}c<{\hspace{-\tabcolsep}\,}>{\hspace{-\tabcolsep}}l}
 \hline
 Telescope & Instrument & Grism/Grating & Slit width &Res. [\textit{R}]& \multicolumn{3}{c}{$\lambda$ range [\AA]}\\
 \hline
\rowcolor{gray!20}  ~~~~~~~~~~~~~~~~~~~~~~~~&~~~~~~~~~~~~~~~~~~~~~~~~& B600 &~~~~~~~~~~~~~~~~~~~~~~~~~~~~ & 1688 & 3600&$-$&6750\\
\rowcolor{gray!20} \multirow{-2}{*}{GEMINI}& \multirow{-2}{*}{GMOS}~& R400 &\multirow{-2}{*}{1.0\arcsec} & 1918 & 5200&$-$&9400\\
 ~~~~~~~~~~~~~~~~~~~~~~~~&~~~~~~~~~~~~~~~~~~~~~~~~& 400/3400 &~~~~~~~~~~~~~~~~~~~~~~~~~~~~~& 650 & 3200&$-$&5700\\
 \multirow{-2}{*}{Keck I}& \multirow{-2}{*}{LRIS}~& 400/8500 &\multirow{-2}{*}{1.0\arcsec} & 1000 & 5500&$-$&10300\\
\rowcolor{gray!20} ~~~~~~~~~~~~~~~~~~~~~~~~&~~~~~~~~~~~~~~~~~~~~~~~~~~~& Y &~~~~~~~~~~~~~~~~~~~~~~~~~~~~ & 3388 &  9716&$-$&11250\\
\rowcolor{gray!20} ~~~~~~~~~~~~~~~~~~~~~~~~&~~~~~~~~~~~~~~~~~~~~~~~~~~~& J &~~~~~~~~~~~~~~~~~~~~~~~~~~~~ & 3318 & 11530&$-$&13520\\
\rowcolor{gray!20} ~~~~~~~~~~~~~~~~~~~~~~~~&~~~~~~~~~~~~~~~~~~~~~~~~~~~& H &~~~~~~~~~~~~~~~~~~~~~~~~~~~~ & 3660 & 14680&$-$&18040\\
\rowcolor{gray!20} \multirow{-4}{*}{Keck I}& \multirow{-4}{*}{MOSFIRE}~& K &\multirow{-4}{*}{1.0\arcsec} & 3610 & 19540&$-$&23970\\
  Keck II~&~DEIMOS~~& 600ZD ~~~~& 1.0\arcsec & 1400 & 3900&$-$&9100\\
\rowcolor{gray!20} Keck II~&~ESI~~~~~&~175~line/mm & 1.0\arcsec & 4000 & 3900&$-$&11000\\
  LCO-2\,m~~&~FLOYDS~~&235~line/mm&~1.6\arcsec & 700 & 3300&$-$&10100\\
\rowcolor{gray!20}  MMT ~~~~&~Binospec&270~line/mm&~1.0\arcsec & 1340 & 3900&$-$&9240\\
  NOT~~~~~&~ALFOSC~~& Grism \#4 & 1.0\arcsec & 360 & 4000&$-$&9000\\
\rowcolor{gray!20} P60~~~~~&~SEDM\footnote{Integral field spectrograph.}~~~~& $-$ & $-$ & 100 & 3750&$-$&7700\\
 ~~~~~~~~~~~~~~~~~~~~~~~&~~~~~~~~~~~~~~~~~~~~~~~~&B830/3460&~~~~~~~~~~~~~~~~~~~~~~~~~~~~&~~~~~~~~~~~~~~~~~~~~~&3400&$-$&4500\\
  ~~~~~~~~~~~~~~~~~~~~~~~&~~~~~~~~~~~~~~~~~~~~~~~~&R1200/5000&~~~~~~~~~~~~~~~~~~~~~~~~~~~~&\multirow{-2}{*}{3200}&5500&$-$&7000\\
 ~~~~~~~~~~~~~~~~~~~~~~~&~~~~~~~~~~~~~~~~~~~~~~~~&B452/3306&~~~~~~~~~~~~~~~~~~~~~~~~~~~~&~~~~~~~~~~~~~~~~~~~~~&3300&$-$&6200\\
\multirow{-4}{*}{Shane}& \multirow{-4}{*}{Kast}~&R300/7500&\multirow{-4}{*}{2.0\arcsec}&\multirow{-2}{*}{650}&6300&$-$&9800\\
\hline
\end{tabular}
\label{tab: instr2}
\end{table*}
\setlength{\tabcolsep}{5pt}

\begin{table*}
\centering
\caption{Spectroscopic Log}
\begin{tabular}{ccccc}
\hline 
UT Date obs. & MJD & Tel.+Inst. & Slit & Grism/Grating\\ 
\hline 
2020-07-17 & 59,047.27 & P60+SEDM ~~~~~ & $-$ & $-$ \\
2020-07-17 & 59,047.32 & Keck II+DEIMOS & 1.0\arcsec & 600ZD ~~~~~\\
2020-07-18 & 59,048.37 & Shane+Kast ~~~ & 2.0\arcsec & B600 + R300\\
2020-07-19 & 59,049.50 & Shane+Kast ~~~ & 2.0\arcsec & B452 + R300\\
2020-07-20 & 59,050.30 & Keck I+LRIS ~~ & 1.0\arcsec & 400/3400 + 400/8500 \\
2020-07-22 & 59,052.26 & Keck II+ESI ~~ & 1.0\arcsec & 175 line/mm~~\\
2020-07-23 & 59,053.33 & GEMINI+GMOS ~~ & 1.0\arcsec & B600 + R400\\
2020-07-24 & 59,054.30 & Keck I+LRIS ~~ & 1.0\arcsec & 400/3400 + 400/8500 \\
2020-07-25 & 59,055.00 & NOT+ALFOSC ~~~ & 1.0\arcsec & Grism \#4 \\
2020-07-25 & 59,055.26 & Keck I+LRIS ~~ & 1.0\arcsec & 400/3400 + 400/8500 \\
2020-07-28 & 59,058.26 & Shane+Kast ~~~ & 2.0\arcsec & B600 + R300\\
2020-07-28 & 59,058.27 & Keck I+MOSFIRE & 1.0\arcsec & $Y + J + H + K$\\
2020-07-30 & 59,060.00 & Shane+Kast ~~~ & 2.0\arcsec & B830 + R1200\\
2020-08-01 & 59,062.26 & LCO-2\,m+FLOYDS  & 1.6\arcsec & 235~line/mm \\
2020-08-04 & 59,065.26 & LCO-2\,m+FLOYDS  & 1.6\arcsec & 235~line/mm \\
2020-08-09 & 59,070.33 & Shane+Kast ~~~ & 2.0\arcsec & B452 + R300\\
2020-08-10 & 59,071.25 & LCO-2\,m+FLOYDS  & 1.6\arcsec & 235~line/mm \\
2020-08-11 & 59,072.19 & Shane+Kast ~~~ & 2.0\arcsec & B600 + R300\\
2020-08-22 & 59,083.28 & LCO-2\,m+FLOYDS  & 1.6\arcsec & 235~line/mm \\
2020-08-25 & 59,086.18 & MMT+Binospec ~ & 1.0\arcsec & 270~line/mm \\
2020-09-07 & 59,099.23 & Shane+Kast ~~~ & 2.0\arcsec & B452 + R300\\
2020-09-17 & 59,109.53 & Keck I+LRIS ~~ & 1.0\arcsec & 400/3400 + 400/8500 \\

\hline 
\label{fig: spec_log}
\end{tabular}
\end{table*}

\begin{table}
\centering
    \caption{Radio Observations of \sn{}.}
    \label{Tab:radio}
    \begin{tabular}{cccccc}
    \hline
    \hline
  Start Date & Frequency & Bandwidth & $3\sigma$ Upper Limits & Telescope \\
 (MJD) & (GHz) & (GHz) & ($\mu$Jy/beam) & \\
    \hline
59,082.8351 & 10.00 & 4.096 & $\leq19$ & VLA\\
59,174.5161 & 10.00 & 4.096 & $\leq12$ & VLA\\
59352.0790 & 10.00 & 4.096 & $\leq30$ & VLA\\
\hline
\label{tbl:radio}
\end{tabular}
\end{table}

\end{document}